\documentclass[traditabstract]{aa}
\usepackage{graphicx}
\usepackage{enumerate}
\usepackage{epsfig}
\usepackage{color}
\usepackage{txfonts}
\usepackage{natbib}
\usepackage{float}
\usepackage{threeparttable}
\bibpunct{(}{)}{;}{a}{}{,} 

\begin{document}

\title{A Milky Way with a massive, centrally concentrated thick disc: \\ new Galactic mass models for orbit computations}

\titlerunning{A Milky Way with a massive, centrally concentrated thick disc}

\author{E. Pouliasis\inst{1,2,3} \and P. Di Matteo\inst{1} \and M. Haywood \inst{1}}

\authorrunning{E. Pouliasis et al.}

\institute{GEPI, Observatoire de Paris, PSL Research University, CNRS,
Univ Paris Diderot, Sorbonne Paris Cit\'e, Place Jules Janssen, 92195
Meudon, France\\
\email{ektoras.pouliasis@obspm.fr}
\and IAASARS, National Observatory of Athens, 15236 Penteli, Greece
\and Department of Astrophysics, Astronomy \& Mechanics, Faculty of Physics, University of Athens, 15783 Athens, Greece
}
\date{Accepted, Received}

\abstract{In this work, two new axisymmetric models for the Galactic mass distribution are presented. Motivated by recent results, these two models include the contribution of a stellar thin disc and of a thick disc, as massive as the thin counterpart but with a shorter scale-length.   
Both models satisfy a number of observational constraints: stellar densities at the solar vicinity, thin and thick disc scale lengths and heights, rotation curve(s), and the absolute value of the perpendicular force Kz as a function of distance to the Galactic centre. We numerically integrate into these new models the motion of all Galactic globular clusters for which distances, proper motions, and radial velocities are available, and the orbits of about one thousand stars in the solar vicinity. The retrieved orbital characteristics are compared to those obtained by integrating the clusters and stellar orbits in pure thin disc models.
 We find that, due to the possible presence of a thick disc, the computed orbital parameters 
of disc stars can vary by as much as 30-40\%. We also show that the systematic uncertainties that affect the rotation curve still 
plague computed orbital parameters of globular clusters by similar amounts.
}


\maketitle

\section{Introduction}

The study of the orbits of stars and stellar systems,  like globular and open clusters in the Milky Way, is essential to understand the properties of the different Galactic stellar populations (thin and thick discs, stellar halo) and their mode of formation. To integrate stellar orbits, realistic models of the mass distribution in the Milky Way are needed. Because of the facility of implementation and usage, it is also highly desirable  that these models are fully analytical.  Several mass models of this type have been developed in the last two decades. Among them, the most widely used is certainly the axisymmetric model proposed by \citet{allen91}, that consists of a stellar thin disc, a central bulge, whose mass is about 15\% of that of the disc, and a dark matter halo which guarantees a nearly flat rotation curve at large radii. This model has been recently revised by \citet{irrgang13}, who used the most recent observational constraints to update masses and mass distributions of the thin stellar disc, bulge, and dark matter component. Asymmetries, like the central stellar bar and spiral arms, can also be added to an axisymmetric model, and their related effect on stellar orbits can be quantified, as done by \citet{pichardo03, pichardo04}.  In all cases, these models are all based on the assumption that most ($> 80\%$) of the stellar mass of the Galaxy is redistributed in a thin stellar disc, and that the remaining fraction is contained in a centrally concentrated bulge. In particular, although the discovery of the presence of an additional stellar component - the thick disc - in the Milky Way dates back to more than 30 years ago (see \citet{yoshii79, gilmore83, reid93}), this component has been neglected in all the previously cited mass models. This is attributed to the fact that it has long been considered that the thick disc is a minor component of the Galaxy, contributing by up to 10-20\% to the total stellar budget. These are the fractions found by assuming the same scale length for the two populations, standard values for the scale height, and local densities, with the thin disc having a scale height equal to 250pc, and containing 90-98 \% of the total local stellar mass, 
and the thick disc having a scale height between  600 and 1200 pc and containing 2-10\% of the total local stellar mass. Most recently, only the models by \citet{smith15} and \citet{barros16} have added the presence of a thick stellar disc to the Galactic potential. However, their modelled thick disc contributes still only marginally to the total stellar budget, the thick-to-thin disc mass ratio adopted in these models being between 10\% and 20\%.  \\

In the last years, however, a number of pieces of evidence have accumulated that may lead to a revision of the aforementioned  picture, suggesting that the mass budget of the Galactic stellar components  may be significantly different from what has been previously thought. The classical bulge appears to be very limited or not existent \citep{shen10, kunder12, dimatteo14, dimatteo15, dimatteo16, kunder16} and the $\alpha$-enhanced thick disc appears to be as massive as the thin disc \citep{snaith14, snaith15}. The latter was obtained \citep[see][]{snaith14, snaith15} by reconstructing the star formation history of the Milky Way disc by fitting a chemical evolution model to the age-$\rm [Si/Fe]$ relation found on solar vicinity data by \citet{haywood13}. The fact that the $\alpha$-enhanced stellar thick disc may be massive finds also an independent confirmation in the revised estimates of thick disc radial density. According to \citet{bensby11, bovy12a}, and \citet{bovy15}, indeed, $\alpha-$abundant thick disc stars have a short scale length of 1.8-2~kpc, about a factor of two smaller than what was suggested by previous works, mainly based on colour selections \citep[see, for example,][]{juric08}. 
With such a short scale length, that turns out to be also a  factor of two smaller than that of the corresponding thin disc \citep{bovy12a}, it appears that the thick-to-thin disc mass ratio at the solar vicinity is not representative of the global thick-to-thin disc ratio and that indeed most of the $\alpha$-enhanced thick disc is present in the inner regions of the Galaxy.  A confirmation of the dominant role of the thick disc in the inner Galaxy can be found also in the Apache Point Observatory Galactic Evolution Experiment (APOGEE) data, which are now revealing the presence of a substantial amount of  stars with thick disc metallicities in the inner disc, that is, inside the solar circle \citep[see for example the results by][]{anders14, hayden15}.

Motivated by these recent discoveries, in this work we propose two new Galactic mass models that include a massive stellar thick disc with properties similar to those observed for the Galactic $\alpha-$enhanced thick disc. Mostly because of the current uncertainties still affecting our knowledge of the Galactic rotation curve, these two models differ in the presence (or not) of a classical bulge in the inner Galactic regions, and in the mass of the dark matter halo at large radii. After presenting the characteristics of these models (Sect.~\ref{models}), we discuss how they fit the most recent observational constraints (Sect.~\ref{data}). We then integrate the orbits of a sample of stars at the solar vicinity obtained by \citep{adibek12} for which proper motions and parallaxes are available from the Hipparcos mission,  and all the Galactic globular clusters for which positions, radial velocities, and proper motions are available (Casetti-Dinescu main catalogue of Galactic globular clusters).  As a reference and comparison, we also integrated all the orbits of stars and globular clusters into the model of \citet{allen91}, to quantify the differences found when a massive thick disc component is included (see Sects.~\ref{stellarorbits} and ~\ref{clusterorbits}), also taking into account 
the remaining significant uncertainties in the rotation curve of the Milky Way. Finally, the conclusions of our work are presented in Sect.~\ref{conclusions}.

\section{Galaxy models with a massive thick disc}\label{models}

Before entering into the description of our models, we would like to emphasize that 
our aim is not to provide a best fit mass model of the Milky Way, but to quantify the 
difference that the adoption of a massive thick disc makes compared to a widely used model
(in this case, the Allen \& Santillan (1991) model) without a thick disc. 

Among the main observational constraints that any Galactic mass model needs to reproduce, there is of course the rotation curve of the Galaxy. As we will see in the following, our knowledge of the Galactic rotation curve still suffers from severe uncertainties : the profile in the inner 5 kpc from the Galactic centre, as well as the value of the rotation curve at the solar radius and beyond, can vary considerably from one study to another \citep{burton78, bovy12b, sofue12, bovy13, reid14}. Because some of the differences among the rotation curves available in the literature are difficult to reconcile, particularly in the inner regions of the Galaxy where a massive centrally concentrated thick disc has its strongest impact, in this paper we choose to develop two different mass models :
\begin{enumerate}

\item a model (hereafter Model I) that consists of a thin disc, a thick disc (as massive as the thin),  a central bulge,  and a dark matter halo, and that - as we will see - generates a rotation curve compatible with the estimates of \citet{sofue12};
\item a model (hereafter Model II) which also contains a thick disc as massive as the thin counterpart, but which lacks a central bulge, and whose rotation curve is compatible with that obtained by \citet{reid14} using maser sources.
\end{enumerate}
In both models, as we will see, the thick disc scale length is about a factor of two shorter than that of the thin disc, in agreement with the results by \citet{bovy12a}. 
The choice of presenting two mass models for the mass distribution of our Galaxy is mainly dictated by two reasons. First, the need to add a central bulge to the global gravitational potential to reproduce the rotation curve in the inner kpcs of the Milky Way strongly depends on the observational data with which one compares the theoretical curve: to reproduce the rise observed in the inner kpcs \citep[see the observational data adopted by][]{caldwell81},  \citet{allen91}  introduced a central mass concentration, whose mass is about 15\% of the disc mass. However, the central rise observed in the rotation of the molecular gas in the inner Galaxy \citep[for more recent estimates see, for example, ][]{sofue12} may be an effect of non circular motions generated by large scale asymmetries like the bar, as has been shown  recently by \citet{chemin15}. Moreover, this feature is not reported in all the observational studies \citep[see, for example,][]{reid14}. Secondly, there is growing evidence that the mass of any classical bulge, if present in the Milky Way, must be small \citep{shen10, kunder12, dimatteo14, dimatteo15, kunder16}. For these reasons, we prefer to present a second model, our Model II, which does not include any spherical central component, and which is still compatible with the rotation curve of the Galaxy, as given by \citet{reid14}. Because it has been widely used in the last decades, and due to the facility of its implementation, we explicitly aim at generating Galactic models similar to the \citet{allen91} model, so to make any implementation of these new models, and any comparison with \citet{allen91}, straightforward.
As for the model proposed by \citet{allen91}, Models I and II are axisymmetric and time-independent, and do not include stellar asymmetries such as a bar or spiral arms. No truncation is assumed for the discs, while the halo is truncated at 100 kpc, in agreement with the choice of \citet{allen91}. As we describe in the following section, the analytic forms for the discs, halo, and bulge potentials are the same as those adopted by \citet{allen91}. To allow an easy comparison with the \citet{allen91} model, in the following we will make use of the same system of units adopted by these authors:  the potential is given in units of $\rm{100 km^2/s^2}$, lengths are in kpc,  masses in units of $\rm{2.32x10^7 M_{\odot}}$, time in units of $\rm{0.1Gyr}$, velocities in units of $\rm{10 km s^{-1}}$ , and the vertical force in units of $\rm{10^{-9} cm s^2}$. In these units, the gravitational constant G is equal to 1 and the mass volume density is in units of  $\rm{2.32x10^7 M_{\odot}/kpc^3}$.\\

Finally, a few words on the functional form of the stellar disc(s) adopted in this paper. We maintained the Miyamoto-Nagai density distribution adopted by \citet{allen91} to model both the thin and thick discs. As discussed in the following lines, the characteristic scale lengths and heights were chosen so that the resultant stellar disc, as well as its constituent -- the thin and the thick -- can be fitted by exponential profiles with scale heights and lengths similar to those found by \citet{bovy12a} in the radial range where these  have been measured to a useful precision with current spectroscopic surveys -- typically a few kpc from the Sun. We are aware that the Miyamoto-Nagai density profile cannot be fitted with a single exponential over the whole radial range, but in the case of the 
Milky Way, there is no evidence that the disc should be represented by a single exponential over its whole extent as done, for example, by \citet{dehnen98}. In fact there may be some evidence to the contrary: the outer disc ($R > 8-10$~kpc) has a longer scale length than the inner disc at $R < 8-10$~kpc \citep[see][]{bovy12a, golubov13}, and it may be difficult for a single exponential to represent a good fit over the whole radial extent. More generally, given the complexity of the radial distribution 
of the different ``mono-abundance" populations as illustrated by the analysis of recent surveys such as the Sloan Extension for Galactic Understanding and Exploration (SEGUE) or APOGEE \citep[see][]{bovy15}, 
it is yet to be demonstrated that a single exponential can give a satisfactory fit on more than a few scale lengths, or at least 
a better fit than a Miyamoto-Nagai density profile.

In the following part of this section, for the sake of completeness, we firstly recall the main features of the \citet{allen91} model (Sect.~\ref{allenpot}, hereafter Model A\&S), then we present Models I and II (Sects.~\ref{modIpot} and \ref{modIIpot}, respectively), and finally we compare the predictions of these two models to observational data (Sect. ~\ref{data}).

\begin{table}
\centering
  \begin{tabular}{  l| c | c | c }
   Parameters & Model I  & Model II & Model A\&S \\ \hline \hline
    $\rm M_{bulge}$ & 460.0 & - & 606.0 \\
    $\rm M_{thin}$ & 1700.0 & 1600.0 & 3690.0 \\
    $\rm M_{thick}$ & 1700.0 & 1700.0 & - \\
    $\rm M_{halo}$ & 6000.0 & 9000.0 & 4615.0 \\
    $\rm a_{thin}$ & 5.3000 & 4.8000 & 5.3178 \\
    $\rm a_{thick}$ & 2.6 & 2.0 & - \\
    $\rm a_{halo}$ & 14.0 & 14.0 & 12.0 \\
    $\rm b_{bulge}$ & 0.3000 & - & 0.3873 \\
    $\rm b_{thin}$ & 0.25 & 0.25 & 0.25 \\
    $\rm b_{thick}$ & 0.8 & 0.8 & - 
  \end{tabular} 
\caption{Adopted parameters of the two new Galactic mass models. Masses are in units of $2.32\times10^7M_{\odot}$, distances in units of kpc, following \citet{allen91}. For comparison, the parameters of the \citet{allen91} model are also given. 
Both models I\&II are designed to have  stellar discs with masses similar to the \citet{allen91} stellar disc, but with half of this total mass contained in a thick disc.
\label{parameters}}
\end{table}

\subsection{Model A\&S}\label{allenpot}
The model of A\&S consists of the sum of an axisymmetric potential,  $\Phi_{thin}(R,z)$, for the stellar disc of Miyamoto-Nagai type  \citep{miyamoto}, a Plummer potential \citep{binney},  $\Phi_{bulge}(r)$, for the central bulge, and a spherical potential, $\Phi_{halo}(r)$, truncated at R=100 kpc,  for the dark matter halo, that is :
\begin{equation}
\Phi_{tot}(R,z)=\Phi_{thin}(R,z)+\Phi_{bulge}(r)+\Phi_{halo}(r),
\end{equation}
where $r=\sqrt{R^2+z^2}$, being $R$ the in-plane distance and $z$ the height above the plane. \\
The analytical forms of these potentials are, respectively :

\begin{equation}\label{disc}
\Phi_{thin}(R,z)=\frac{-GM_{thin}}{(R^2+[a_{thin}+\sqrt{z^2+b_{thin}^2}]^2)^{1/2}}
\end{equation}

\begin{equation}\label{bulge}
\Phi_{bulge}(r)=\frac{-GM_{bulge}}{(r^2+b_{bulge}^2)^{1/2}}
\end{equation}

\begin{equation}\label{halo}
\begin{split}
\Phi_{halo}(r)=&\frac{-GM_{halo}}{r}-\frac{M_{halo}}{1.02a_{halo}}\\
& \Bigg[\frac{-1.02}{1+(\frac{r}{a_{halo}})^{1.02}}+ln{(1+(\frac{r}{a_{halo}})^{1.02})}\Bigg]_R^{100}
\end{split}
\end{equation}

, where $M_{thin}, M_{bulge}$ and $M_{halo}$, $a_{thin}, b_{thin}, b_{bulge}, a_{halo}$ are the disc, bulge, and halo constants, respectively (see Table 1 in \citet{allen91} and Table  \ref{parameters} of this paper, where these parameters are recalled).

The corresponding densities, related to these potentials by means of the Poisson equation, are: 
\begin{equation}
\begin{split}
\rho_{thin}(R,z)=&\frac{b_{thin}^2 M_{thin}}{4\pi}\times\\
&\frac{(R^2 a_{thin}+3(z^2+b_{thin}^2)^{1/2})(a_{thin}+(z^2+b_{thin}^2)^{1/2})^2}{({{R^2}+[a_{thin}+(z^2+b_{thin}^2)^{1/2}]^2})^{5/2}(z^2+b_{thin}^2)^{3/2}}
\end{split}
\end{equation}

\begin{equation}
\rho_{bulge}(r)=\frac{3b_{bulge}^2 M_{bulge}}{4\pi(r^2+b_{bulge}^2)^{5/2}}
\end{equation}

\begin{equation}
\rho_{halo}(r)=\frac{M_{halo}}{4\pi a_{halo}r^2}\Big(\frac{r}{a_{halo}}\Big)^{1.02}\Bigg[\frac{2.02+(r/a_{halo})^{1.02}}{(1+(r/a_{halo})^{1.02})^2}\Bigg].
\end{equation}   
                                                                                                                         
The total mass of the system is $9\times10^{11} M_{\odot}$, considering that the halo is truncated at $100$~kpc.


\begin{figure}[h!]
\includegraphics[trim=0.cm 1.55cm 0cm 0cm,width=0.5\textwidth]{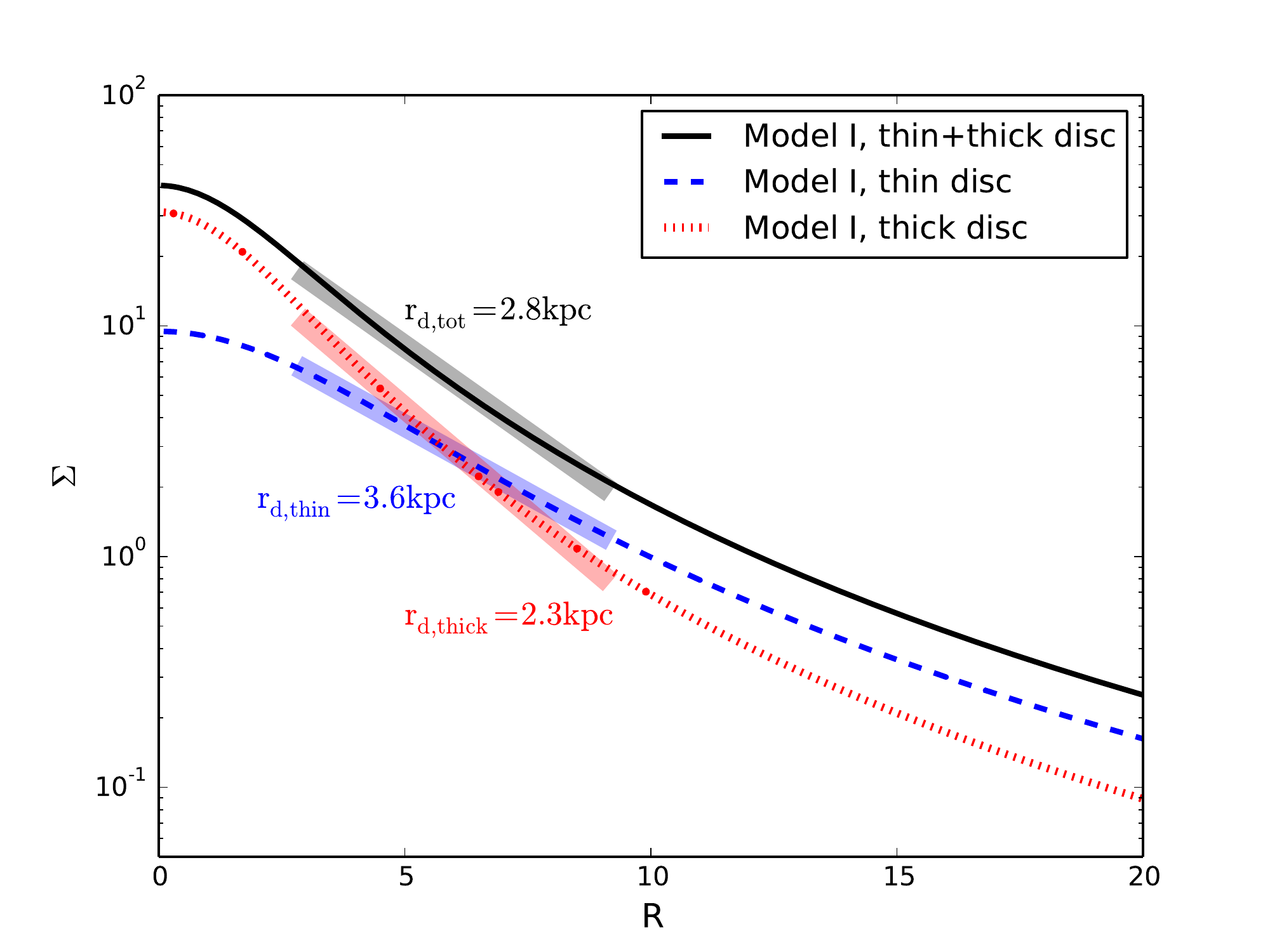}
\includegraphics[trim=0.cm 5.cm 0cm 5.5cm, clip=true,width=0.5\textwidth]{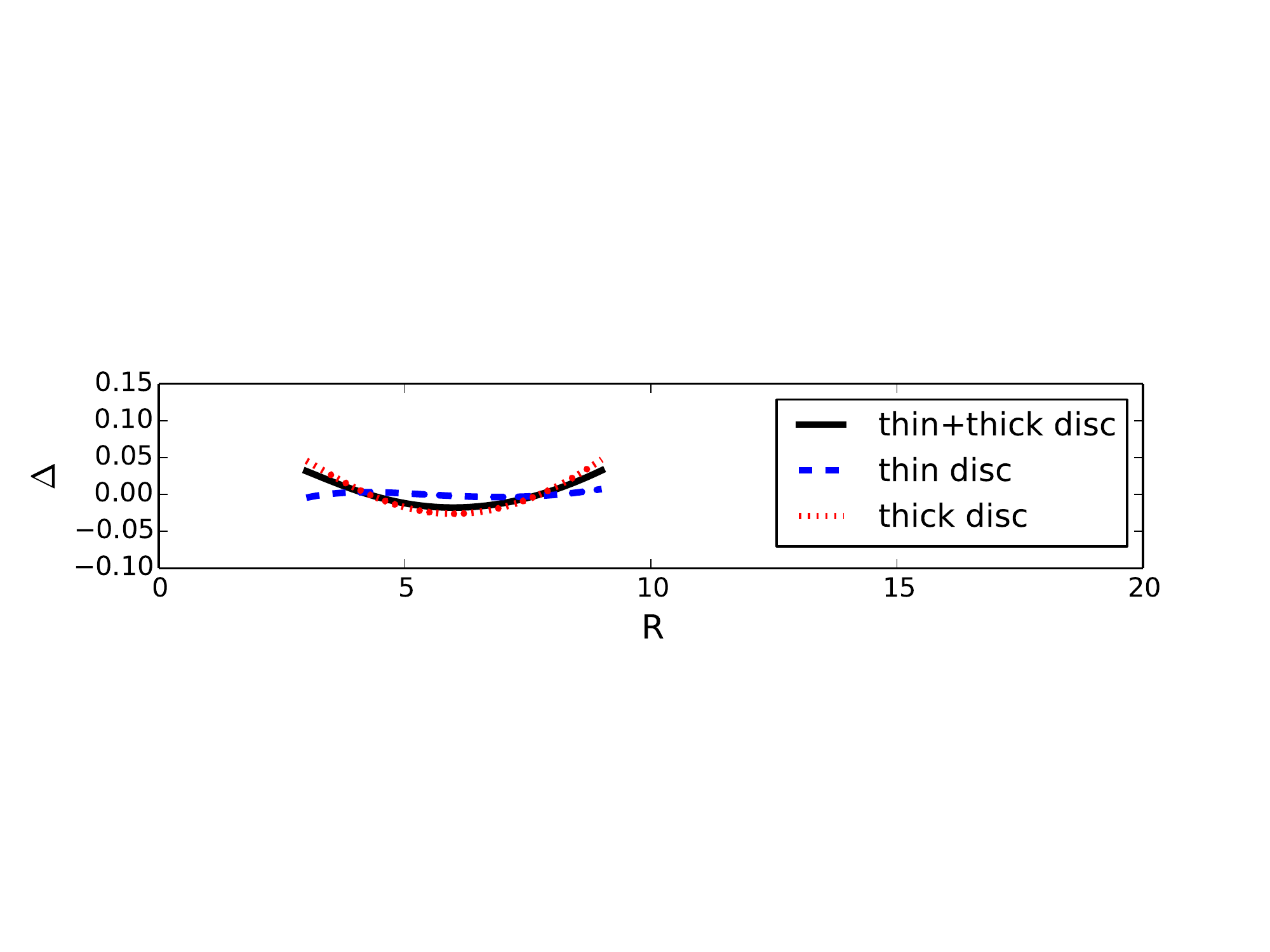}
\includegraphics[trim=0.cm 1.42cm 0cm 0cm,width=0.5\textwidth]{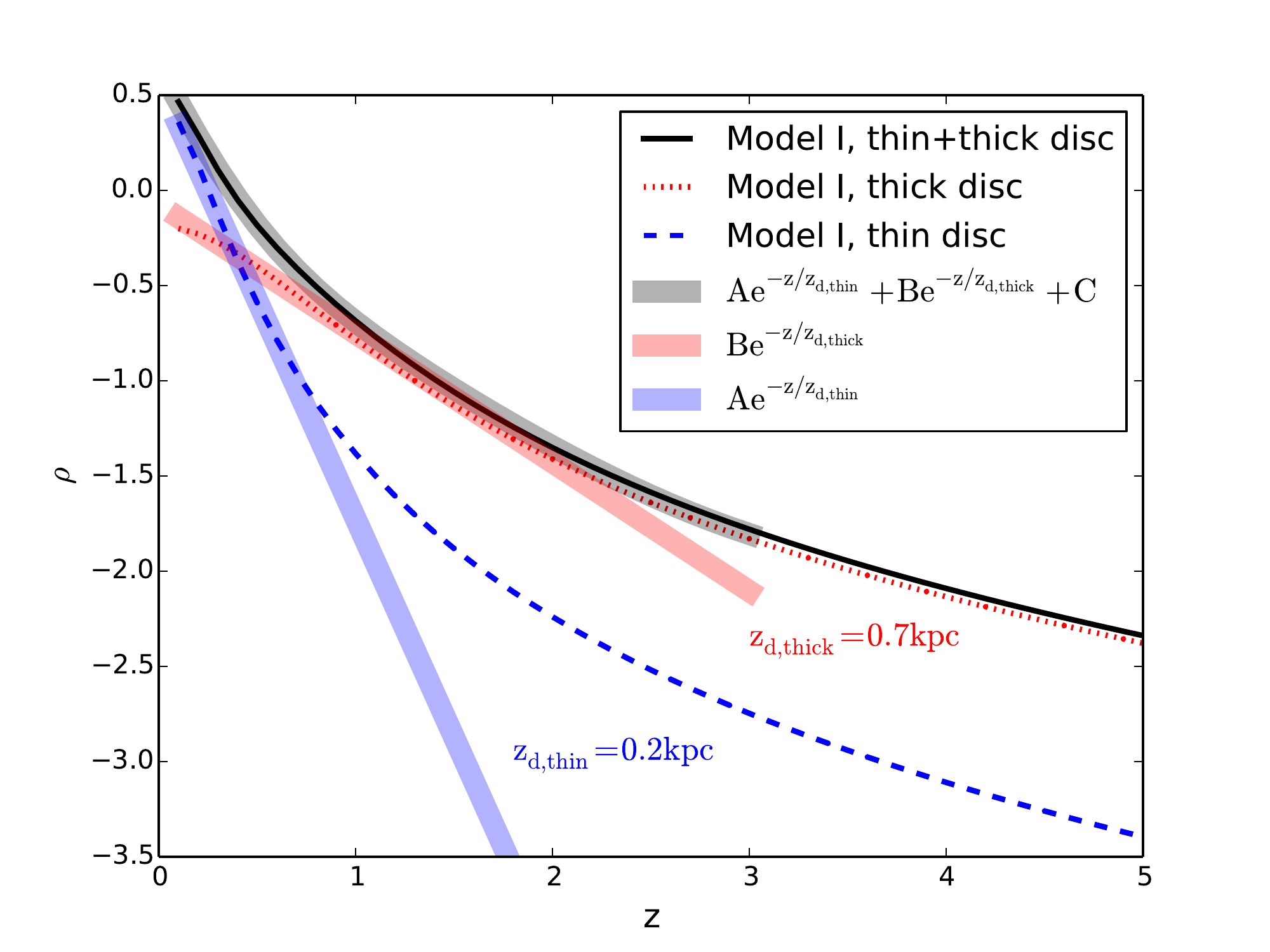}
\includegraphics[trim=0.cm 4cm 0cm 5.cm, clip=true,width=0.5\textwidth]{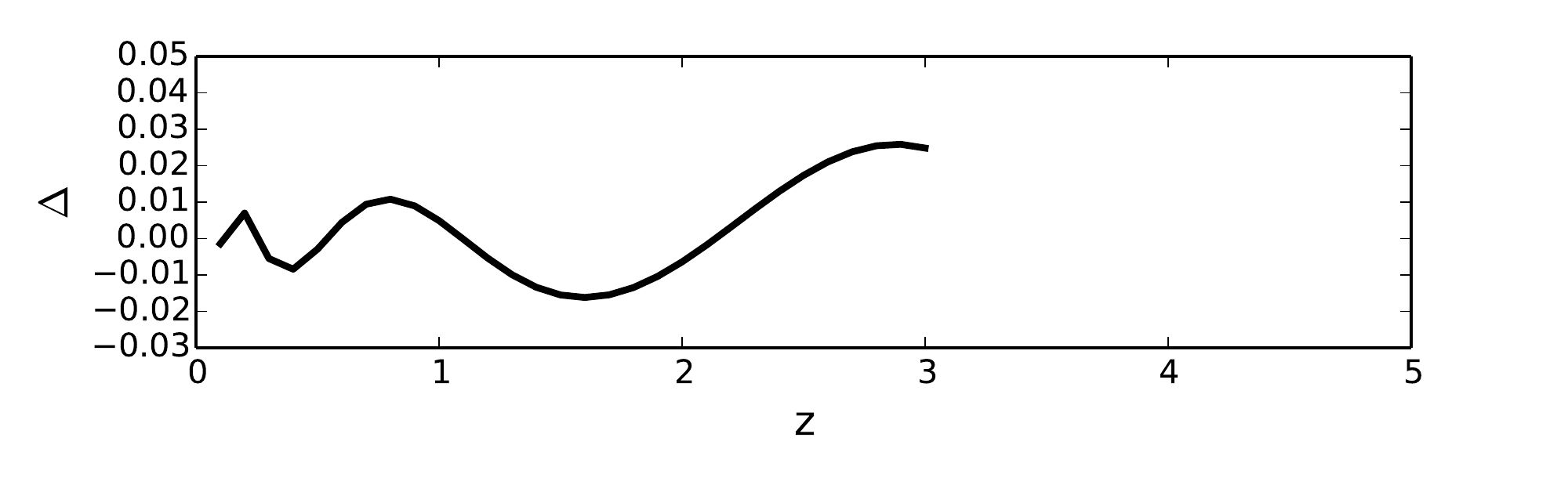}
\caption{From top to bottom, first panel: Total radial surface density (solid black curve), thin disc surface density (dashed blue curve), and thick disc surface density (dotted red curve) of Model I. For each curve, the shaded area shows the exponential fit to the curve, in the distance range of 3--9~kpc. The corresponding exponential disc scale lengths of the total, thin, and thick discs are reported in the plot. Densities and distances are in model units. \emph{Second panel:} Residuals of the exponential fit  to the density curves in the 3--9~kpc radial range.  \emph{Third panel:} Total vertical volume density (solid black curve), thin disc volume density (dashed blue curve), and thick disc volume density (dotted red curve) of Model I. The black shaded line shows the two-exponential fit to the curve, as given in the legend, in the $z$ range of 0--3~kpc. The blue and red shaded lines show the contribution of the exponential thin and thick discs, respectively, to the total vertical density. Bottom panel: Residuals of the two exponentials fit  to the total vertical density  in the 0--3~kpc $z-$range. The residuals in this and in the second panel are defined as the difference between the densities and the fit functions, both expressed in logarithmic scale.\label{densityMI}}
\end{figure}

\subsection{Model I}\label{modIpot}

\par The gravitational potential of Model I is the sum of a spherical bulge, $\Phi_{bulge}(r)$, and a spherical halo, $\Phi_{halo}(r)$, with the same functional form as Model A\&S (Eqs.~\ref{bulge} and \ref{halo}, respectively), and two stellar discs, whose potentials are indicated by $\Phi_{thin}(R,z)$ and $\Phi_{thick}(R,z)$, respectively, and which correspond to the thin and the thick disc components. The total potential in this model is given by : 
\begin{equation}
\Phi_{tot}(R,z)=\Phi_{bulge}(r)+\Phi_{thin}(R,z)+\Phi_{thick}(R,z)+\Phi_{halo}(r),
\end{equation}
where the functional form of $\Phi_{thin}(R,z)$ and $\Phi_{thick}(R,z)$ are still described by a Miyamoto \& Nagai profile (Eq.~\ref{disc}), with characteristic masses, scale lengths and scale heights given by $M_{thin}, M_{thick}$, $a_{thin}, a_{thick}$, and $b_{thin}, b_{thick}$ (see column 1 in Table \ref{parameters}).

\begin{figure}
\includegraphics[trim=0.cm 1.55cm 0cm 0cm,width=0.5\textwidth]{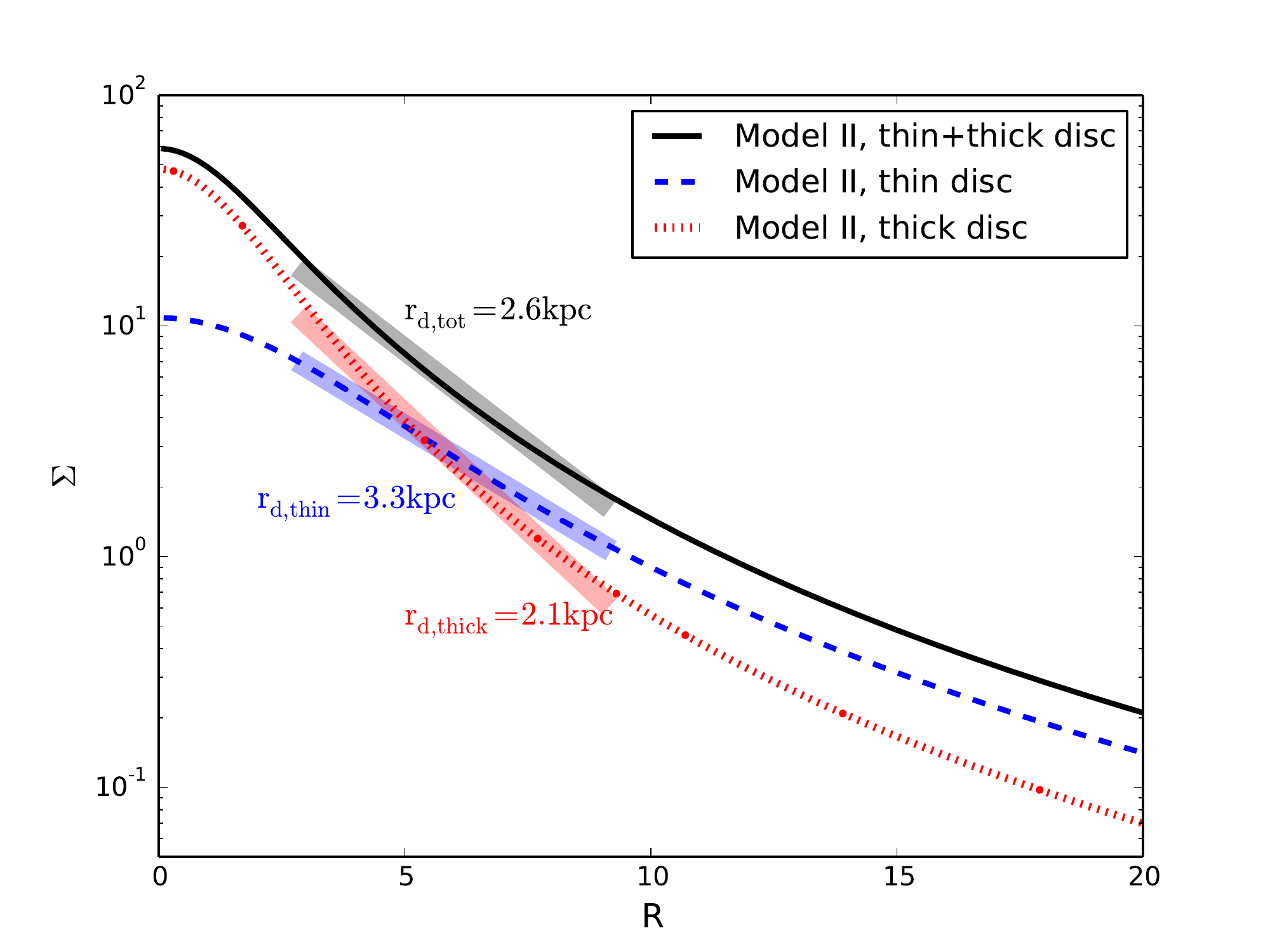}
\includegraphics[trim=0.cm 5.cm 0cm 5.5cm, clip=true,width=0.5\textwidth]{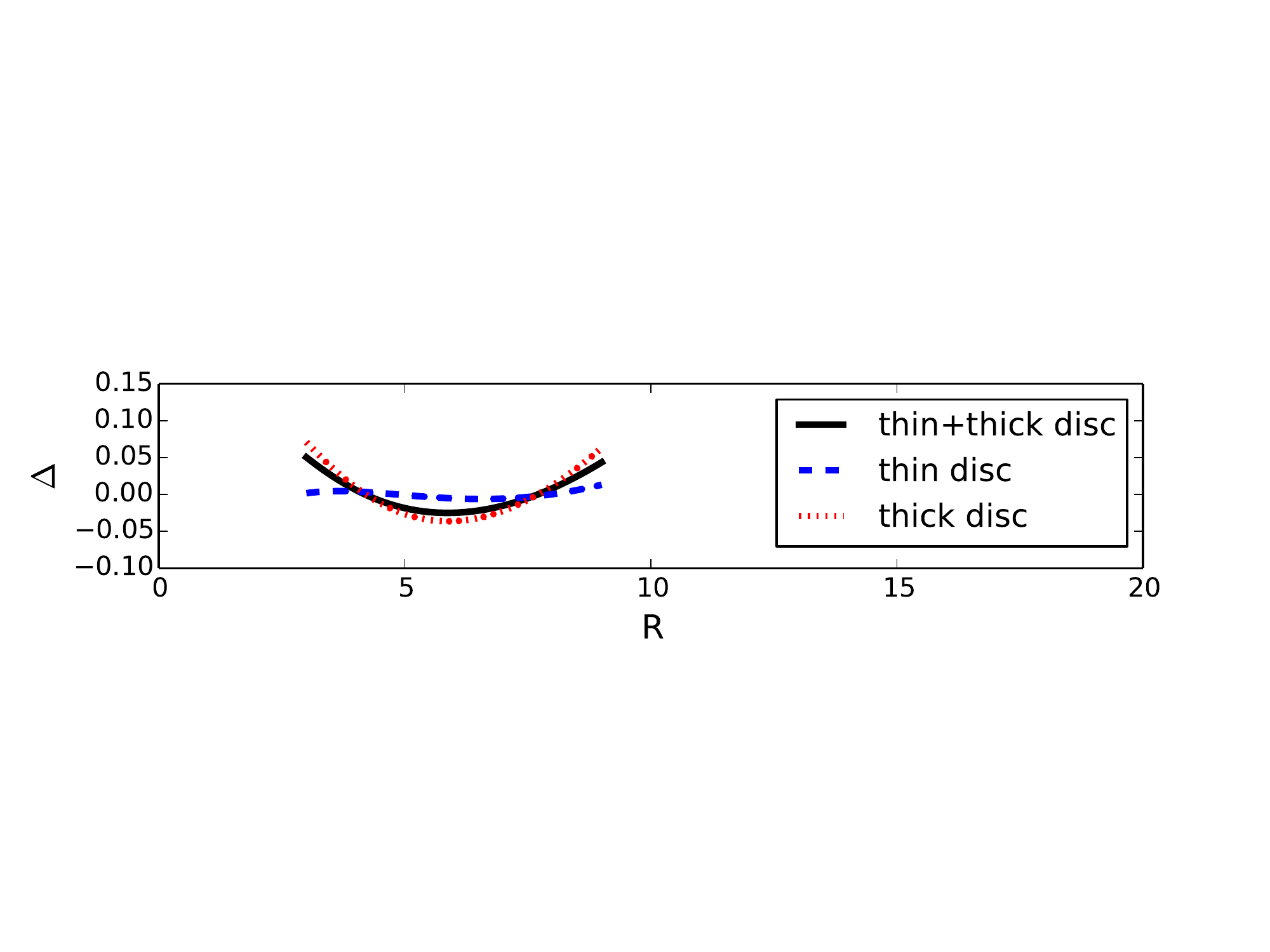}
\includegraphics[trim=0.cm 1.42cm 0cm 0cm,width=0.5\textwidth]{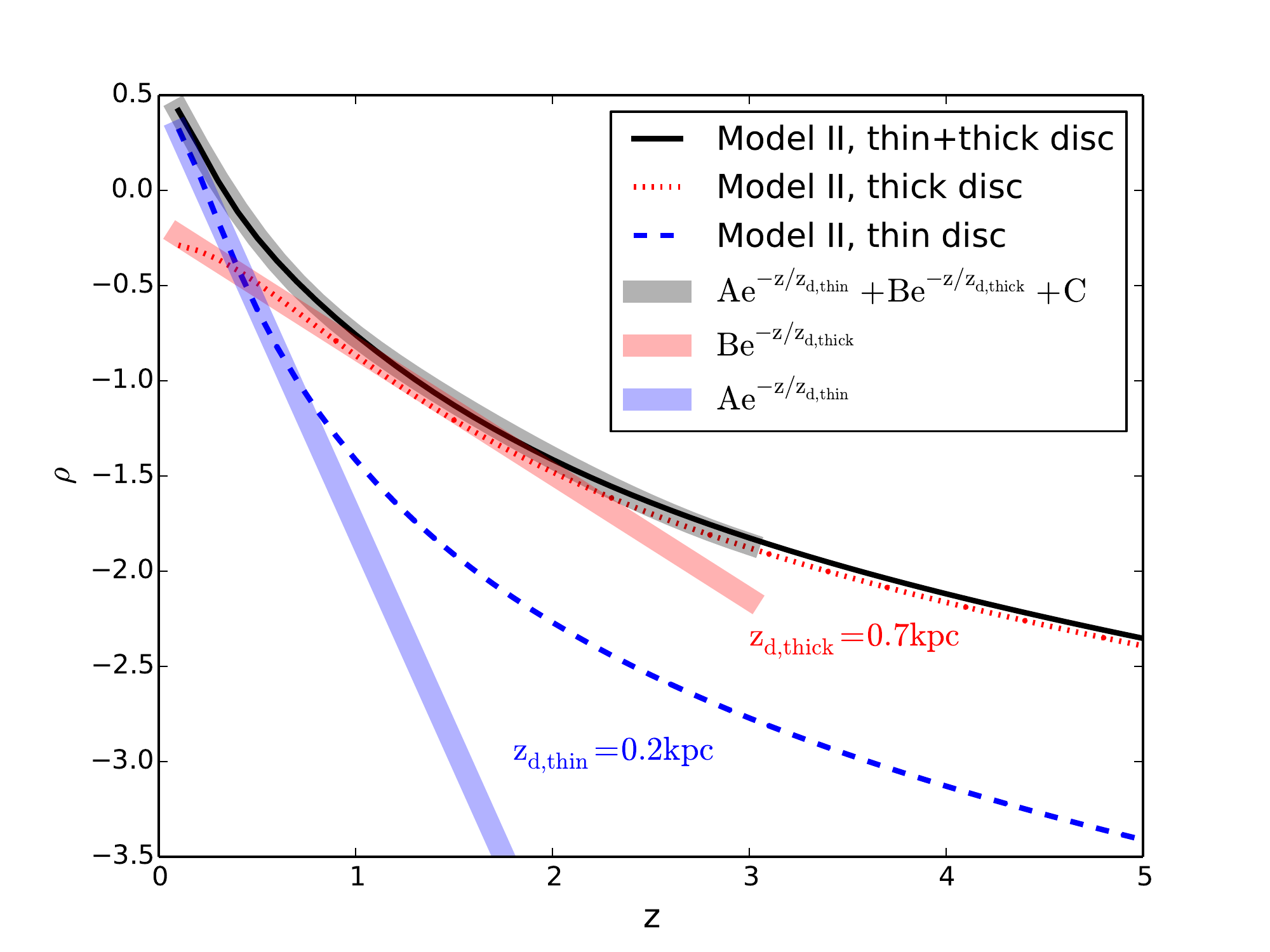}
\includegraphics[trim=0.cm 4.cm 0cm 5.cm, clip=true,width=0.5\textwidth]{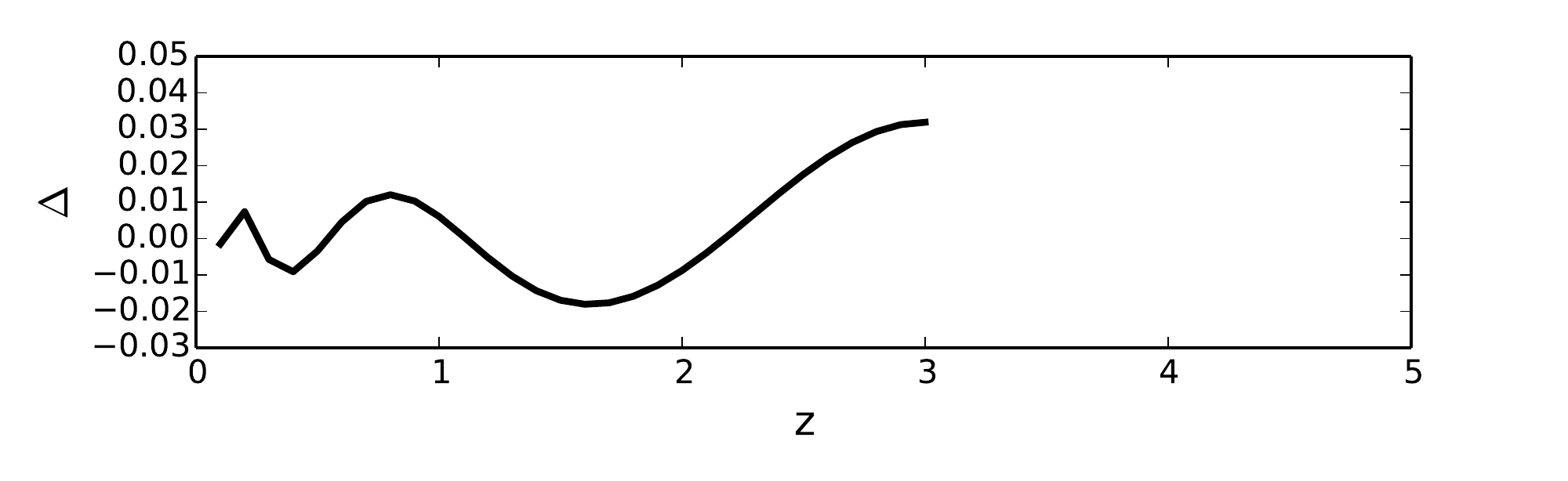}
\caption{Same as in Fig. 1, but for Model II. \label{densityMII}}
\end{figure}

\begin{figure}
\includegraphics[trim=0.cm 0cm 0cm 1cm, clip=true,width=0.5\textwidth]{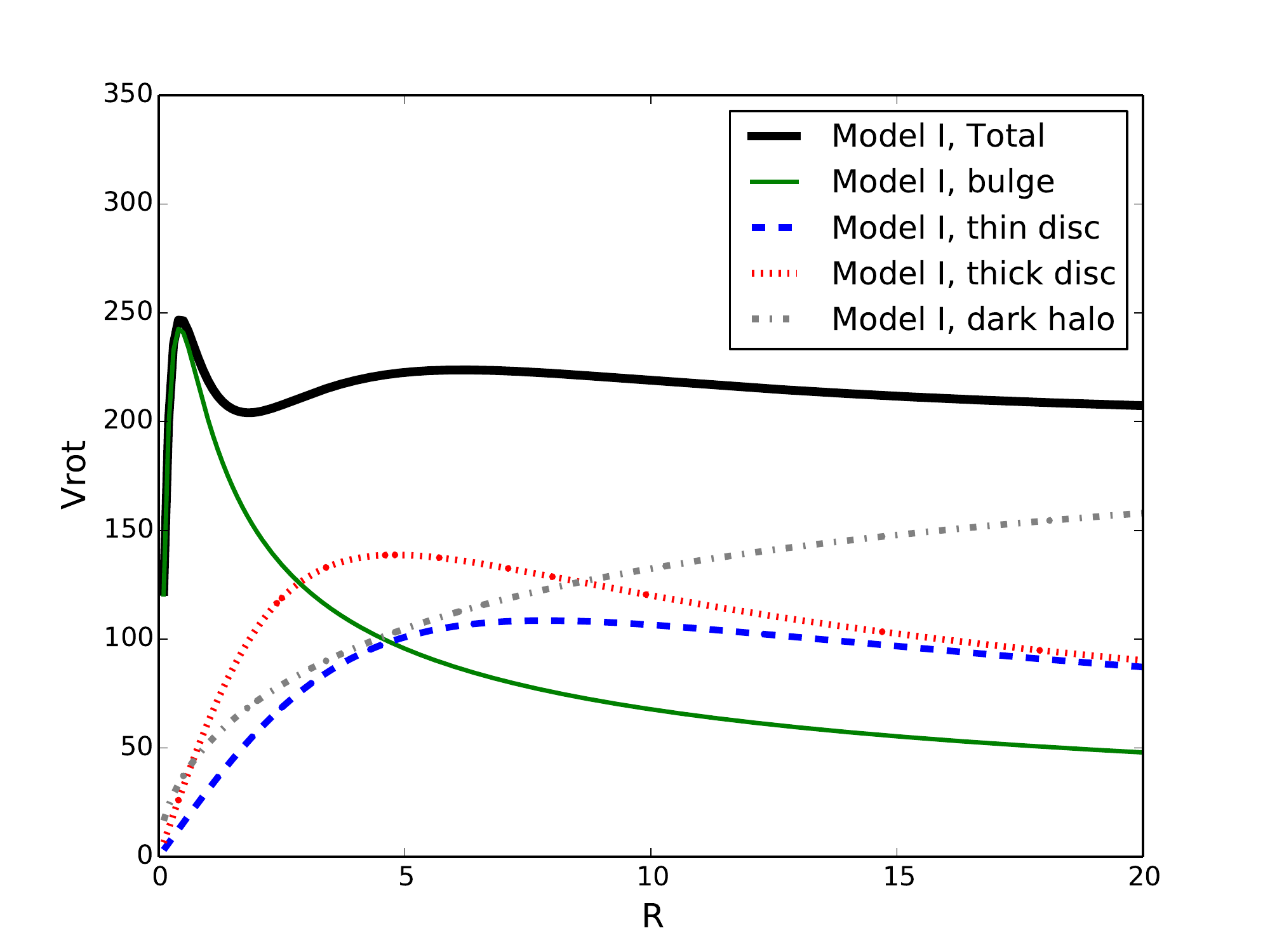}
\includegraphics[trim=0.cm 0cm 0cm 1cm, clip=true,width=0.5\textwidth]{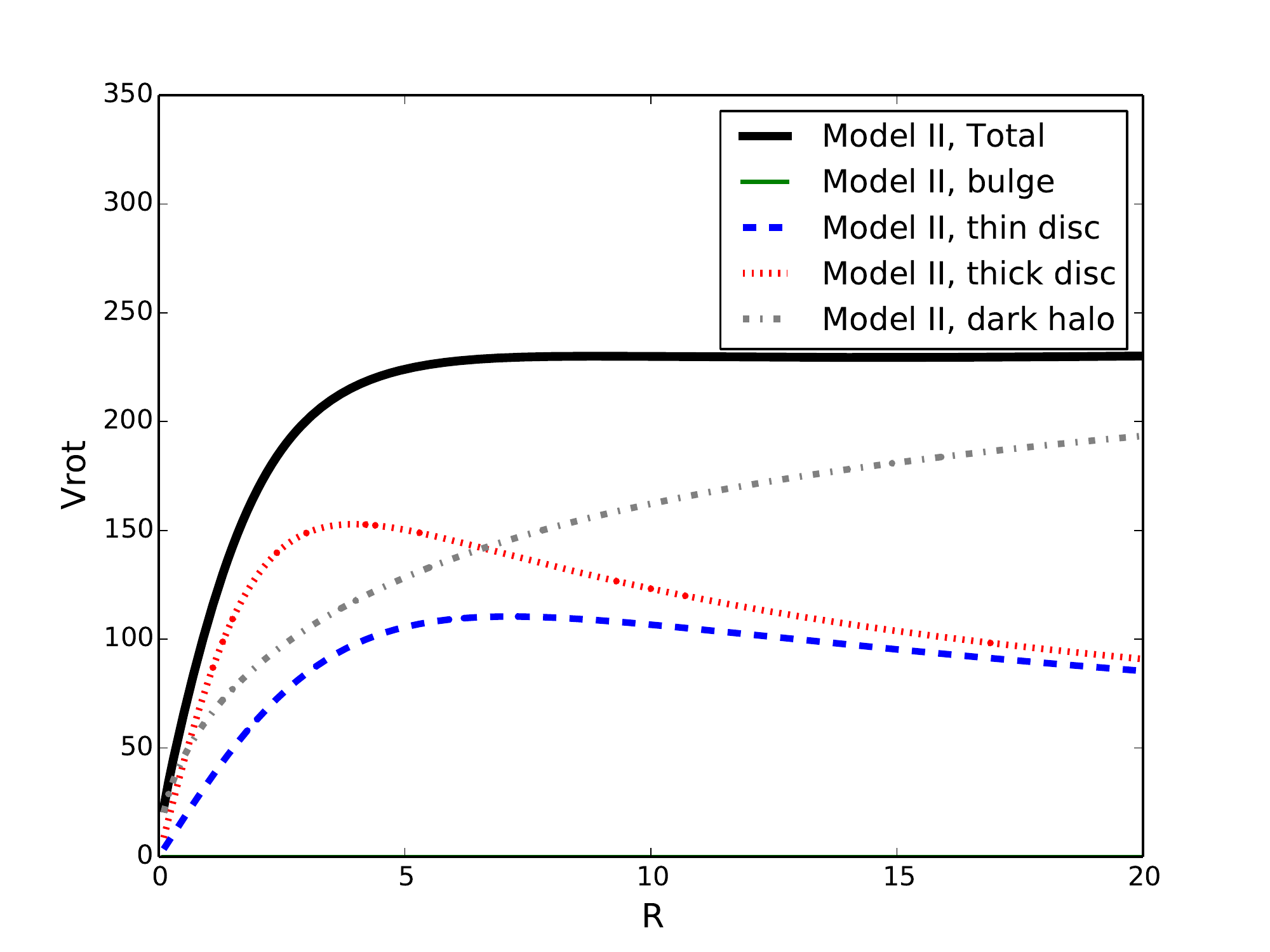}
\caption{Top panel: Total rotation curve of Model I (black line). The contribution to the total rotation curve of the bulge, thin disc, thick disc, and dark matter halo are indicated by the solid green,  dashed blue, dotted red, and dot-dashed grey curves, respectively. Velocities are in units of km/s, distances in kpc. Bottom panel: Same as in the top panel, but for Model II. \label{rotcurveMI-II}}
\end{figure}

In this model, the mass of the central bulge has been reduced with respect to the value adopted in Model A\&S, and the total stellar mass included in the two discs - the thin and the thick - is slightly lower than that of the single stellar disc in Model A\&S (see Table 1, column 3). The mass of the thin and the thick discs are both equal to 1700, in mass units. This implies a total stellar disc mass (at infinity) of $\rm 7.8\times10^{10} M_{\odot}$, in agreement with estimates of the Milky Way stellar disc mass \citep[see, for example,][]{kafle14}. The characteristic lengths of the discs have been chosen in such a way that their corresponding exponential scale lengths,  found by fitting the radial surface density profiles of the discs for $\rm 3~kpc\le R \le 9~kpc$, are compatible with those given by \citet{bensby11, bovy12a}, that is, $\sim$ 2~kpc for the thick disc and $\sim$ 3.6~kpc for the thin disc (see Fig.~\ref{densityMI}, top panels).
For the characteristic heights of the thin and thick discs we have adopted $b_{thin}=0.25$~kpc and $b_{thick}=0.8$~kpc, respectively.  The corresponding exponential scale heights, found by fitting the vertical density profile at the solar radius for $\rm 0\le z \le 3$~kpc, are $z_{d,thin}=0.2$~kpc and $z_{d,thick}=0.7$~kpc, respectively (see Fig.~\ref{densityMI}, bottom panels).
These values are well within the range of estimated parameters found in the literature (see, for example, \citet{bland16}).\\ Our modelled Miyamoto-Nagai discs can be well fitted by exponentials both in the radial and in the vertical direction, for the selected radial and vertical ranges, with the residuals $\Delta$ defined as the difference between the densities and the fit functions, both expressed in logarithmic scale, never exceeding 0.05 (again, see Fig.~\ref{densityMI}).  This corresponds to differences in densities of 10\% at most, well within the uncertainties in the current estimates.\\
A point that we want to emphasize is that the choice to represent the stellar disc by two discs, a thin and a thick, is probably a simplification with respect to the results  of \citet{bovy12c}, who found that the thick disc is not distinct from the thin disc, but rather the Galactic thick stellar disc is a continuum of decreasing scale heights with decreasing $\alpha-$abundances. In our models, we are discretizing, and representing by means of a unique component, a thick disc whose properties, and in particular the scale height, may vary more continuously with the chemistry. We consider this a first approximation before moving to more complex models. 

With this choice of parameters, at the solar vicinity, the thin and thick disc volume densities are respectively 6.63$\rm \times 10^7M_{\odot}/kpc^3$ and 1.50$\rm \times 10^7M_{\odot}/kpc^3$ , and their corresponding surface densities 3.32$\rm \times 10^7M_{\odot}/kpc^2$ and 2.51$\rm \times 10^7M_{\odot}/kpc^2$. The corresponding total volume and surface densities are respectively 0.08$\rm  M_{\odot}/pc^3$ and 58.3$\rm M_{\odot}/pc^2$, in good or reasonable agreement with the recent estimates of the baryonic (gas and stars) density from \citet{mckee15} (0.084$\pm$0.04$\rm  M_{\odot}/pc^3$ and 47$\pm$3$\rm M_{\odot}/pc^2$). These values are also in agreement with the baryonic (gas and stars) volume and surface densities at the solar vicinity \citep[e.g.][their Table~2]{flynn06}. Because our models do not contain any gaseous disc, we take into account its mass and corresponding density in the stellar discs. Figure~\ref{rotcurveMI-II} (top panel) shows the total rotation curve of Model I, together with the contribution of all the different components to it.

\subsection{Model II}\label{modIIpot}

Model II consists of a spherical dark matter halo $\Phi_{halo}(r)$, with the same functional form adopted in the \citet{allen91} model (Eq.~\ref{halo}), but more massive, and two disc components (a thin and a thick disc) both described by Miyamoto \& Nagai potentials, as for Model I. Differently from Model I and from Model A\&S, this model does not include any central spheroid, that is, this is a bulge-less model whose total potential is the sum of three components only:
\begin{equation}
\Phi_{tot}(R,z)=\Phi_{thin}(R,z)+\Phi_{thick}(R,z)+\Phi_{halo}(r).
\end{equation}

The  total stellar disc mass (at infinity) is equal to $7.6\times10^{10} M_{\odot}$, the corresponding exponential scale lengths of the discs, found by fitting the radial surface density profiles in the radial range $\rm 3~kpc\le R \le 9~kpc$,  are compatible with those given by \citet{bensby11, bovy12a}, that is, $\sim$ 2~kpc for the thick disc and $\sim$ 3.6~kpc for the thin disc (see Fig.~\ref{densityMII}, top panel).
For the  characteristic heights of the thin and thick discs we have, again, adopted $b_{thin}=0.25$~kpc and $b_{thick}=0.8$~kpc, respectively. The corresponding exponential scale heights, found by fitting to the total vertical density profile at the solar radius in the $z-$range of 0--3~kpc, are $z_{d, thin}=0.2$~kpc and $z_{d,thick}=0.7$~kpc, respectively (see Fig.~\ref{densityMII}). As for Model I, also in this case the residuals of the radial and vertical fits never exceed 0.05 (see Fig.~\ref{densityMII}, second and fourth panels), implying differences between the modelled densities and exponential distributions of less than 10\%, in the spatial range explored. 
With this choice of parameters, at the solar vicinity, the thin and thick disc volume densities are respectively $\rm 6.11\times10^7M_{\odot}/kpc^3$ and $\rm 1.22\times10^7M_{\odot}/kpc^3$, and their corresponding surface densities are $\rm 3.06\times10^7M_{\odot}/kpc^2$ and $\rm 2.09\times10^7M_{\odot}/kpc^2$  (corresponding to 0.073$\rm M_{\odot}/pc^3$ and 51.5$\rm M_{\odot}/pc^2$, also in good agreement with the estimates of \citet{mckee15}).

In Fig.~\ref{densityMII},  the surface and vertical densities of the thin disc, thick disc, and total (= thin+thick) disc for this model are shown. The resulting rotation curve is shown in Fig.~\ref{rotcurveMI-II}.  As for Model A\&S and Model I, all the parameters of Model II are summarized in Table~\ref{parameters} (see column 2).

\begin{figure}[h!]
\includegraphics[width=0.5\textwidth]{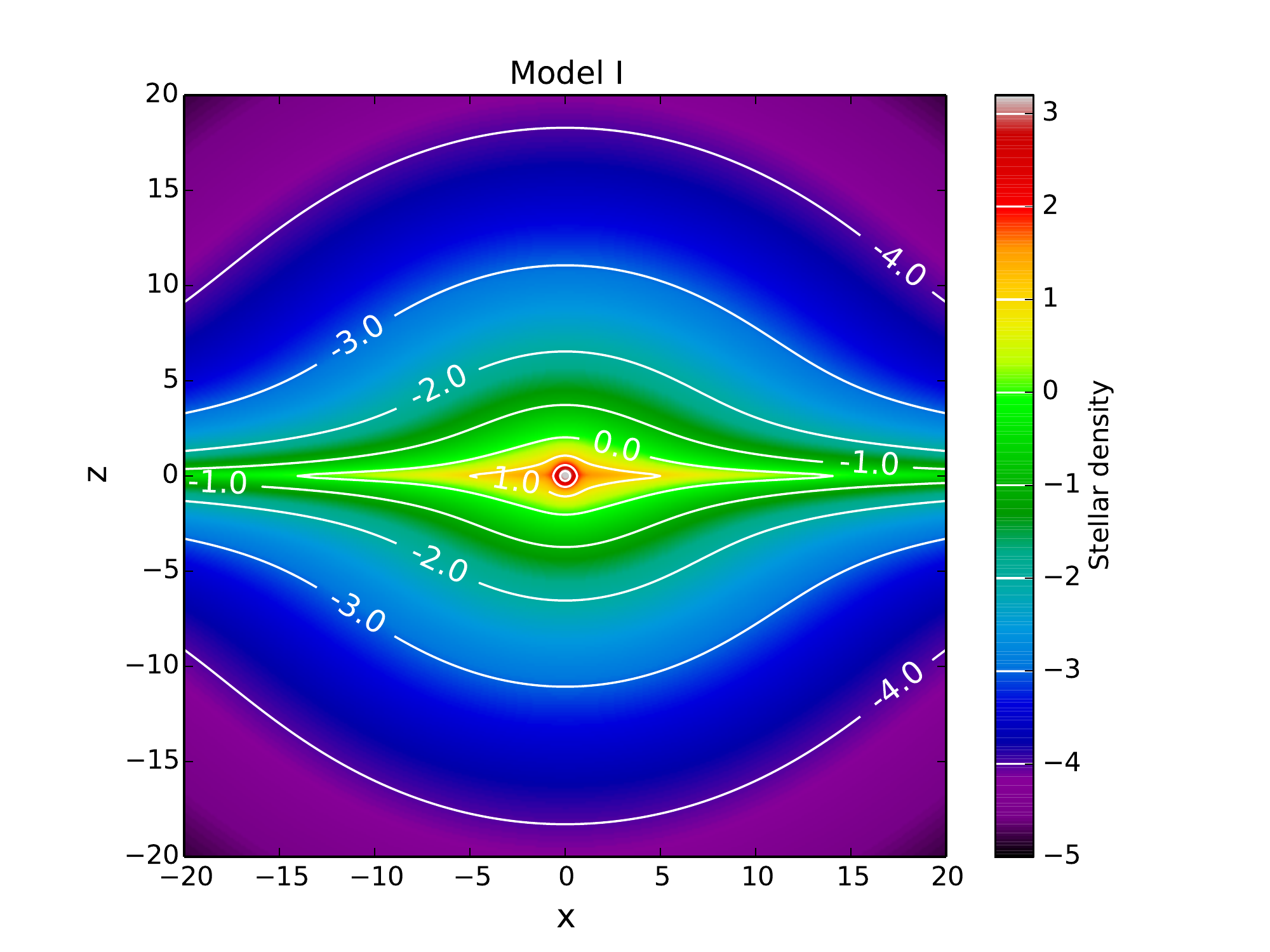}
\includegraphics[width=0.5\textwidth]{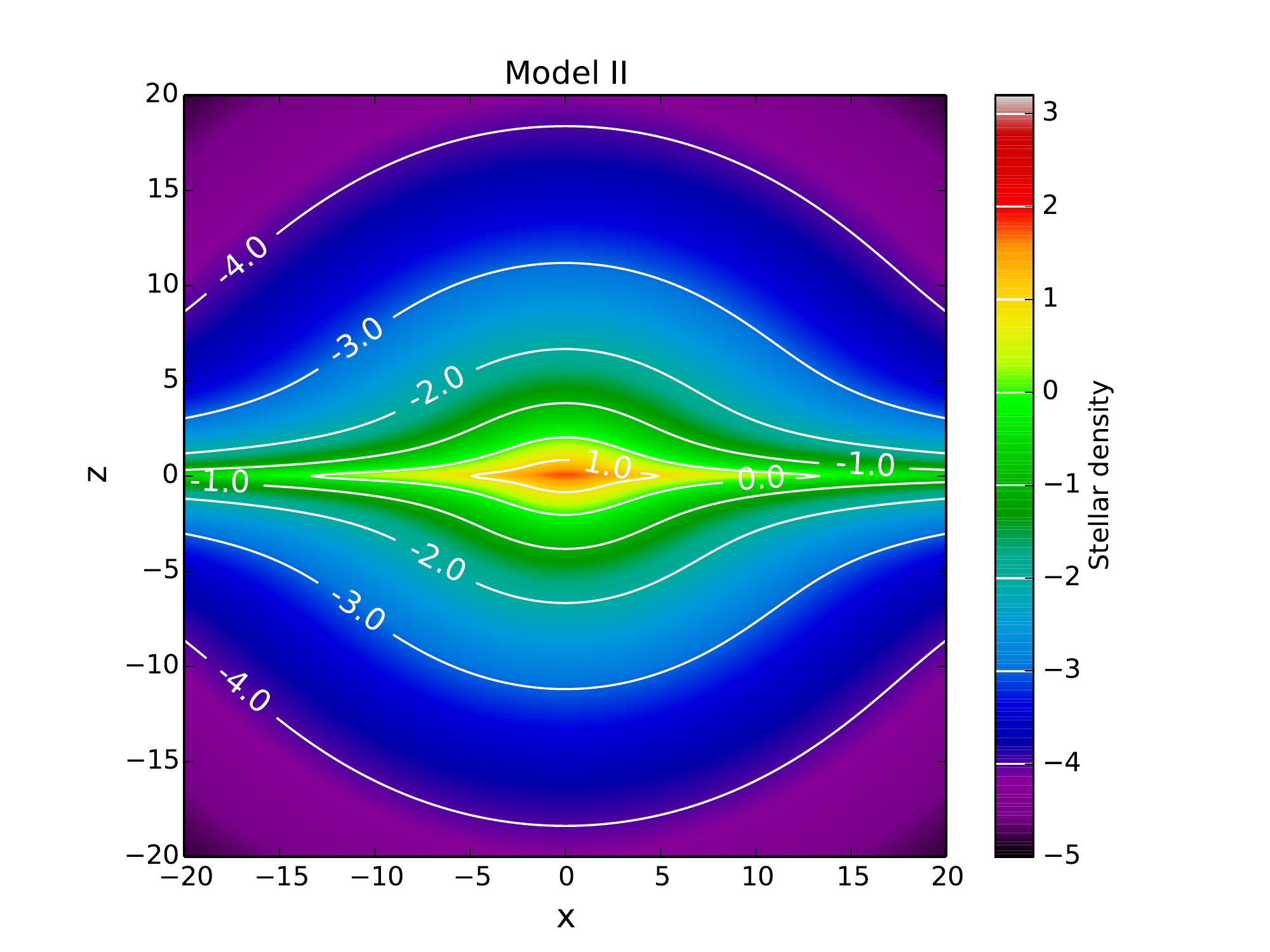}
\includegraphics[width=0.5\textwidth]{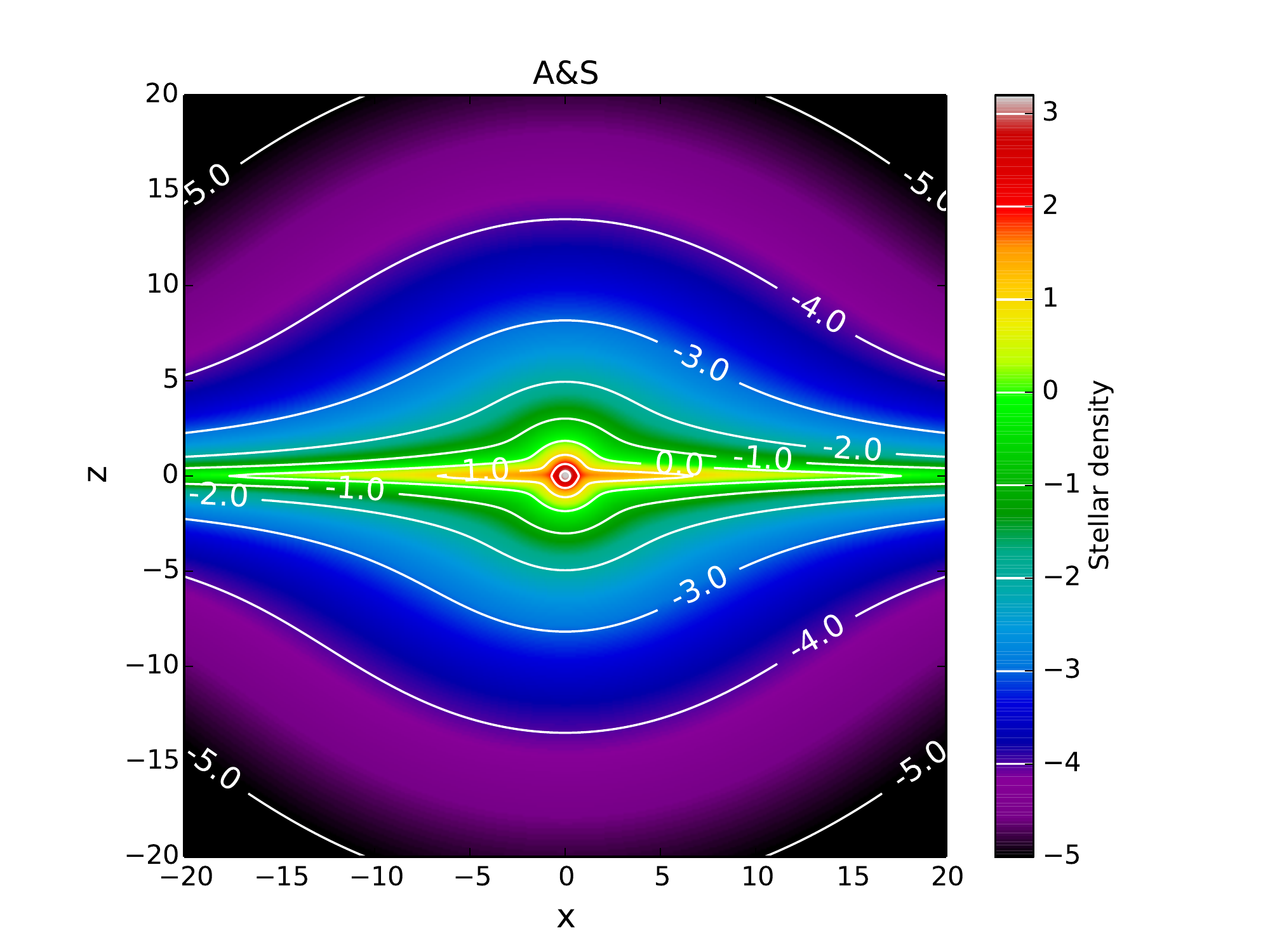}
\caption{Stellar density maps projected into the $x-z$ plane, for Model I (top panel), and Model II (middle panel). For comparison, the corresponding density map of Model A\&S is also shown. Densities are in units of $2.3\times10^7M_{\odot}/kpc^2$, distances in kpc. Densities are in logarithmic scale.}\label{densmap}
\end{figure}

\begin{figure}[h!]
\includegraphics[width=0.5\textwidth]{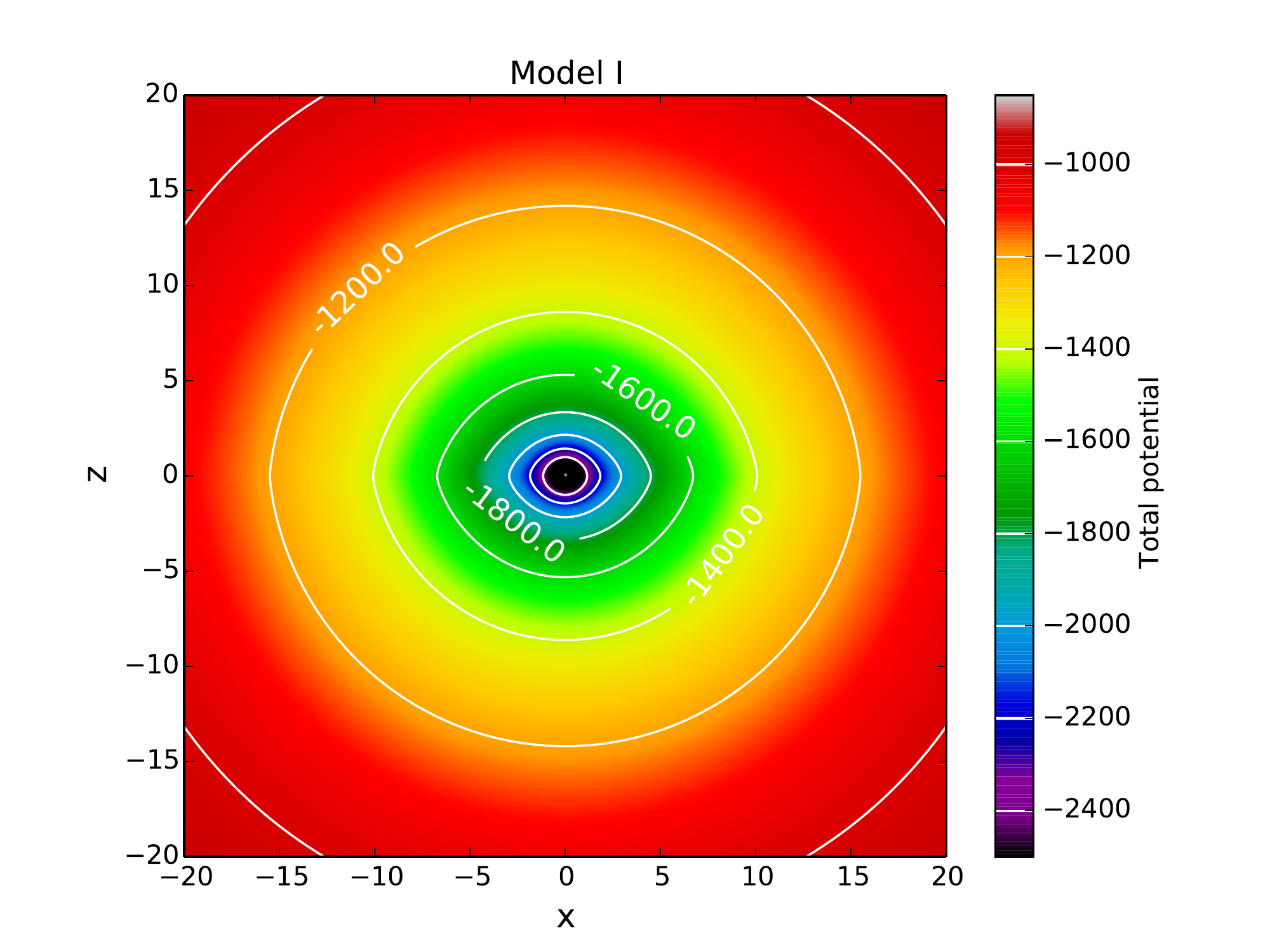}
\includegraphics[width=0.5\textwidth]{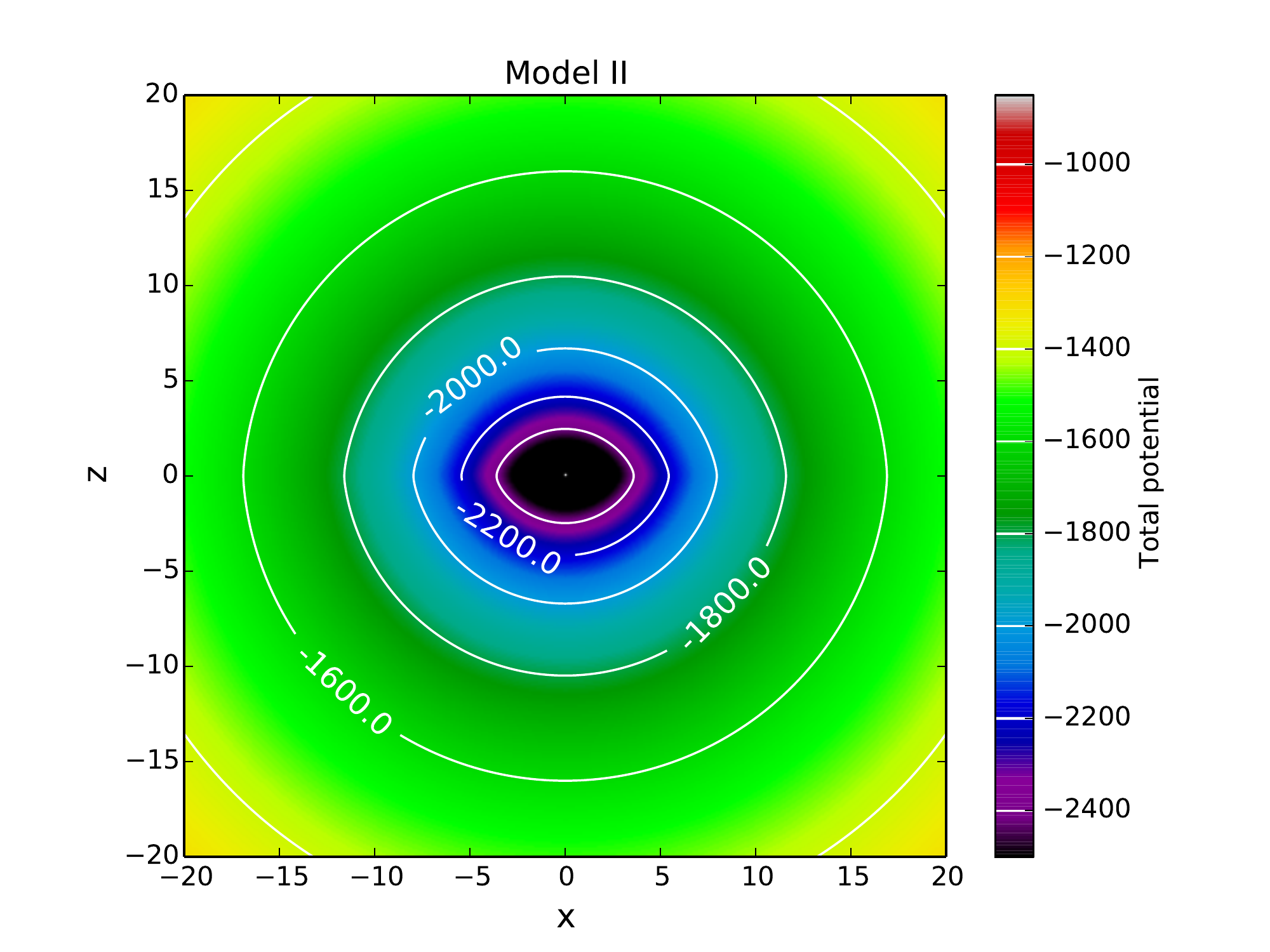}
\includegraphics[width=0.5\textwidth]{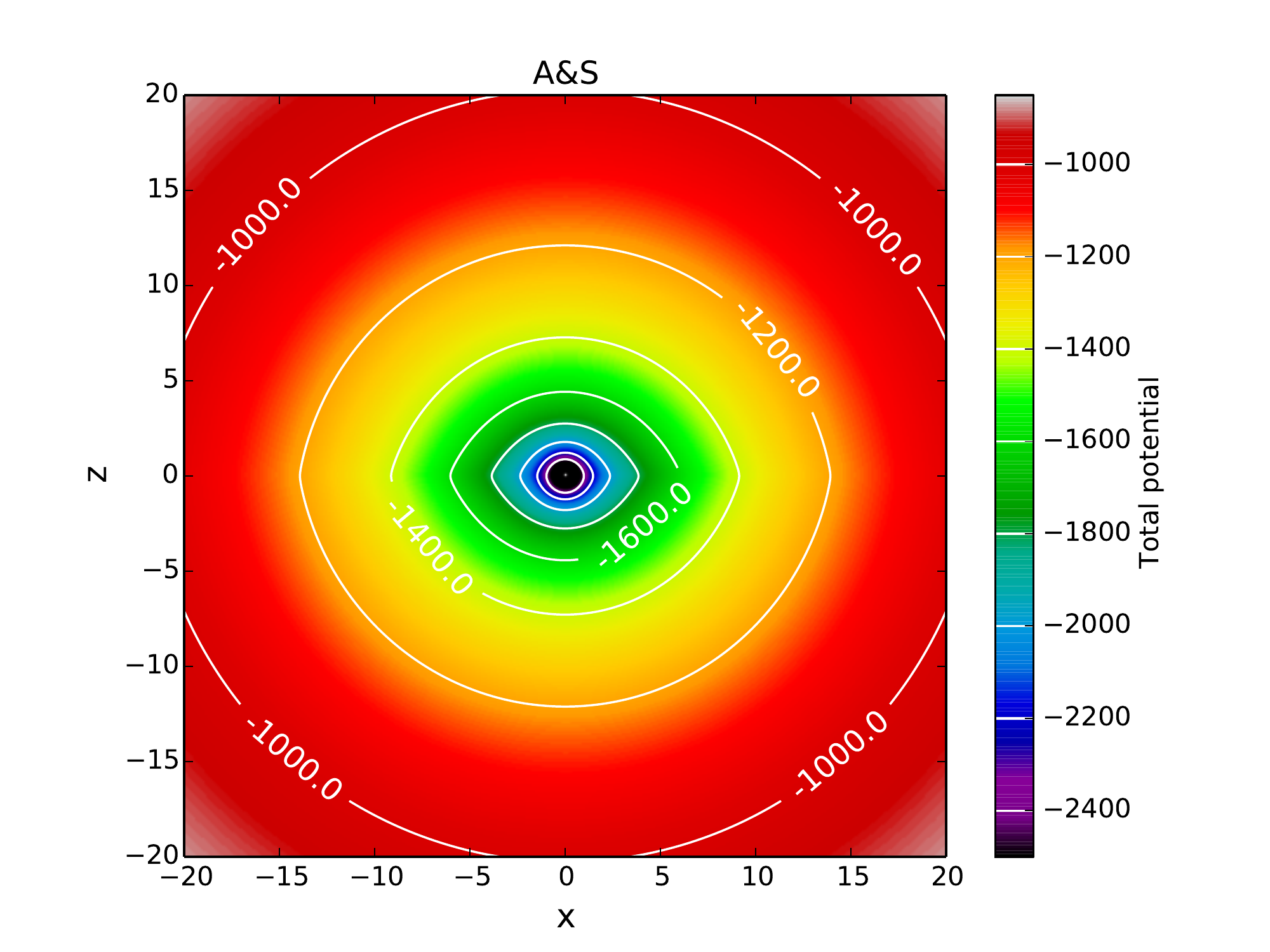}
\caption{Maps of the total (= visible+dark matter) potential energy of Model I (top panel), and Model II (middle panel). For comparison, the corresponding density map of Model A\&S is also shown. Energies are in units of $100 km^2/s^2$, distances in kpc. }\label{potmap}
\end{figure}

The resulting edge-on stellar density maps and total gravitational potential of Model I, Model II, and Model A\&S are given in Figs.~\ref{densmap} and \ref{potmap}. From Fig.~\ref{densmap}, one can see that, as expected, the thickness of the stellar distribution increases when a massive thick disc is added to the mass distribution, and that the gravitational potential of Models I and A\&S are very similar at large radii, while that of Model II is systematically deeper than the two other models. As we will discuss in the following, this has an impact on the orbital characteristics of stars and star clusters reaching large apocentres.

\begin{figure}
\includegraphics[width=3.5in]{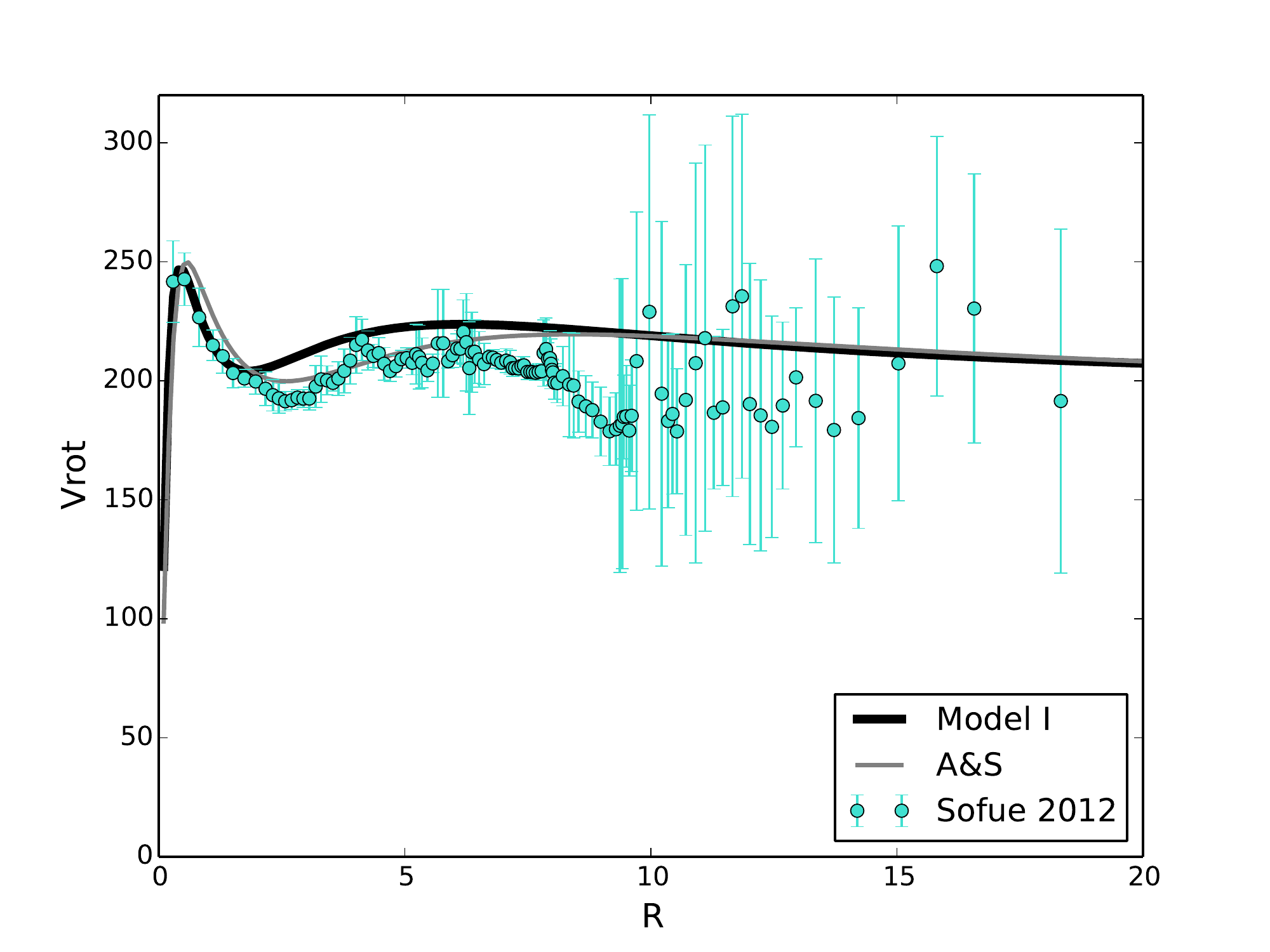} 
\includegraphics[width=3.5in]{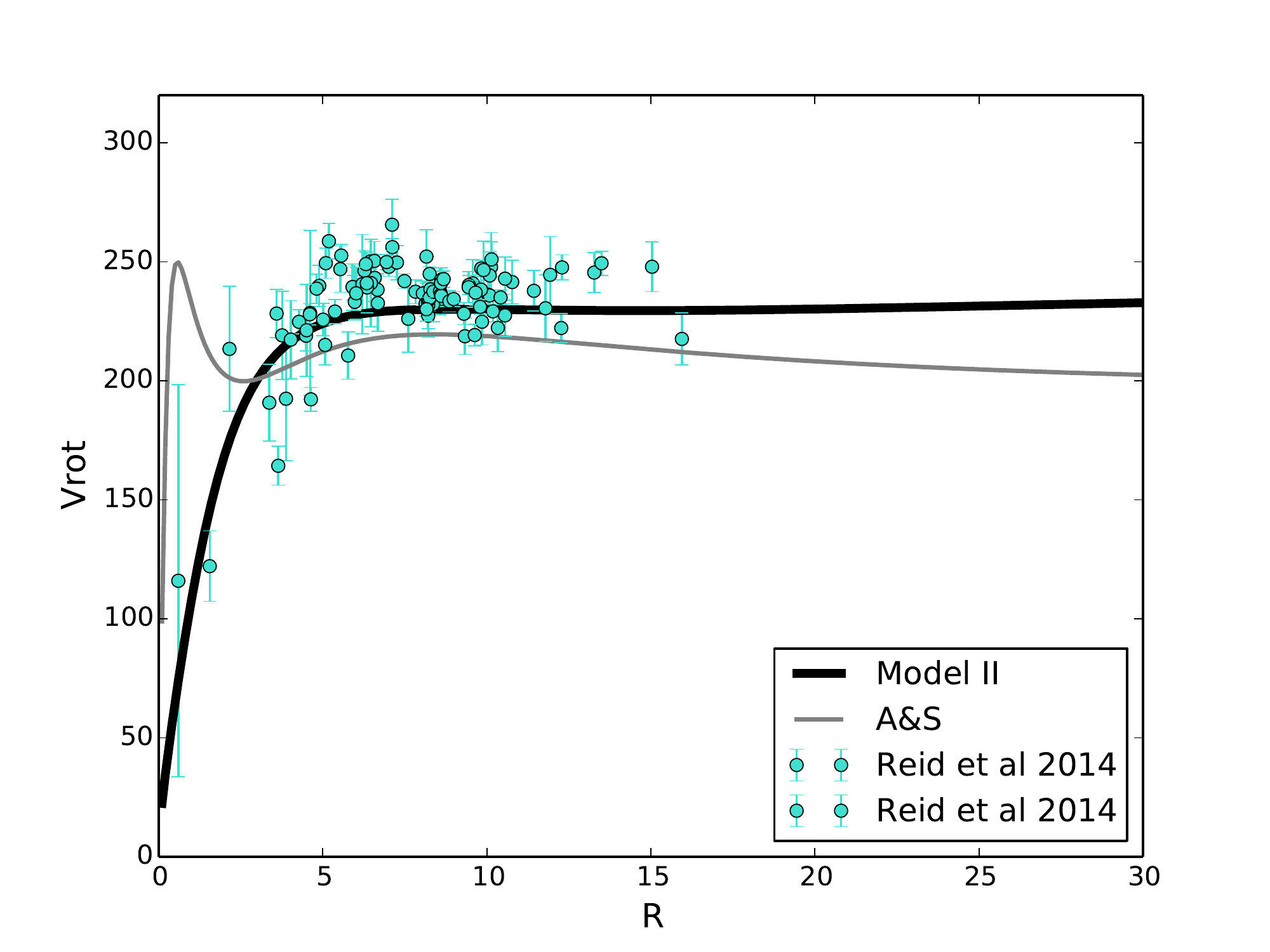}
\caption{Upper panel: Total rotation curve  of Model I in the inner 20 kpc (black curve) compared to the CO/HI data by \citet{sofue12} (cyan points). For comparison, the rotation curve predicted by the \citet{allen91} model is also shown (grey curve). Lower panel: Total rotation curve of Model II (black curve), compared to the Galactic rotation curve obtained with VLBI observations of maser sources by \citet{reid14} (cyan points). The \citet{allen91} rotation curve is shown for comparison (grey curve). 
Models I and II are well bracketed by the two observed rotation curves.}\label{vrot}
\end{figure}

\subsection{Concerning the use of Miyamoto-Nagai density profiles to model the Milky Way disc(s)}

Estimates of the density profiles, scale length and scale height of the Milky Way disc(s) are currently available in a range from about 3 to 9-10 kpc in radius and up to 3-5 kpc from the plane, in the vertical direction. The most recent estimates come from spectroscopic surveys like SEGUE and APOGEE, and in particular from the works of Bovy and collaborators \citep{bovy12a, bovy12b, bovy12c, bovy15}. 
The picture emerging from these works is complex. In a first approximation, the thick disc (defined as the $\alpha$-enhanced stellar disc) has a scale length of $\sim$ ~2~kpc, while the thin disc (defined as the low-$\alpha$ stellar disc) has a scale length of about 3.6~kpc \citep[see][]{bovy12a, bovy15}. These estimates are in agreement with those derived by \citet{bensby11} using a much smaller sample of stars in the inner and outer Galactic disc. As shown in Figs. 1 and 2,  in the same radial range as that covered by the observations, a Miyamoto-Nagai density profile can be fitted by an exponential with a scale length compatible with observations (the uncertainties in the scale lengths derived from observational data are of about 10\%, see Table 2 in \citet{bovy12a} ). In this radial range, the differences between the fitted exponential and the Miyamoto-Nagai density profile correspond to differences in densities smaller than 10\%, thus smaller than the current uncertainties in the estimates of the Galactic disc surface densities. As an example, \citet{flynn06} report  uncertainties  in  the estimates of stellar surface densities at the solar vicinity of 10-15\%. Thus, in the radial range covered by  current observations, it is not possible to make the difference between the two profiles -- a Miyamoto-Nagai profile, as adopted in this work, and an exponential profile, as adopted for example by \citet{dehnen98} -- because the uncertainties in the surface density estimates and in the scale lengths are still too large to disentangle between those two profiles .\\ Looking deeper into the results of \citet{bovy12a}, the situation is even more complex: the stellar disc cannot be represented by two single exponentials - one for the thin and one for the thick disc - but it is rather made of  ``mono-abundance" populations, with different scale lengths. Among the low-$\alpha$ population -- which classically corresponds to the thin disc -- several mono-abundance populations can be identified, with scale lengths varying from $\sim$1.8 kpc to more than 4 kpc \citep[Fig. 5 in][]{bovy12a}. It is difficult to reconcile this finding with a single exponential fit in the radial direction for the whole stellar disc, especially because the populations with large scale length (i.e. greater than 4 kpc) constitute a not negligible fraction of the surface density at the solar vicinity (about 10-20\%, as evaluated from the metallicity distribution function of stars at the solar vicinity). Uncertainties are still too large to predict the fall-off of the stellar density with radius, but Bovy et al.'s findings may suggest that the fall-off of the Galactic disc at large radii is more complex than that expected from a single exponential law.  \\
In the vertical direction, the choice to model the disc with two Miyamoto-Nagai density profiles does a good job in the vertical range where data are available (see Figs.~1 and 2).  When the total vertical density profile is fitted with two exponentials - one representing the thin disc, the second representing the thick disc - the recovered scale heights are in good agreement with observations (again, we refer to \citet{bovy12a} for a comparison). Observational data may suggest an even more complex scenario for the vertical profile of thin and thick disc populations. Citing once again the work of \citet{bovy12a}, they point out that their $\alpha$--poor and $\alpha$--enhanced populations are, both, statistically better fitted by two exponentials in the vertical direction rather than one, with one of the two dominating in mass. Thus, also in the vertical direction the density profile of the thin and of the thick discs may be more complex than a single exponential. Unfortunately, the SEGUE sample is not large enough to understand whether this effect is true, or whether it is affected by small statistics or by the abundance resolution  of the survey. \\
Finally, it is worth remembering here that even for external galaxies, the fall-off of the stellar disc light is not always that of a single exponential. A not negligible fraction of galaxies exist with so called anti-truncated profiles \citep{erwin05, maltby12, elichemoral15}. One of the main questions of the Galactic structure, that will be possible to address in the near future with Gaia and follow-up spectroscopic surveys, is indeed the nature of its stellar disc and  its behaviour at large radii.

\subsection{Comparing to observational data}\label{data}

We have seen in the previous section that Models I and II have, by construction, a number of characteristics compatible with estimates available for the Milky Way disc:  total stellar disc mass, thin and thick disc scale lengths and heights,  and baryon density as measured at the solar vicinity.  In this section we discuss more in detail two other observational constraints: the agreement of the models with the Galactic rotation curve(s) and with the estimates of the perpendicular force in the inner disc, as measured by \citet{bovy13} from APOGEE data.
We reiterate that our objective is not to provide a best fit model, but to assess the effect of a  
thick disc and the uncertainties on the rotation curve of the Milky Way on the orbits computed in a widely used potential. 
This last point is justified by the present state of confusion regarding the rotation curve, with systematic 
differences that largely exceed the error bars, as illustrated in Fig.~\ref{vrot}.

\subsubsection{Rotation curves}
Model I can well reproduce the shape of the Galactic rotation curve, as given by \citet{sofue12} (see Fig.~\ref{vrot}, upper panel): the rise of the rotation curve in the inner few kpcs is well reproduced by adding a centrally concentrated, classical bulge, whose impact on the total rotation curve decreases very rapidly outside the inner 2-3 kpc. From Fig.~\ref{vrot} (upper panel),  one can also see how similar the recovered rotation curve of Model I is to that proposed by \citet{allen91}. At the solar radius \citep[assumed to be at R=8.5~kpc, as in][]{allen91}, Model I predicts a circular velocity of 221.4, very similar to that of the A\&S model at the same radius, which is 219.9 km/s.

With a classical bulge mass set to zero, and a more massive dark matter halo, Model II is in better agreement with the observational data of the rotation curve derived by \citet{reid14}. With respect to  Model A\&S, Model II lacks the central rise of the rotational velocity curve, whose physical significance has been also recently questioned by \citet{chemin15}, and predicts a larger circular velocity (230.0~km/s) at the solar radius (Fig.~\ref{vrot}, lower panel). Outside the inner regions (R$>$2kpc) both Models I \& II are  bracketed by the two observed rotation curves.

\begin{figure}[H]
\centering
\includegraphics[width=3.5in]{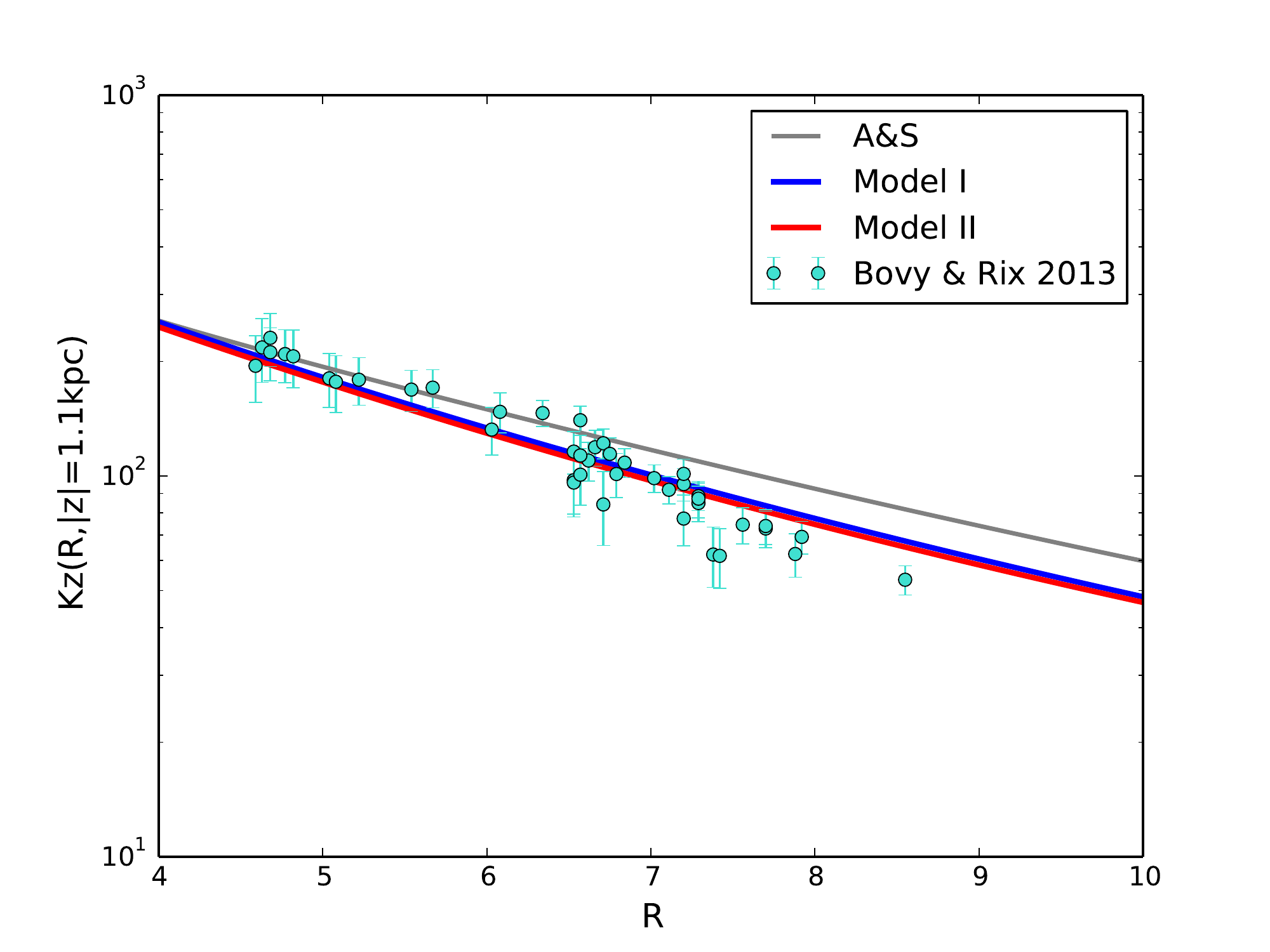}
\caption{Absolute value of the perpendicular force Kz as a function of the Galactic distance R at a vertical distance $|z|$=1.1kpc from the Galactic plane: Model I (blue curve), Model II (red curve), Model A\&S (grey curve). The Kz distribution at $|z|$=1.1kpc, as derived by APOGEE data \citep{bovy13}, is shown for comparison (cyan points).}\label{kz}
\end{figure}

\subsubsection{The perpendicular force in the inner disc}
Figure~\ref{kz} shows the predicted perpendicular force,  Kz, at a vertical distance from the plane of 1.1 kpc., in the region between 4 and 10 kpc, for both Models I and II. For completeness, the corresponding profile of the \citet{allen91} model is also given. These curves are compared to the values derived by \citet{bovy13} for the inner disc, at $|z|=1.1$kpc, with APOGEE. As can be seen, the match between models' predictions and data is very good for all the radial extent covered by the data.

\begin{figure}
\includegraphics[trim=1cm 3.5cm 0cm 4.cm, clip=true,width=0.5\textwidth]{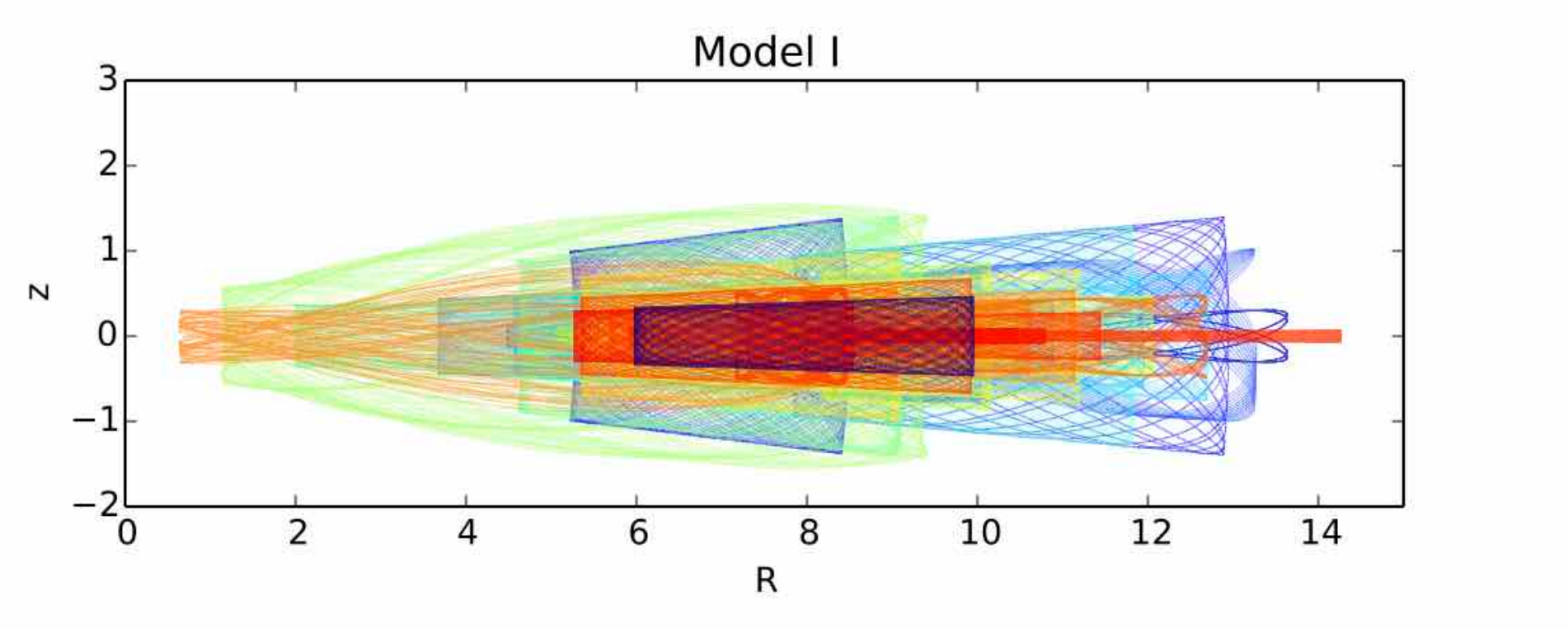} 
\includegraphics[trim=1cm 3.5cm 0cm 4.cm, clip=true,width=0.5\textwidth]{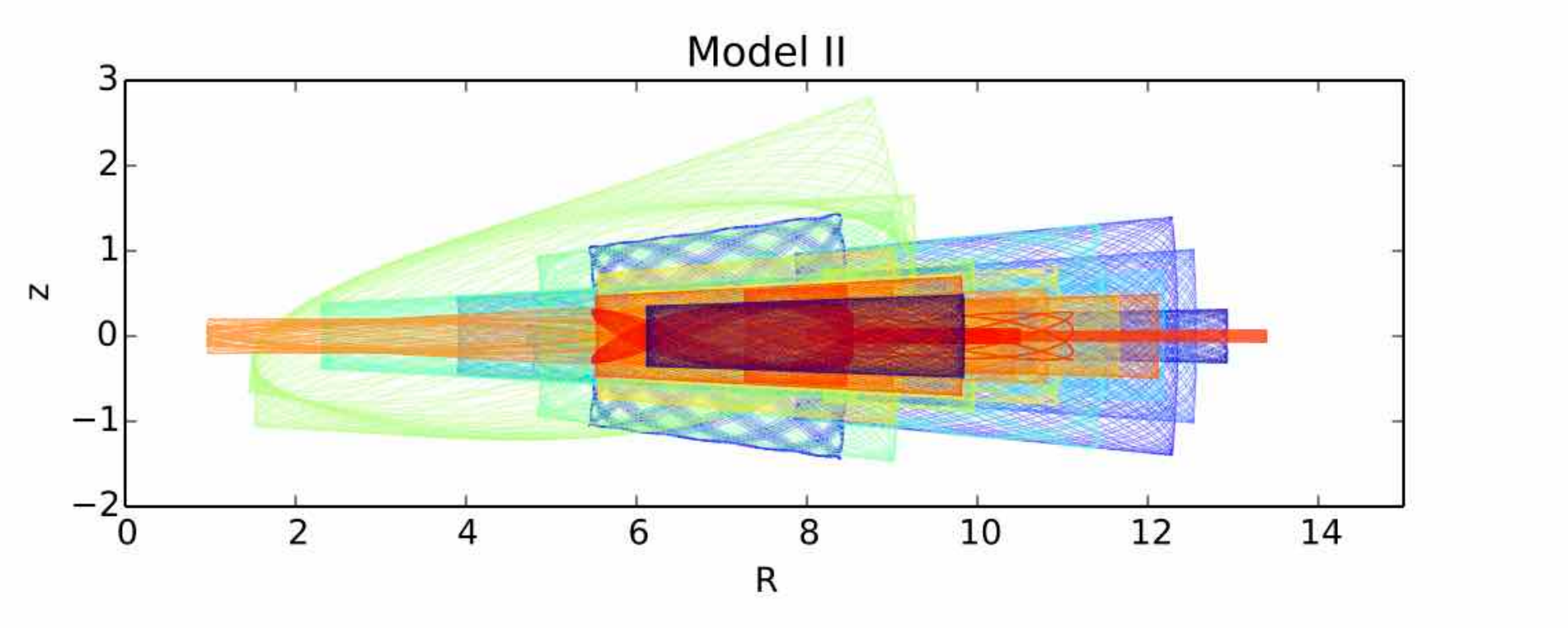} 
\includegraphics[trim=1cm 3.5cm 0cm 4.cm, clip=true,width=0.5\textwidth]{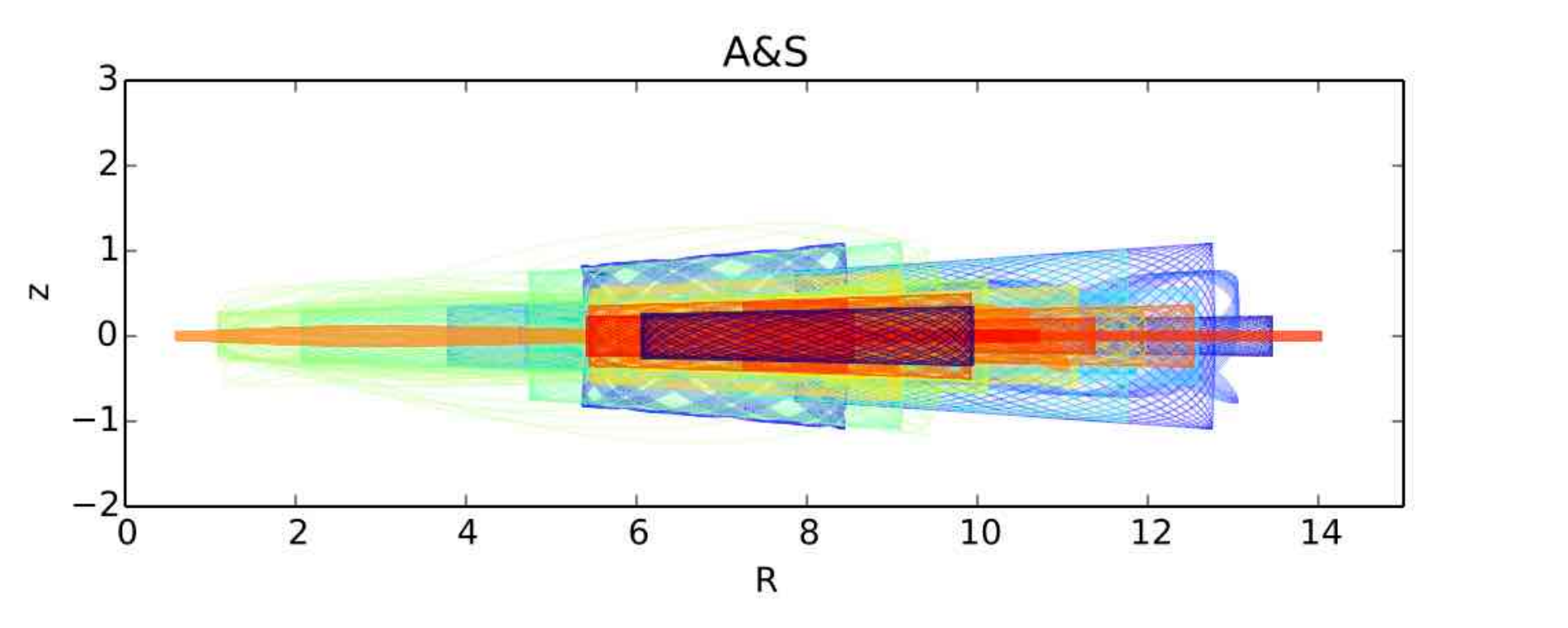} 
\caption{Comparison of the orbits of \citet{adibek12} stars projected in the R-z plane for Model I (top panel), Model II (middle panel), and Model A\&S (bottom panel). Only one out of ten stars of those in the \citet{adibek12} sample is shown. Different colours correspond to different stars. }\label{allstarsRz}
\end{figure}

\begin{figure*}
\centering
\includegraphics[trim=2.5cm 0cm 1.cm 0cm, clip=true, width=0.45\textwidth]{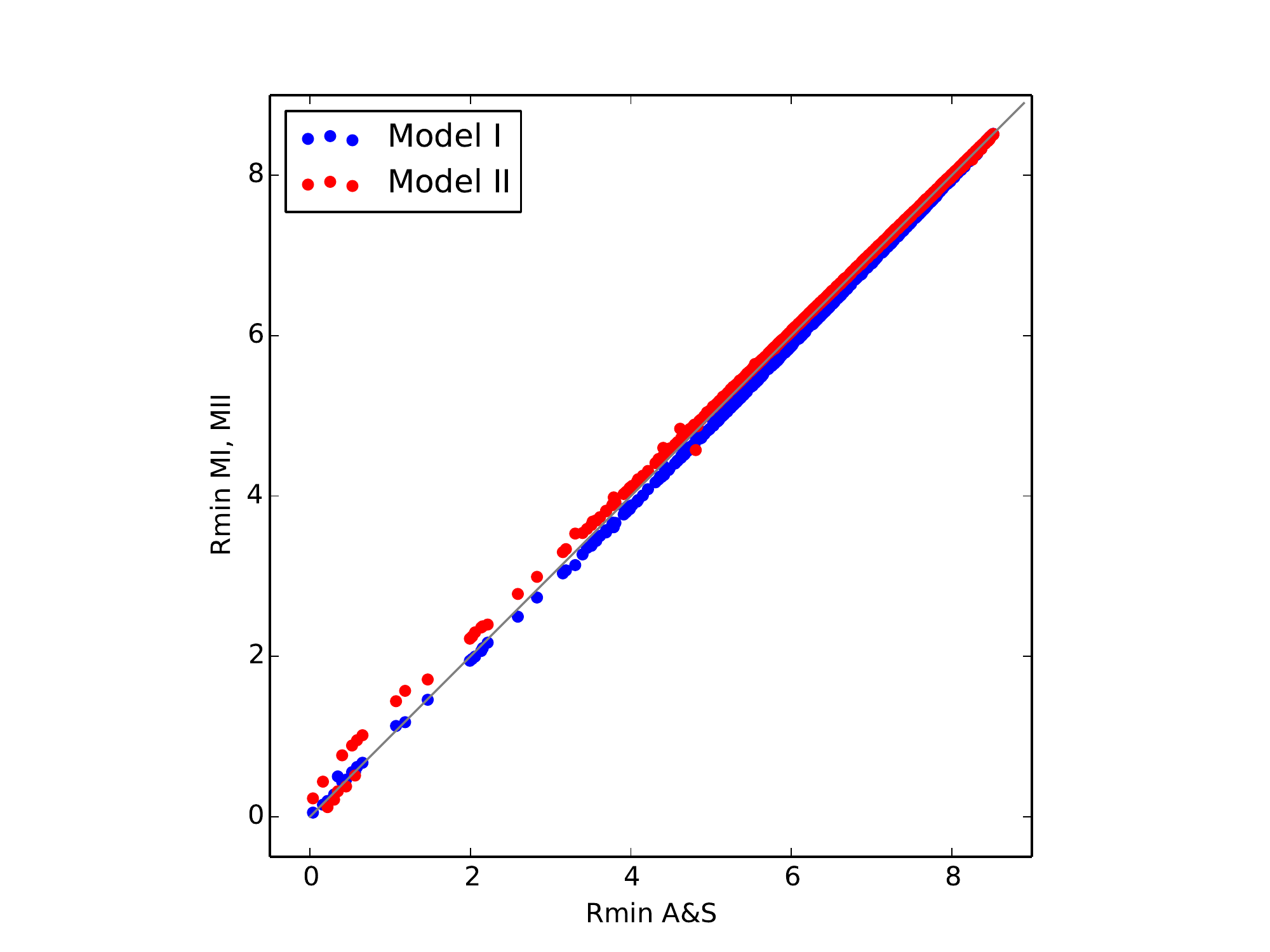}
\includegraphics[trim=2.5cm 0cm 1.cm 0cm, clip=true, width=0.45\textwidth]{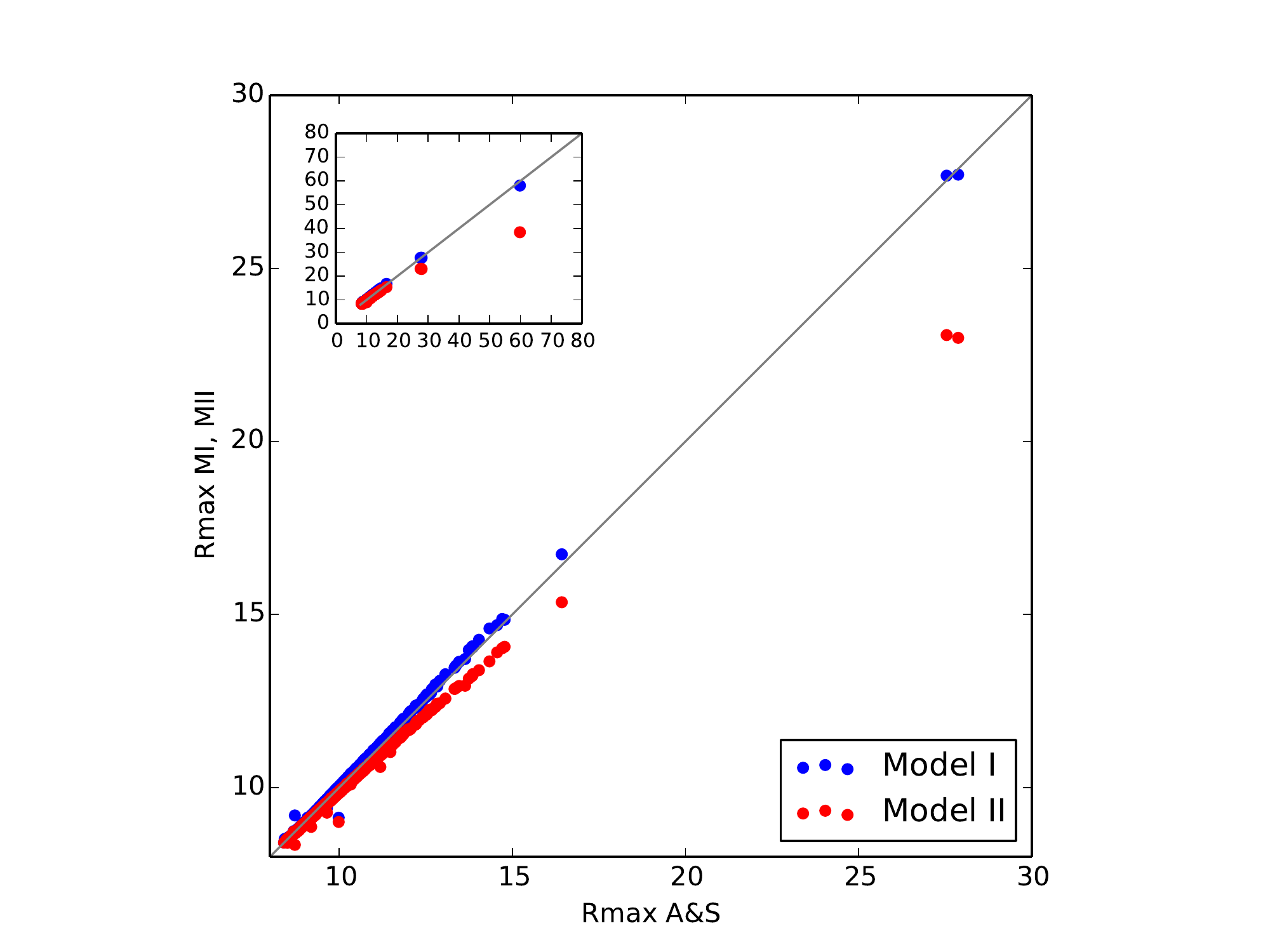}
\includegraphics[trim=2.5cm 0cm 1.cm 0cm, clip=true, width=0.45\textwidth]{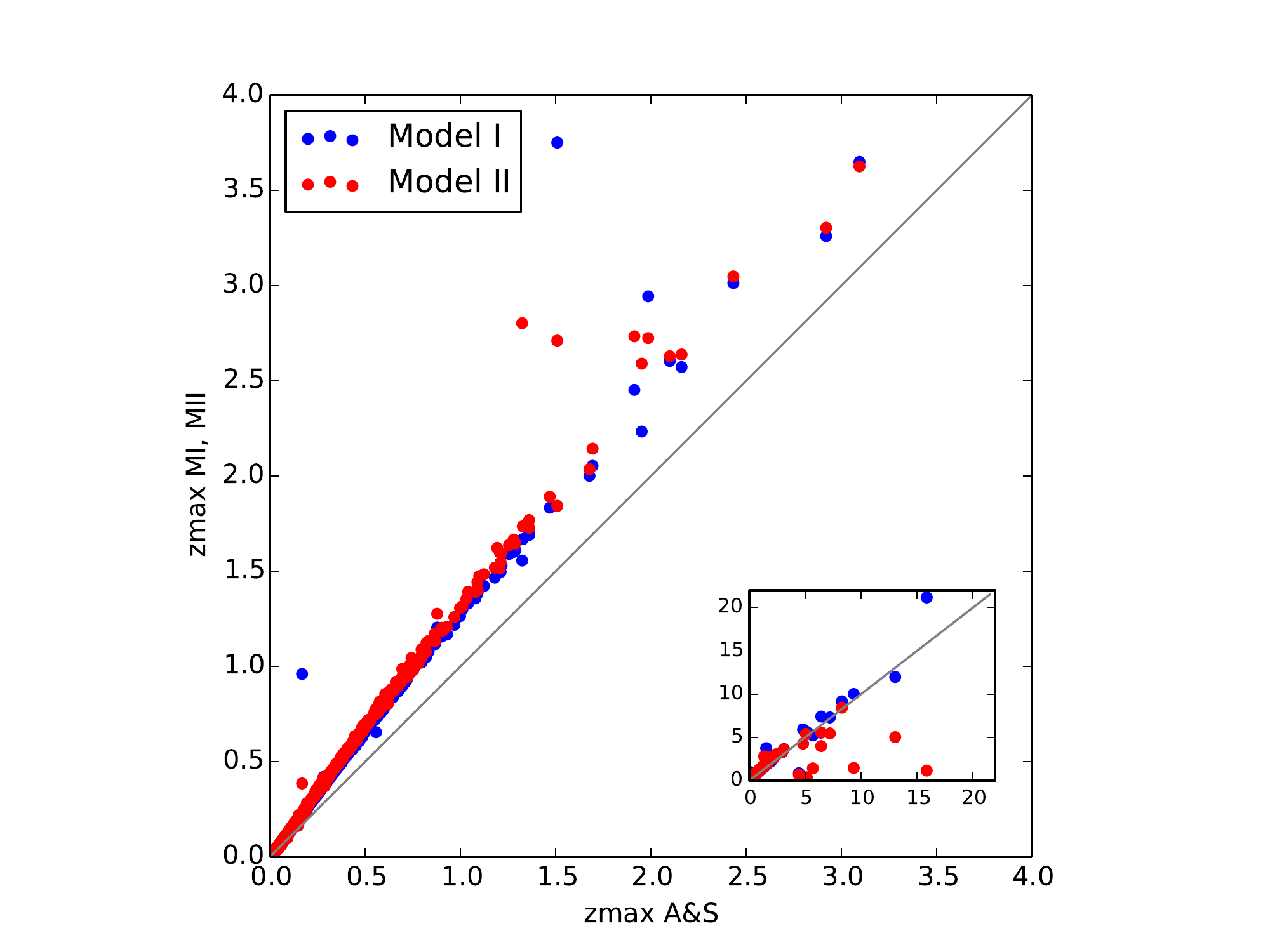}
\includegraphics[trim=2.5cm 0cm 1.cm 0cm, clip=true, width=0.45\textwidth]{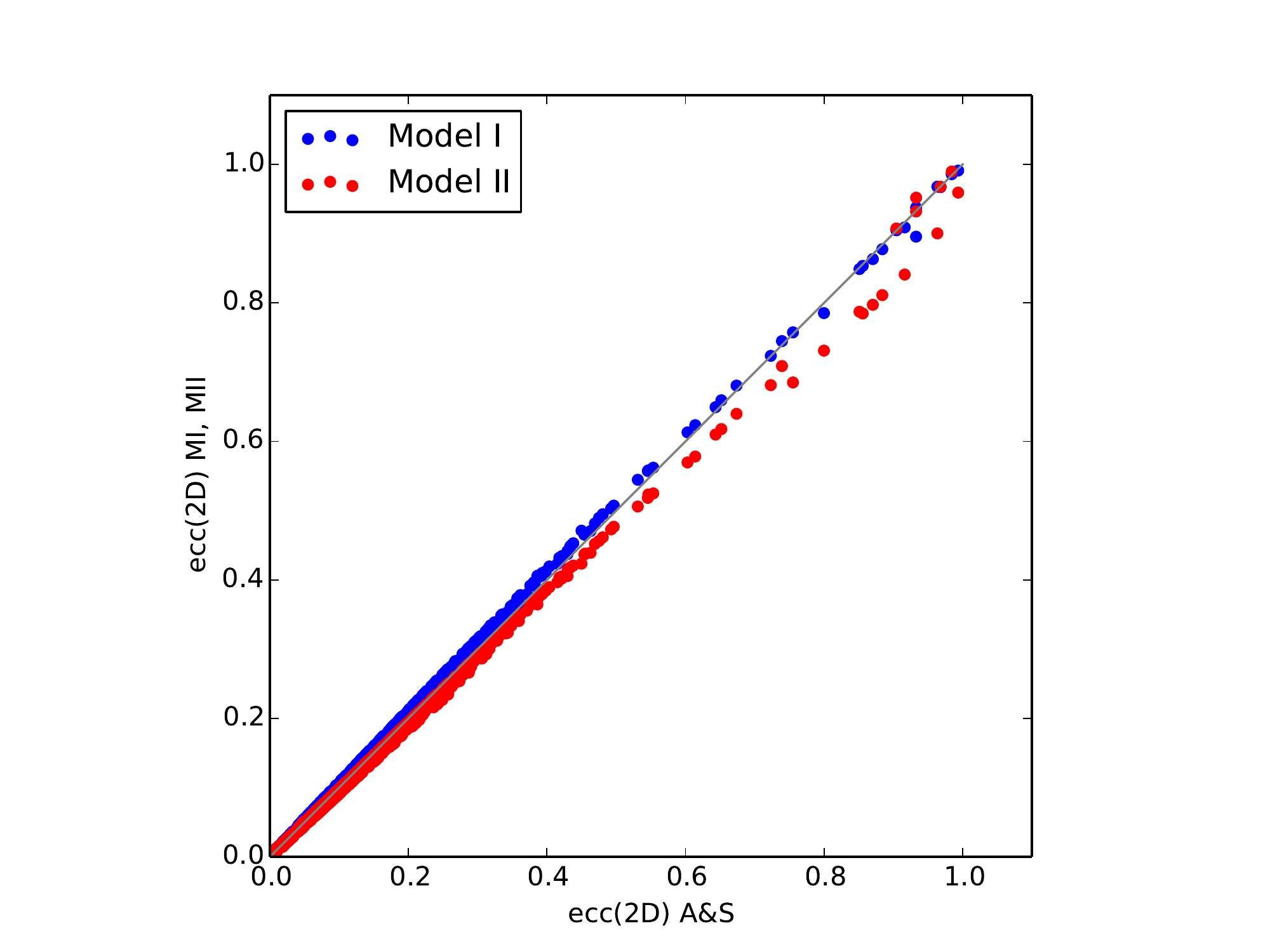}
\includegraphics[trim=2.5cm 0cm 1.cm 0cm, clip=true, width=0.45\textwidth]{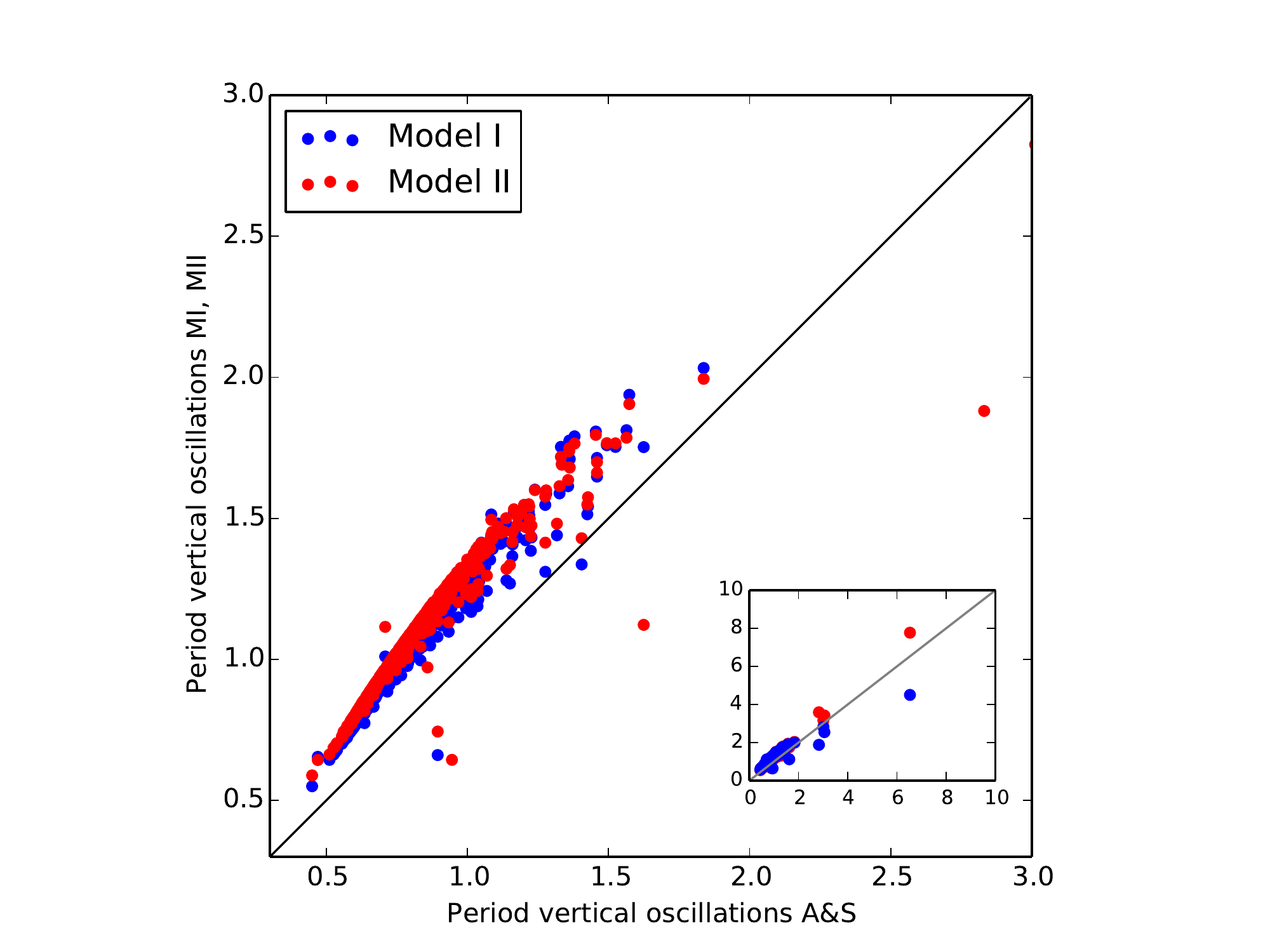}
\includegraphics[trim=2.5cm 0cm 1.cm 0cm, clip=true, width=0.45\textwidth]{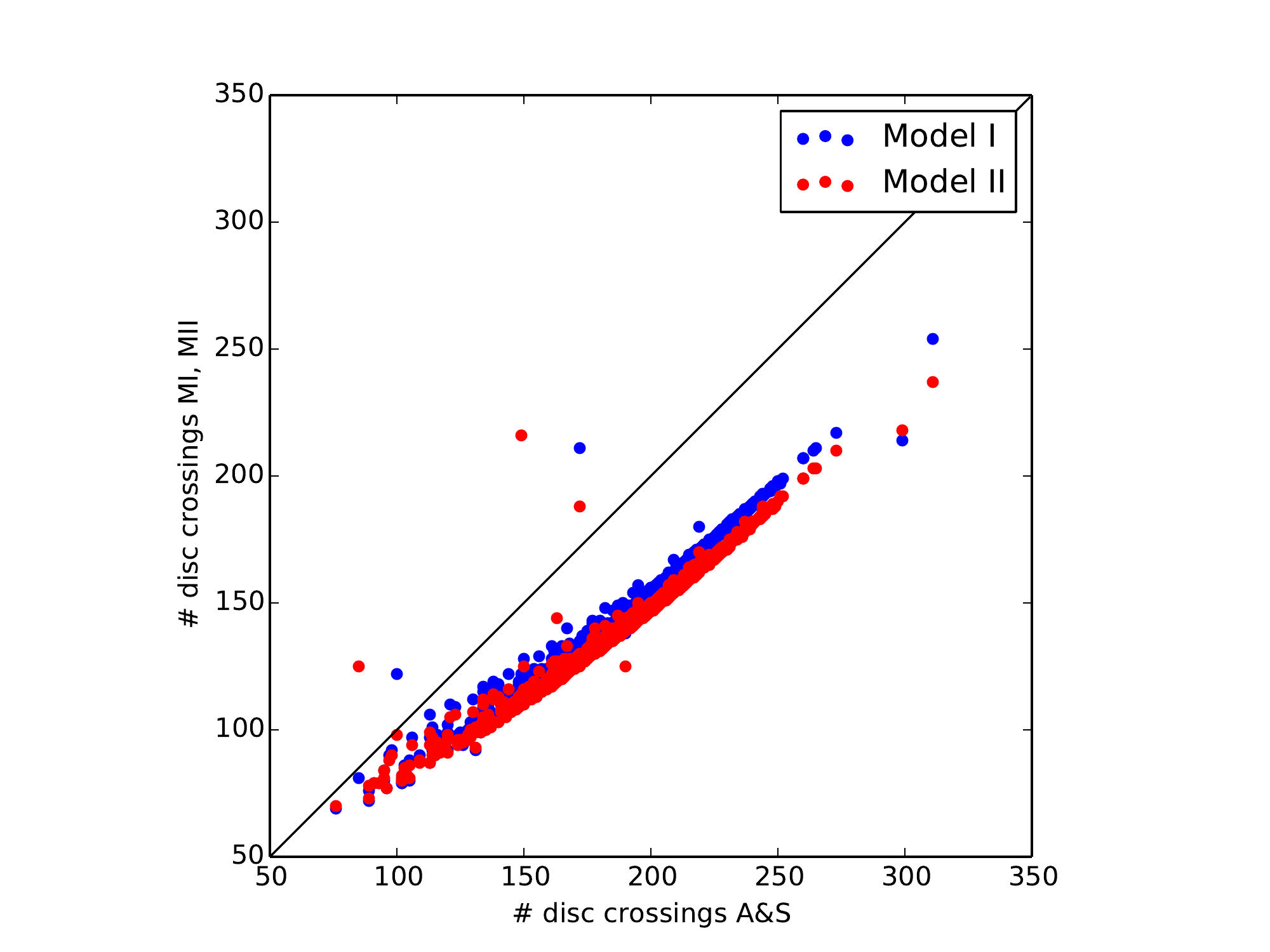}
\caption{Comparison of the orbital characteristics of stars of the \citet{adibek12} sample integrated in Model I (blue points) and Model II (red points) versus the \citet{allen91} model. From top-left to bottom-right: Pericentres $R_{min}$, apocentres $R_{max}$, maximum vertical distance from the plane $z_{max}$,  2D eccentricities, period of vertical oscillations, and number of disc crossings are given. The insets in some plots show a larger range of values than that covered in the corresponding main plots. In all the plots, distances are in kpc, time in units of $10^8$ yr. }\label{allstarscomp}
\end{figure*}

\begin{figure*}
\centering
\includegraphics[trim=0.5cm 0cm 1.cm 0cm, clip=true, width=0.45\textwidth]{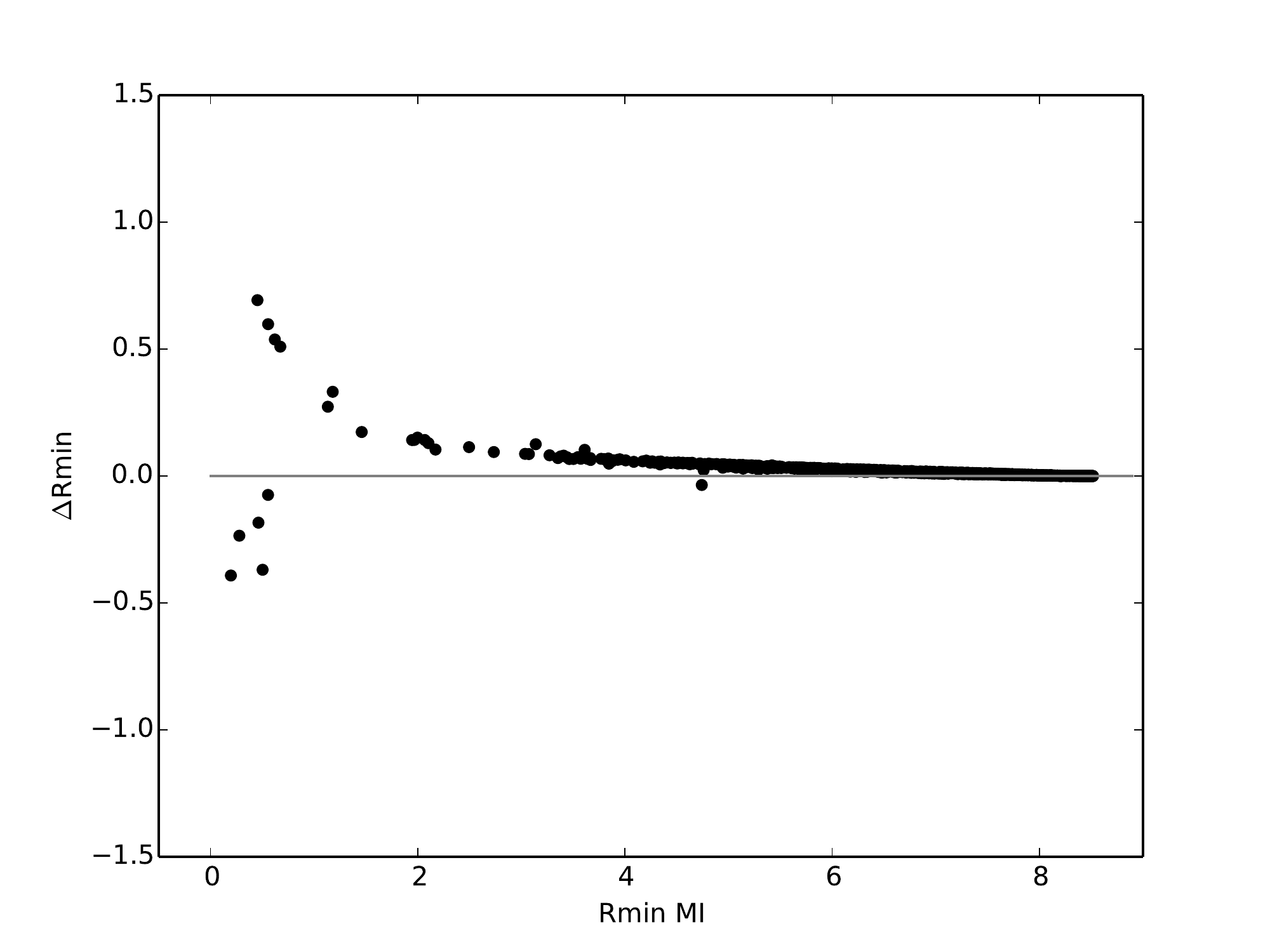}
\includegraphics[trim=0.5cm 0cm 1.cm 0cm, clip=true, width=0.45\textwidth]{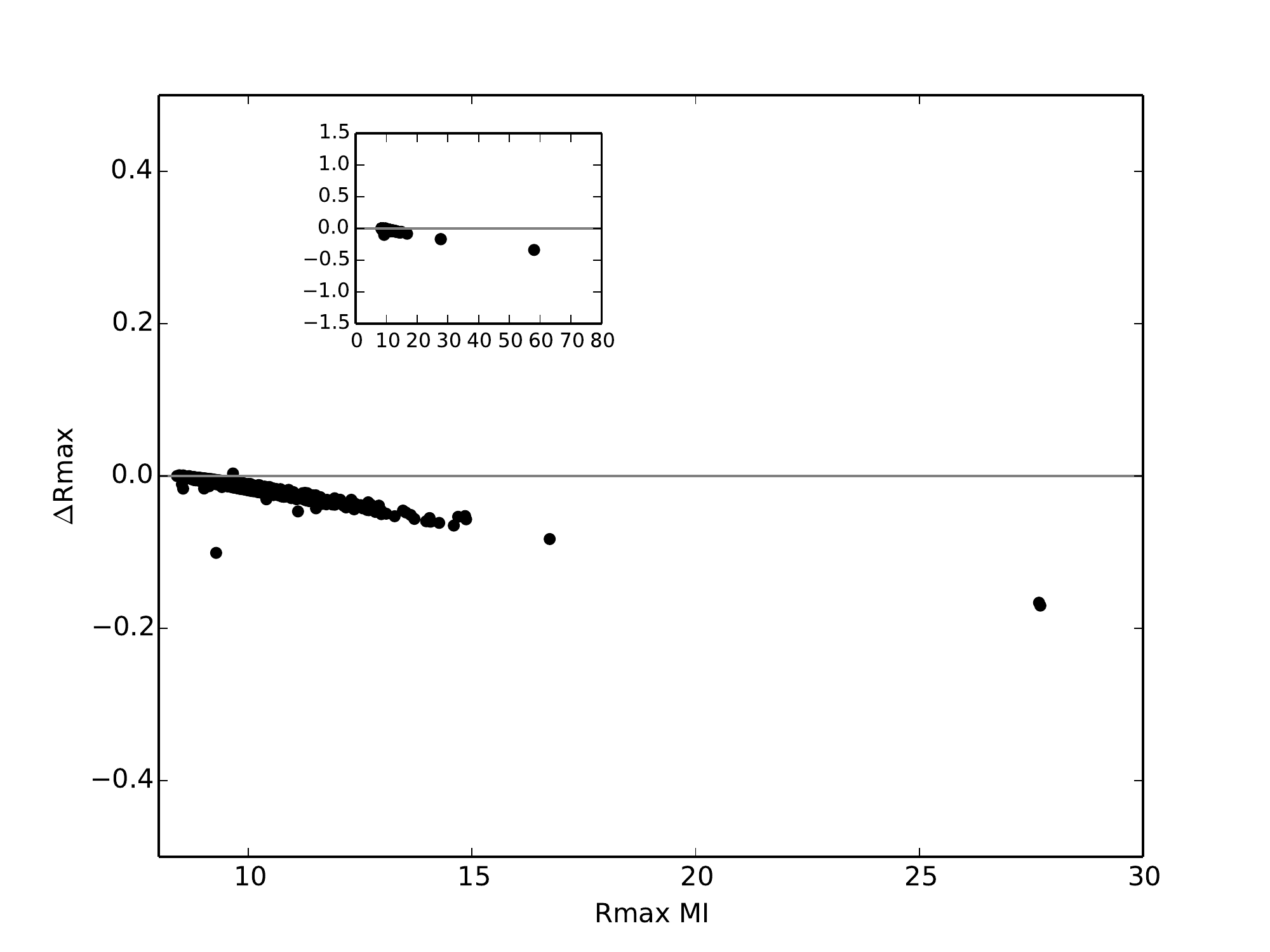}
\includegraphics[trim=0.5cm 0cm 1.cm 0cm, clip=true, width=0.45\textwidth]{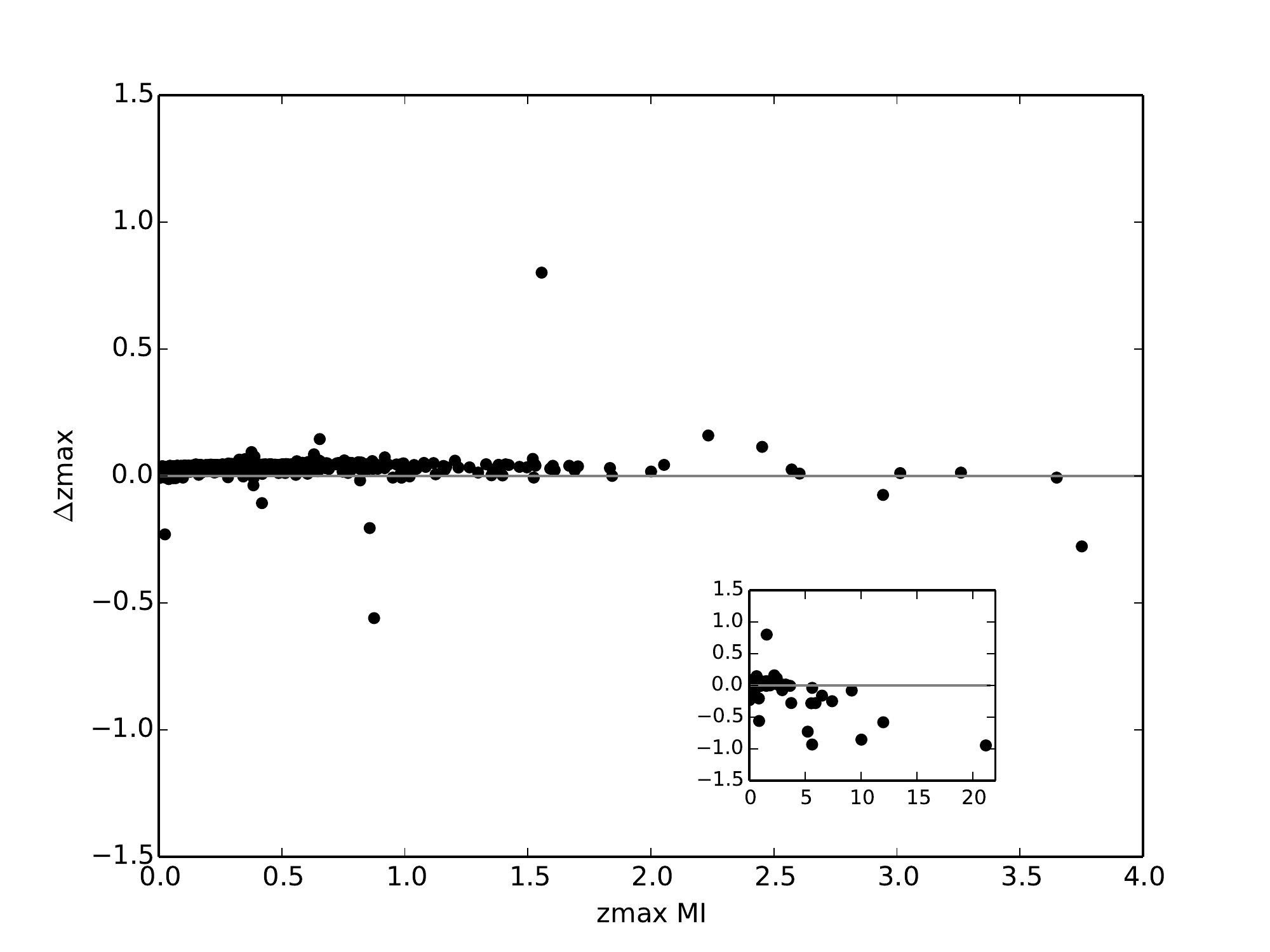}
\includegraphics[trim=0.5cm 0cm 1.cm 0cm, clip=true, width=0.45\textwidth]{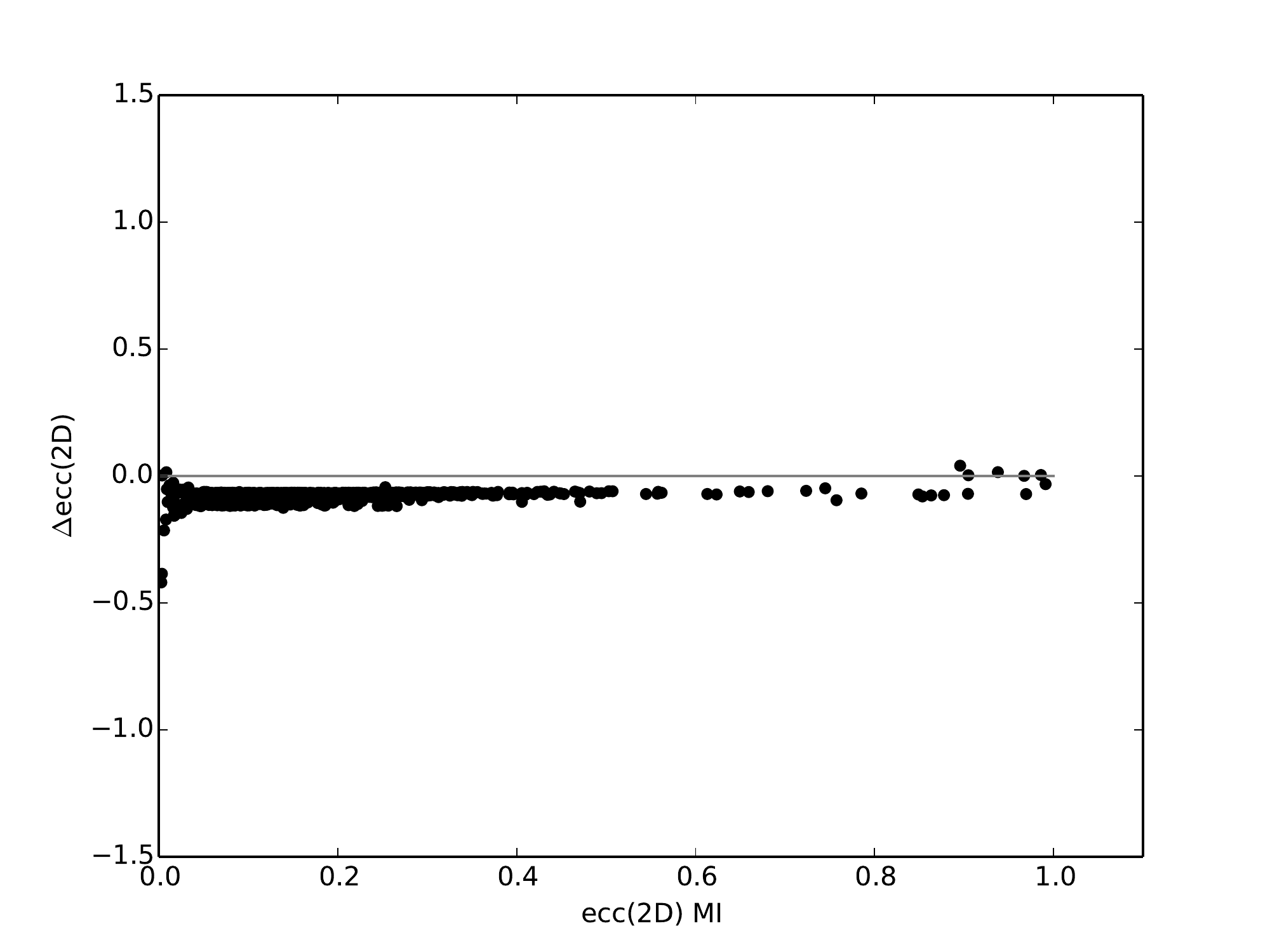}
\includegraphics[trim=0.5cm 0cm 1.cm 0cm, clip=true, width=0.45\textwidth]{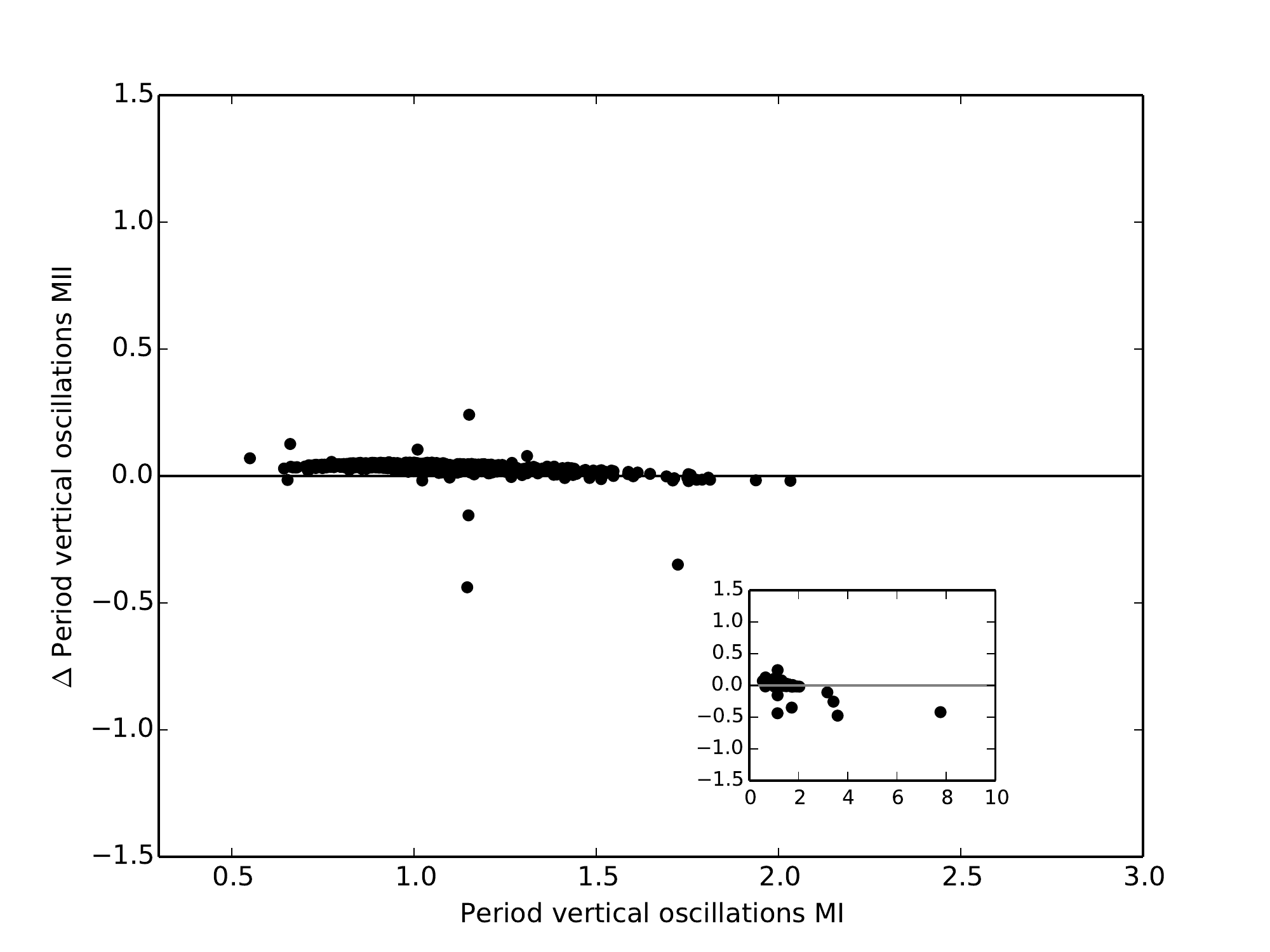}
\includegraphics[trim=0.5cm 0cm 1.cm 0cm, clip=true, width=0.45\textwidth]{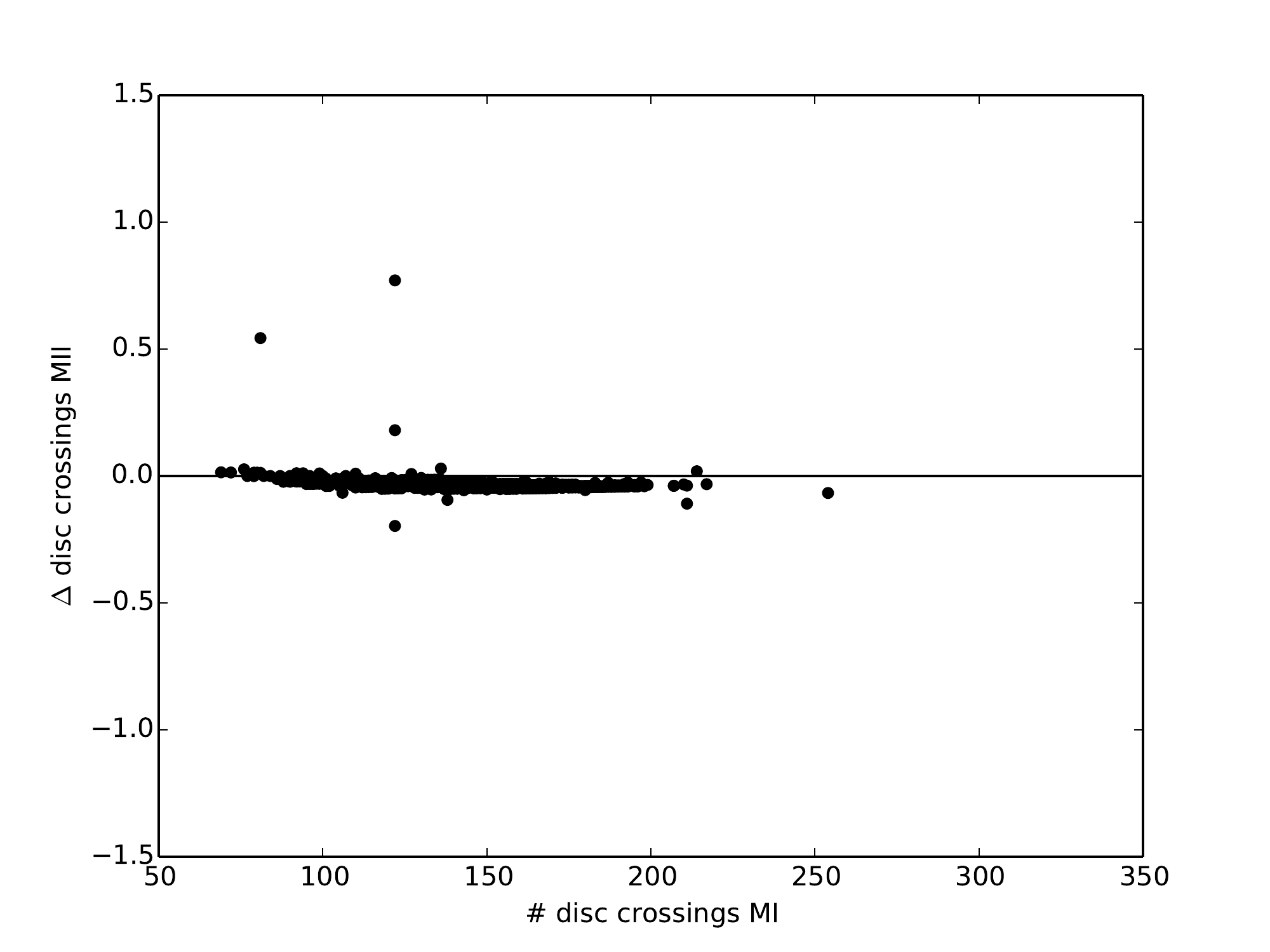}
\caption{Comparison of the orbital characteristics of stars of the \citet{adibek12} sample integrated in Model I versus Model II. From top-left to bottom-right: Pericentres $R_{min}$, apocentres $R_{max}$, maximum vertical distance from the plane $z_{max}$,  2D eccentricities , period of vertical oscillations, and number of disc crossings are given. The insets in some plots show a larger range of values than that covered in the corresponding main plots. In all the plots, the y-axis shows the difference between Model II and Model I, relative to Model I. Distances are in kpc, time in units of $10^8$ yr. }\label{allstarscomp2}
\end{figure*}
\section{Orbit integration}

\subsection{Numerical method}
We integrated in these two new mass models the orbits of  a thousand stars at the solar vicinity from the \citet{adibek12} sample, and the orbits of all the Galactic globular clusters with available positions and 3D velocities in the main catalogue of Casetti-Dinescu.\footnote{see http://www.astro.yale.edu/dana/gl\_2012\_J2000.cat1.} 
With this selection, we have the opportunity to have a limited but still representative sample of orbits of stars and stellar systems in the Galaxy, associated mainly to the thin and thick discs. Some halo stars are also present. 
To integrate the orbits, first of all we need to transform equatorial coordinates, parallaxes, heliocentric radial velocities, and proper motions in a cartesian Galactocentric inertial reference frame, where the $x$-axis coincides with the Sun-Galaxy centre direction and is positive towards the centre, the $y$-axis is oriented parallel to the motion of the Local Standard of Rest (LSR) in the disc, and the $z$-axis is positive towards the North Galactic pole. In this coordinate system, the position of the Sun is ($x_\odot$, $y_\odot$, $z_\odot$)=(-8.5, 0., 0.). To make the transformation, we have adopted the formulas in \citet{johnson87}, with coordinates defined at J2000.0 epoch. For each model we have assumed the velocity of the local standard of rest to be given by the value of the rotation curve at $R_{\odot}=8.5~$kpc. For Model I this choice gives $V_{LSR}=$221.4~km/s and for Model II this corresponds to a slightly higher velocity, $V_{LSR}=$230.0~km/s. For the solar motion with respect to the LSR, we have adopted the values given by  \citet{schonrich10}:  U$_\odot=$11.10~km/s, V$_\odot=$12.24~km/s, and W$_\odot=$7.25~km/s. 
Having defined the coordinate system, as well as the initial positions and velocities of all stars and globular clusters with respect to it, we have integrated the orbits forward in time for 7 Gyr, by using a leap-frog algorithm \citep[see for example][]{heggie03} with a constant time step of 0.1 Myr. Using this  time step, we obtain a very good energy conservation for all the integrated orbits, the average error in energy and angular momentum (that is $\Delta$E/E and $\Delta$L/L) is, respectively, of the order of  $10^{-6}$ and  $10^{-16}$ for stars, and $10^{-5}$ and  $10^{-16}$ for globular clusters.
The fact that the error in energy conservation is in general higher for globular clusters than solar vicinity stars is probably associated to the former larger eccentricities, and consequently larger variations in the range of accelerations that they experience.

\subsection{Orbits of thin and thick disc stars at the solar vicinity: the \citet{adibek12} sample}\label{stellarorbits}
 
 The orbits of a subsample of about one hundred stars extracted randomly from the Adibekyan sample, projected on the meridional plane $R-z$, are shown in Fig.~\ref{allstarsRz}.
 As evident from the plot, the sample contains thin and thick disc stars whose orbits span a range of characteristics: some stars have orbits confined in the Galactic plane, with very small excursions from the solar radius (low eccentricities, thin disc orbits), while some other stars have orbits still confined in the Galactic plane, but with larger radial excursions, leading them to reach larger distances from the Sun (high eccentricity, thin disc orbits). Finally, thick disc stars are in general characterized by mild to high eccentric orbits with moderate to large vertical excursions from the Galactic plane. From this plot, one can already note some differences between the appearance of the orbits in the three models: in Models I and II some orbits appear to be thicker (i.e. larger vertical excursions) than the corresponding orbits in Model A\&S; moreover Model II shows outer thin disc orbits which are slightly less elongated than the corresponding orbits in Models I and A\&S. This last trend is particularly evident for stars reaching larger apogalactica ($R_{max} > 20$),  as we will comment in the following part of this section. 
 
A quantitative comparison between the orbital characteristics recovered in Model I, Model II, and Model A\&S is made in Fig.~\ref{allstarscomp}, where the pericentres $R_{min}$, apocentres $R_{max}$, maximal vertical distances from the plane $z_{max}$,  2D eccentricities\footnote{The 2D eccentricity is defined as $\rm ecc2D=(R_{max}-R_{min})/(R_{max}+R_{min}).$}, number of disc crossings, and period of vertical oscillations are shown and compared to those predicted when integrating in the \citet{allen91} model. While the values of pericentres are very similar in the three models, one can see that some differences appear in the apocentres of stars, $R_{max}$, with Model II predicting apocentres up to 36\% smaller than those recovered by the \citet{allen91} model.  The largest differences are found for stars with large apogalactica ($R_{max} >~15$~kpc), such as HIP71979, HIP87062, HIP34285, and HIP74234. This behaviour is a consequence of the deeper gravitational potential of Model II, compared to Models I and A\&S: at large radii, Fig.~\ref{potmap} indeed shows that Model II has a deeper potential well than that of the two other models, with the consequence that stars  are more bound. This is also visible in the maximum vertical heights, $z_{max}$,  reached by their orbits: stars whose $z_{max}$ is greater than 5~kpc  in Model A\&S,  never reach vertical distances from the plane greater than 5~kpc in Model II. Among those stars are  HIP34285, HIP63918, HIP80837, HIP100568, and HIP116285. This is, again, a consequence of the presence of a more massive dark matter halo in Model II, with respect to Models I and A\&S. 

For stars  whose vertical excursion is closer to the Galactic plane, the inclusion of a (massive) thick disc tends to increase the value of $z_{max}$: Model I and II can indeed lead to maximum vertical heights up to 60\% higher than that predicted by Model A\&S (Fig.~\ref{allstarscomp}, middle left panel).  This is because in the A\&S model, having no thick disc, all the disc mass is more 
concentrated towards the Galactic plane, increasing the restoring force at small heights (see Fig.~\ref{kz}).
Because of the larger vertical excursions in Models I and II, the period of vertical oscillations of those stars in general increases, and as a consequence, the number of disc crossings decreases. No significant differences are found in the 2D eccentricities predicted by the different models.

Finally, Fig.~\ref{allstarscomp2} shows the comparison between Model I and Model II, by showing the differences $\Delta$ between the orbital characteristics of Model I and Model II. For each quantity (i.e. pericentre, apocentre, maximum vertical distance from the plane, etc.. ) $\Delta$ is defined as the difference between the value attained by this quantity in Model II minus the corresponding value for Model I, all normalized to Model I. One sees that for most of the parameters, there is no significant difference between the two models, except for: (1) the pericentre distances $R_{min}$, where the relative difference between Model I and Model II can be as high as 50\% for stars with pericentres in the innermost regions of the Galaxy ($R < 2$~kpc); (2) the apocentres found by integrating the orbits of stars in Model I and Model II can differ by more than 10\% for stars having large $R_{max}$.

\subsection{Orbits of Galactic globular clusters from the Casetti-Dinescu catalogue}\label{clusterorbits}

\begin{figure*}
\centering
\includegraphics[trim=2.5cm 0cm 1.cm 0cm, clip=true, width=0.45\textwidth]{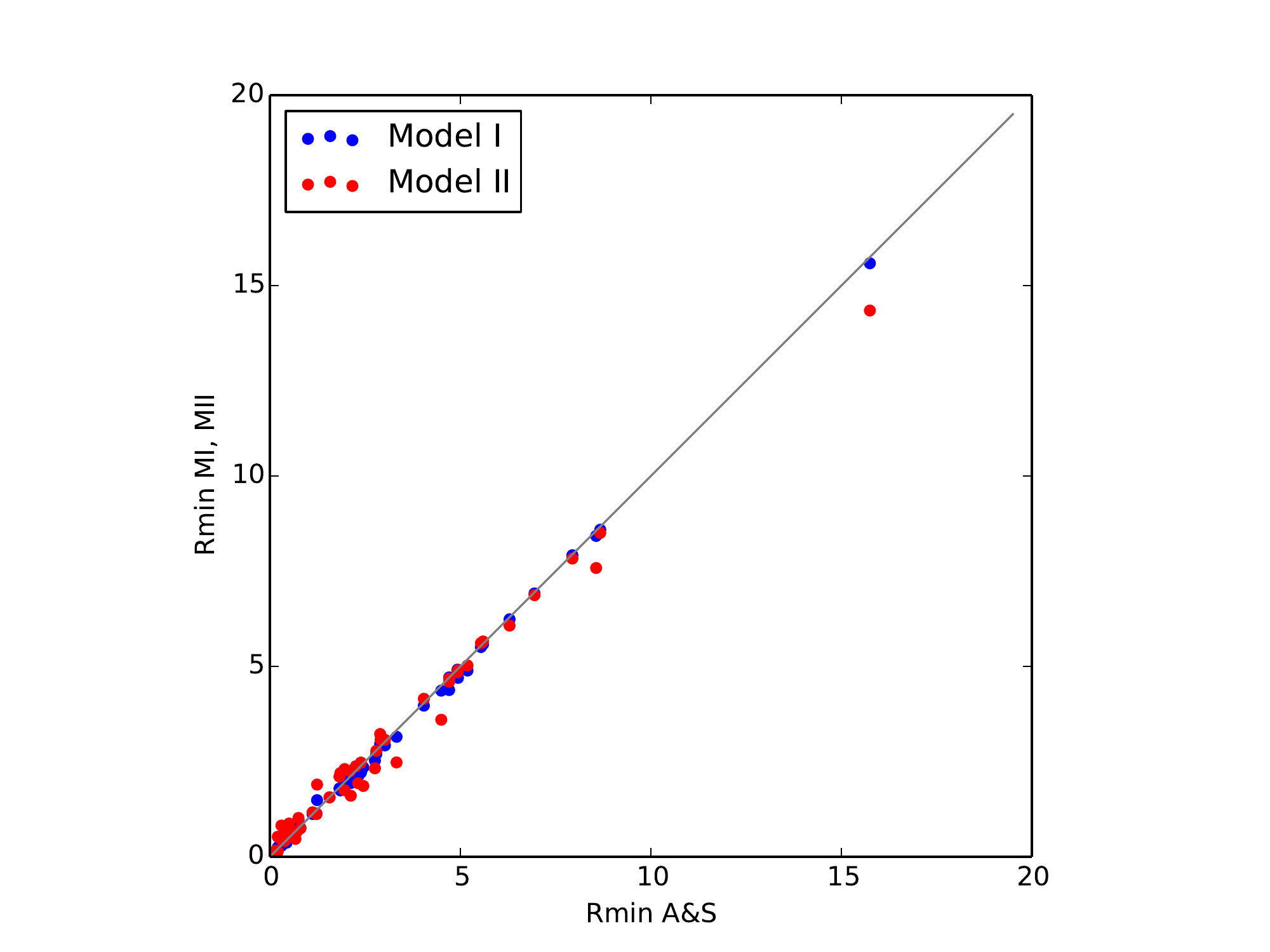}
\includegraphics[trim=2.5cm 0cm 1.cm 0cm, clip=true, width=0.45\textwidth]{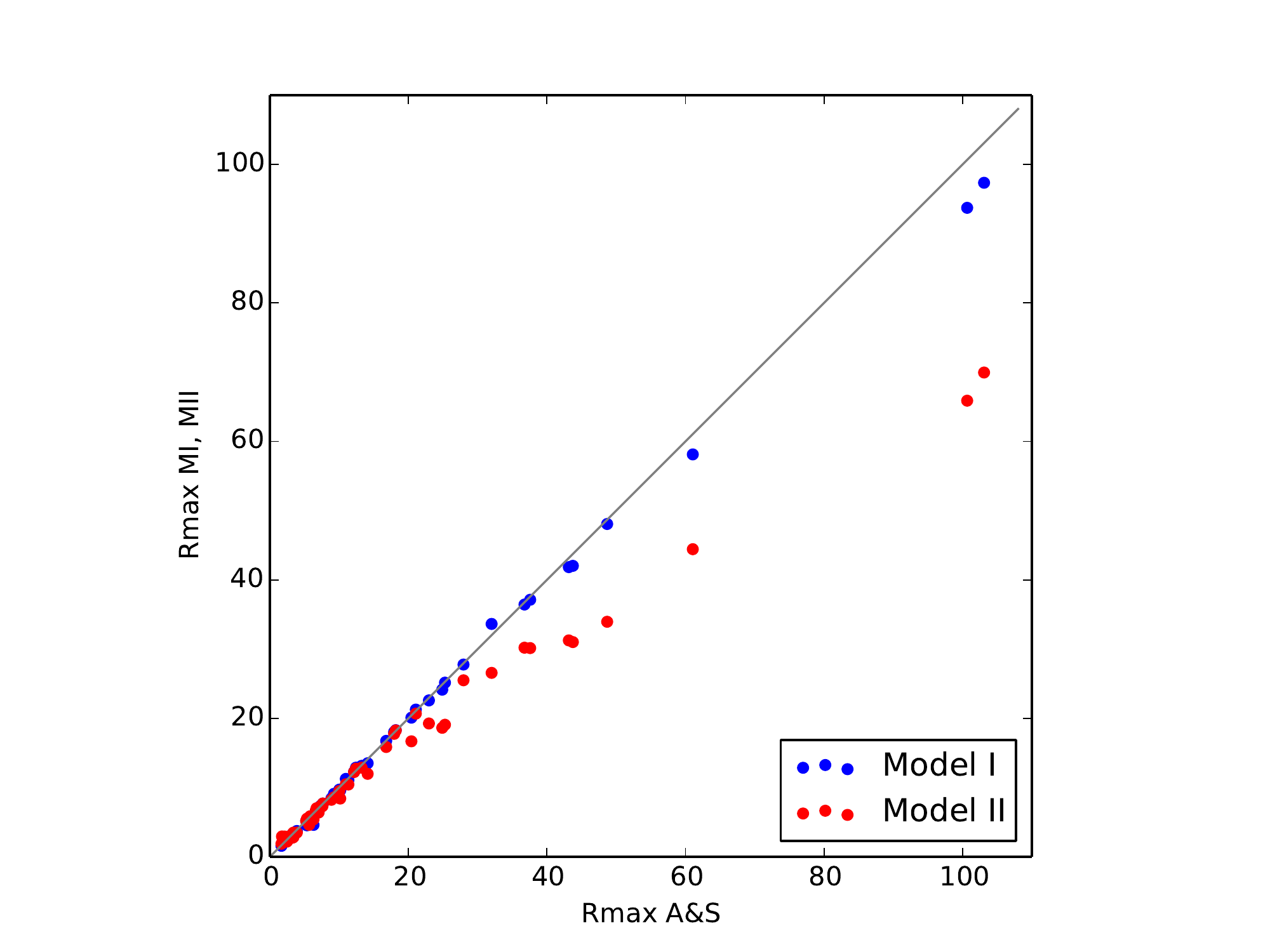}
\includegraphics[trim=2.5cm 0cm 1.cm 0cm, clip=true, width=0.45\textwidth]{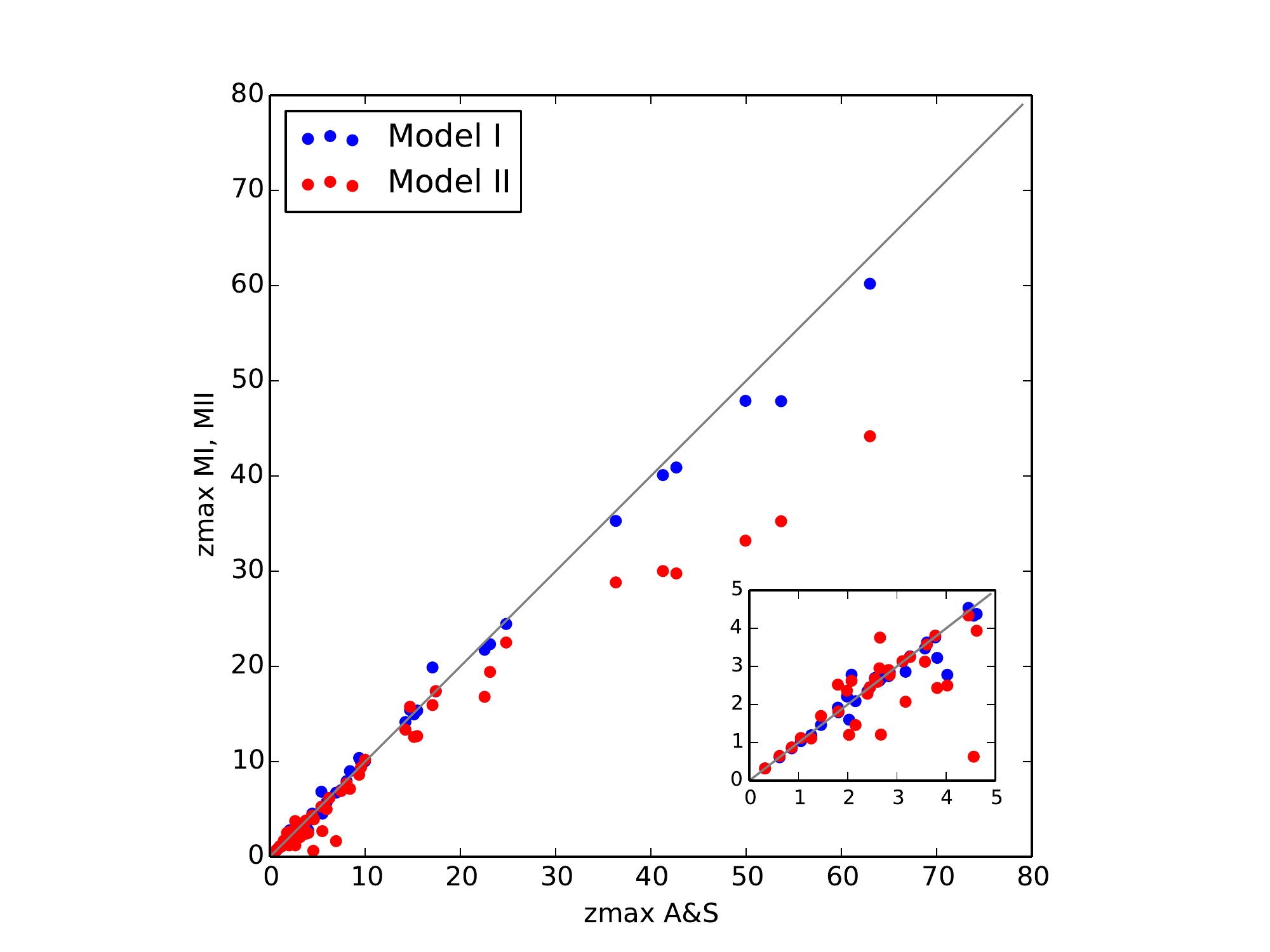}
\includegraphics[trim=2.5cm 0cm 1.cm 0cm, clip=true, width=0.45\textwidth]{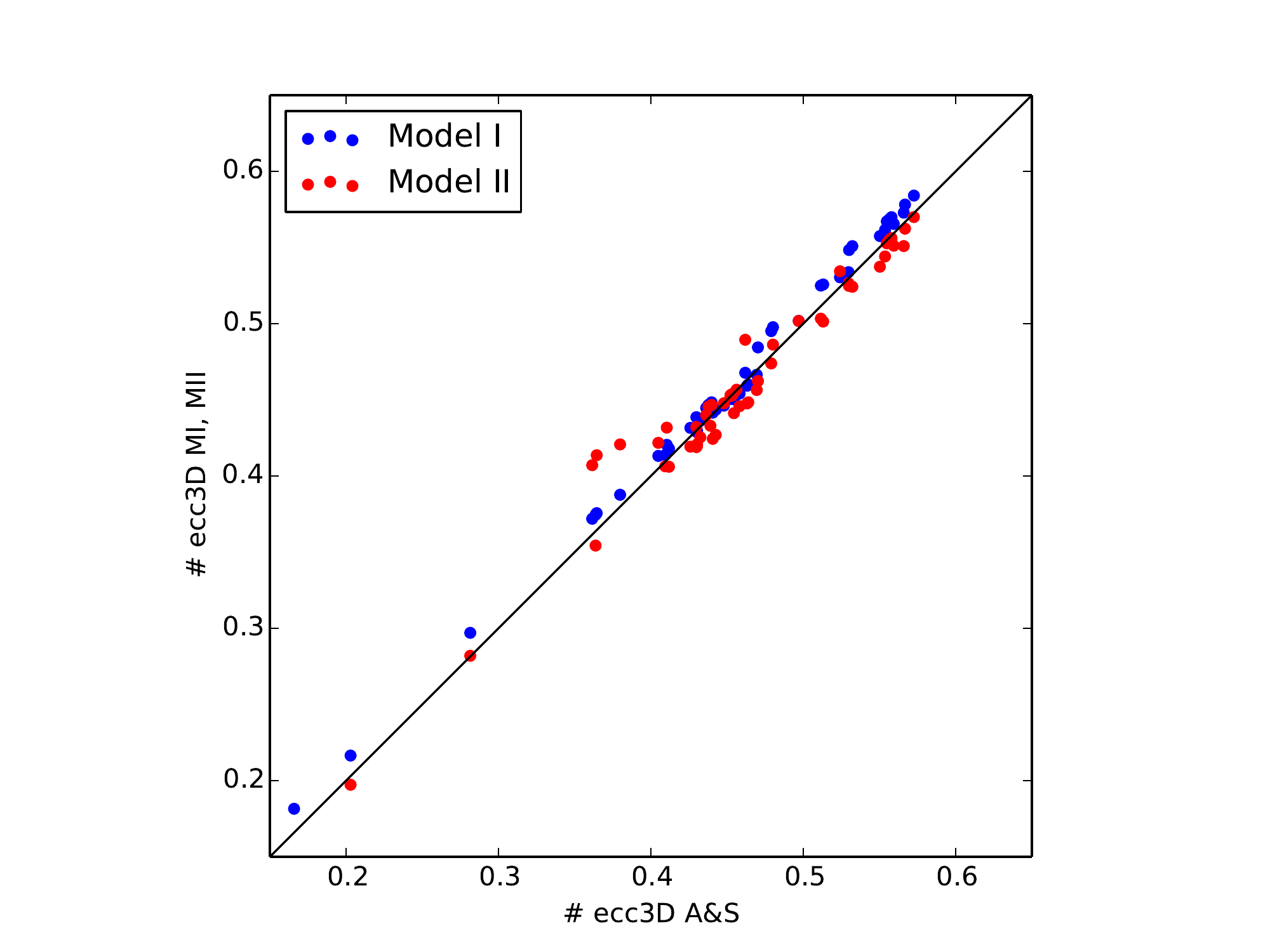}
\includegraphics[trim=2.5cm 0cm 1.cm 0cm, clip=true, width=0.45\textwidth]{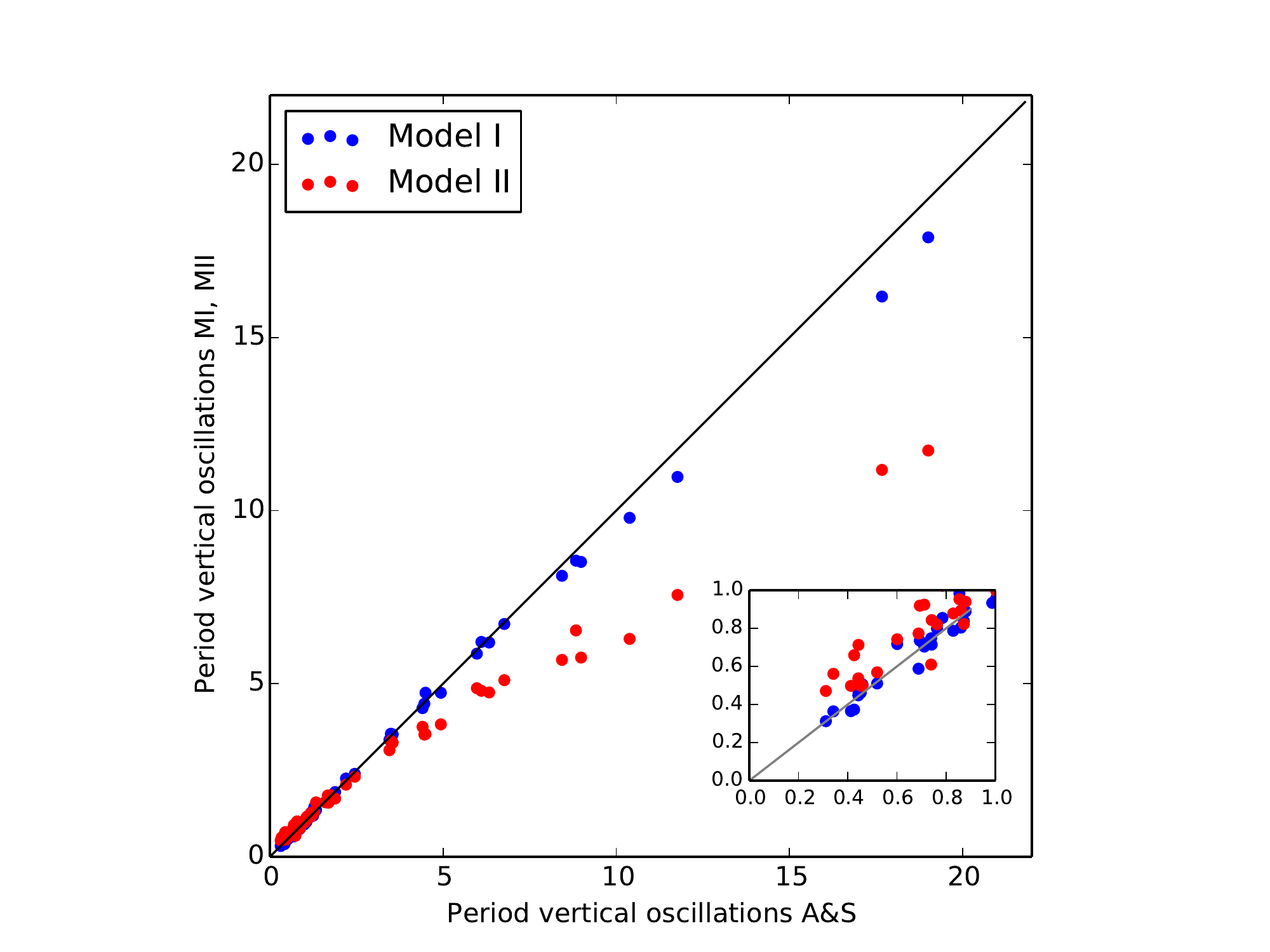}
\includegraphics[trim=2.5cm 0cm 1.cm 0cm, clip=true, width=0.45\textwidth]{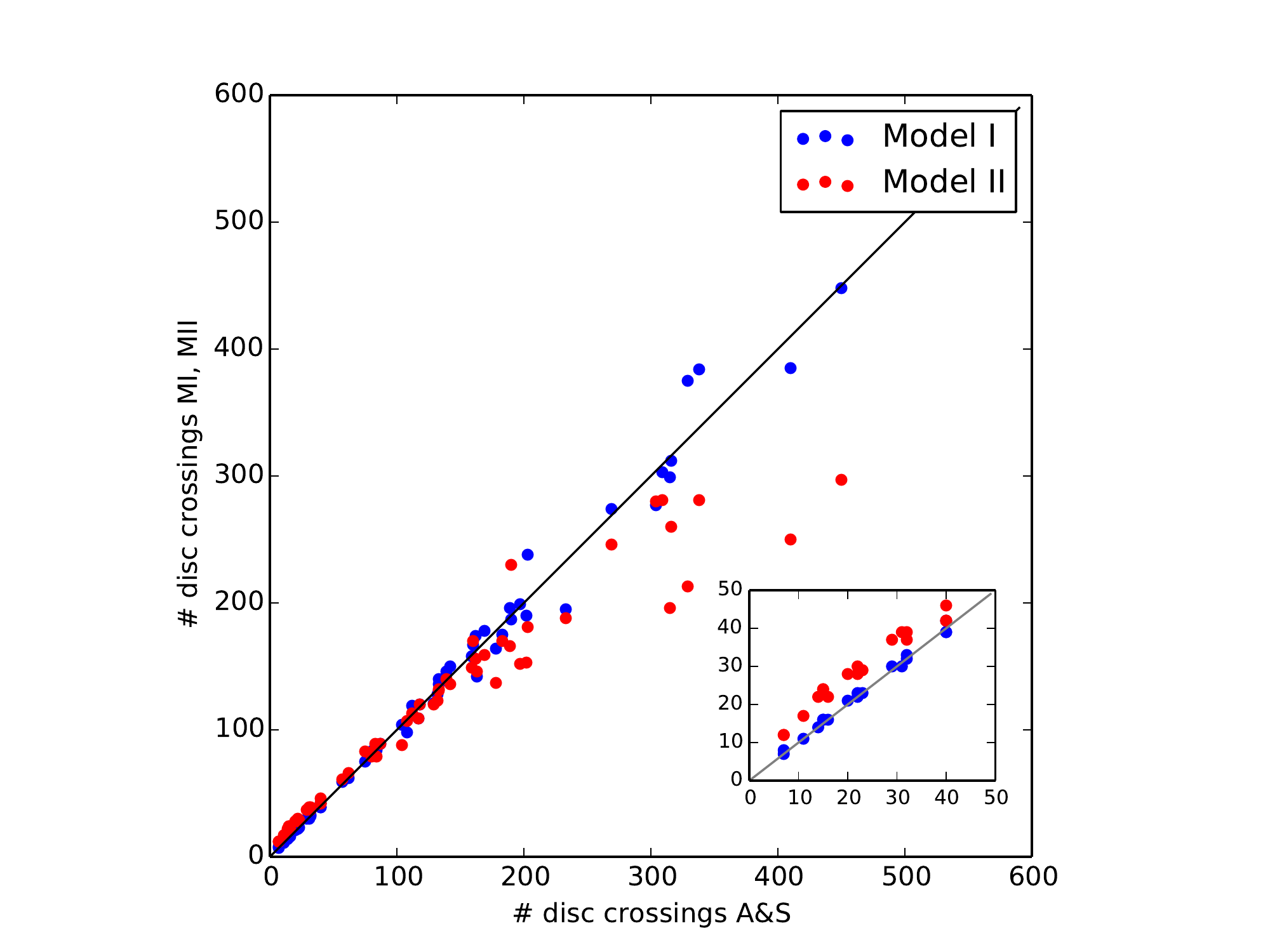}
\caption{Comparison of the orbital characteristics of Galactic globular clusters  integrated in Model I (blue points) and Model II (red points) versus the \citet{allen91} model. From top-left to bottom-right: Pericentres $R_{min}$, apocentres $R_{max}$, maximum vertical distance from the plane $z_{max}$,  period of vertical oscillations, and 3D eccentricities are given.  The insets in some of the plots show a zoom in the inner regions. In all the plots, distances are in kpc, time in units of $10^8$ yr. }\label{allGCcomp}
\end{figure*}

\begin{figure*}
\centering
\includegraphics[trim=0.cm 0cm 2.5cm 0cm, clip=true, width=0.24\textwidth]{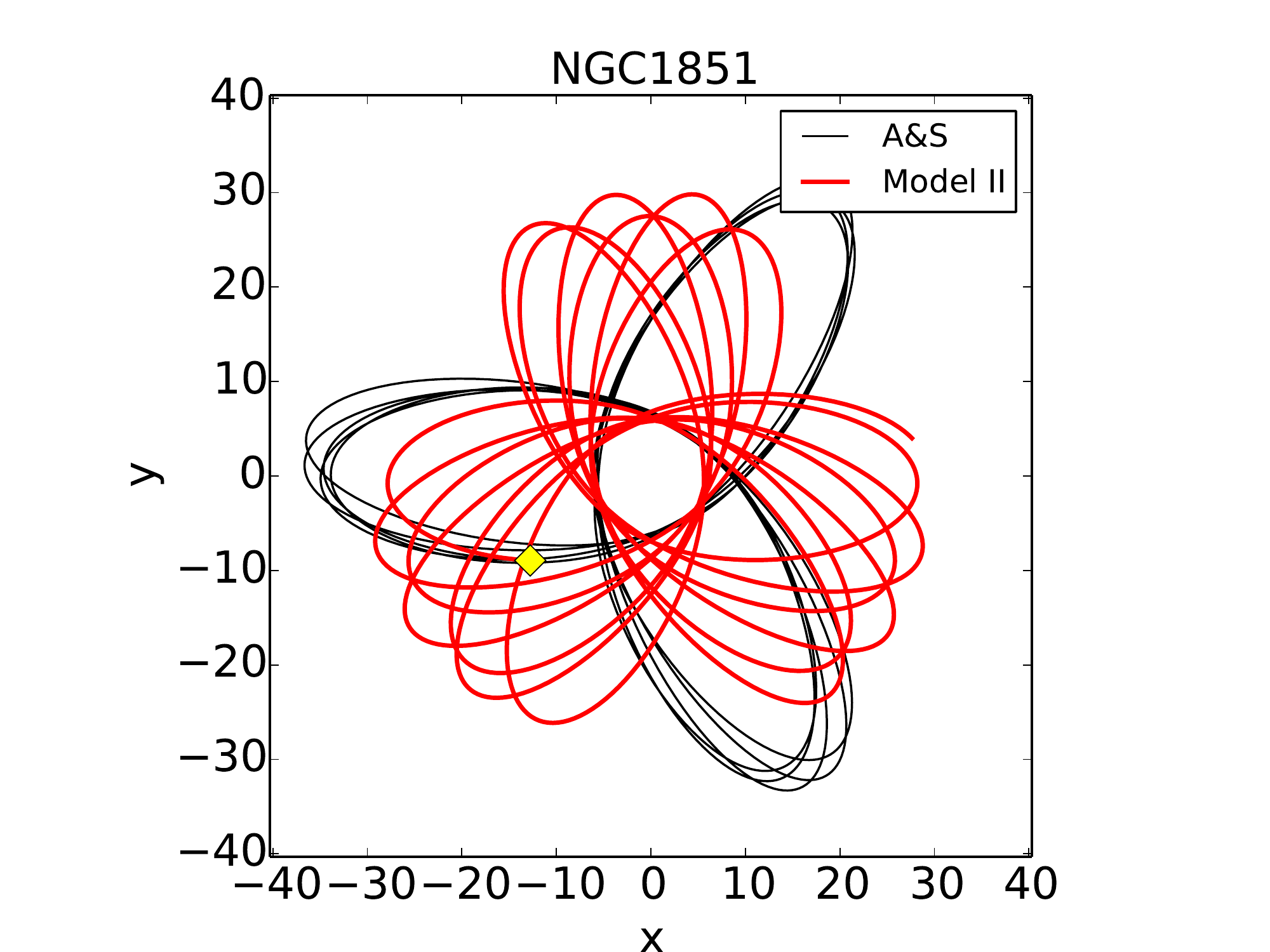}
\includegraphics[trim=0.cm 0cm 2.cm 0cm, clip=true, width=0.24\textwidth]{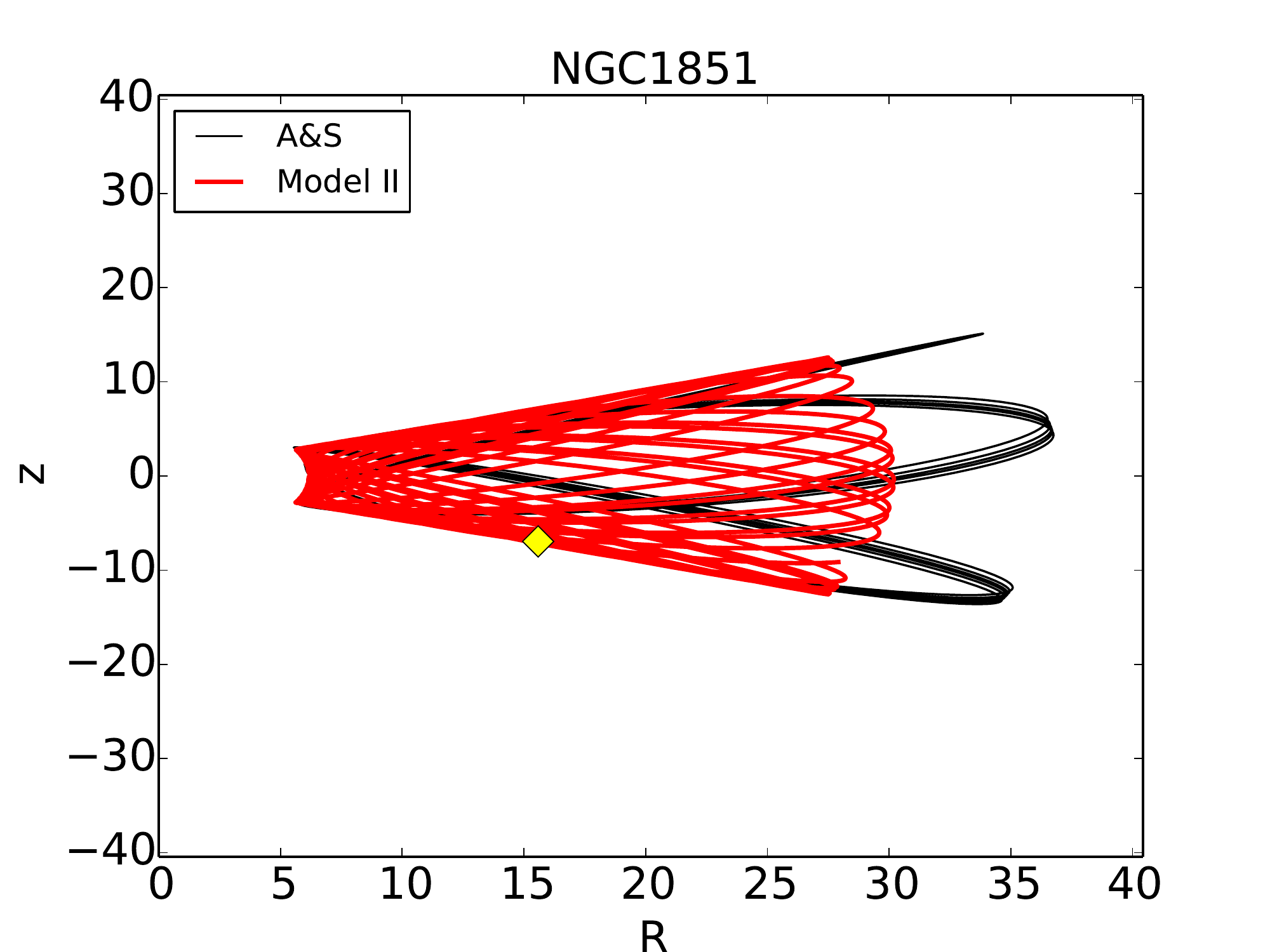}
\includegraphics[trim=0.cm 0cm 2.5cm 0cm, clip=true, width=0.24\textwidth]{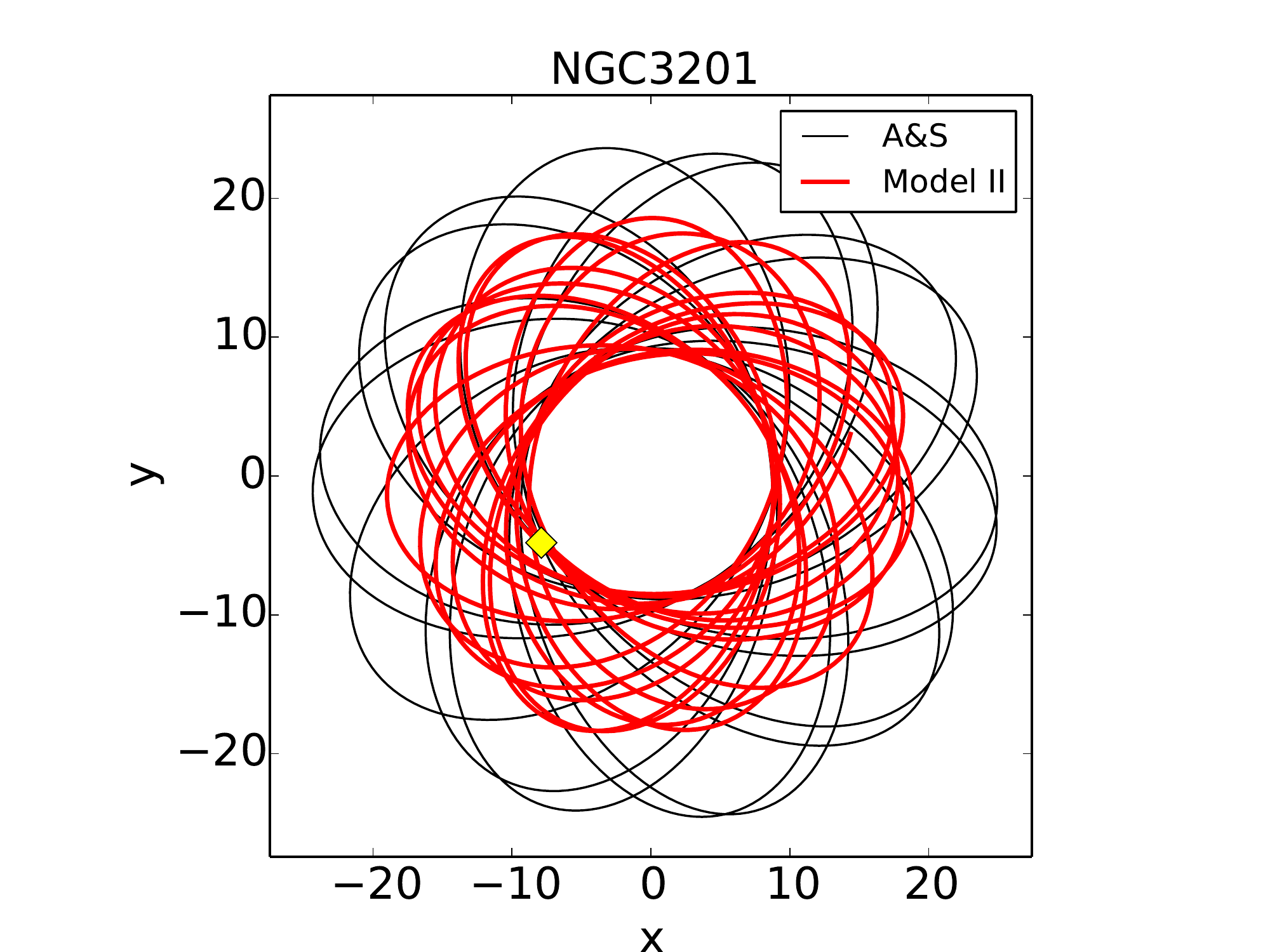}
\includegraphics[trim=0.cm 0cm 2.cm 0cm, clip=true, width=0.24\textwidth]{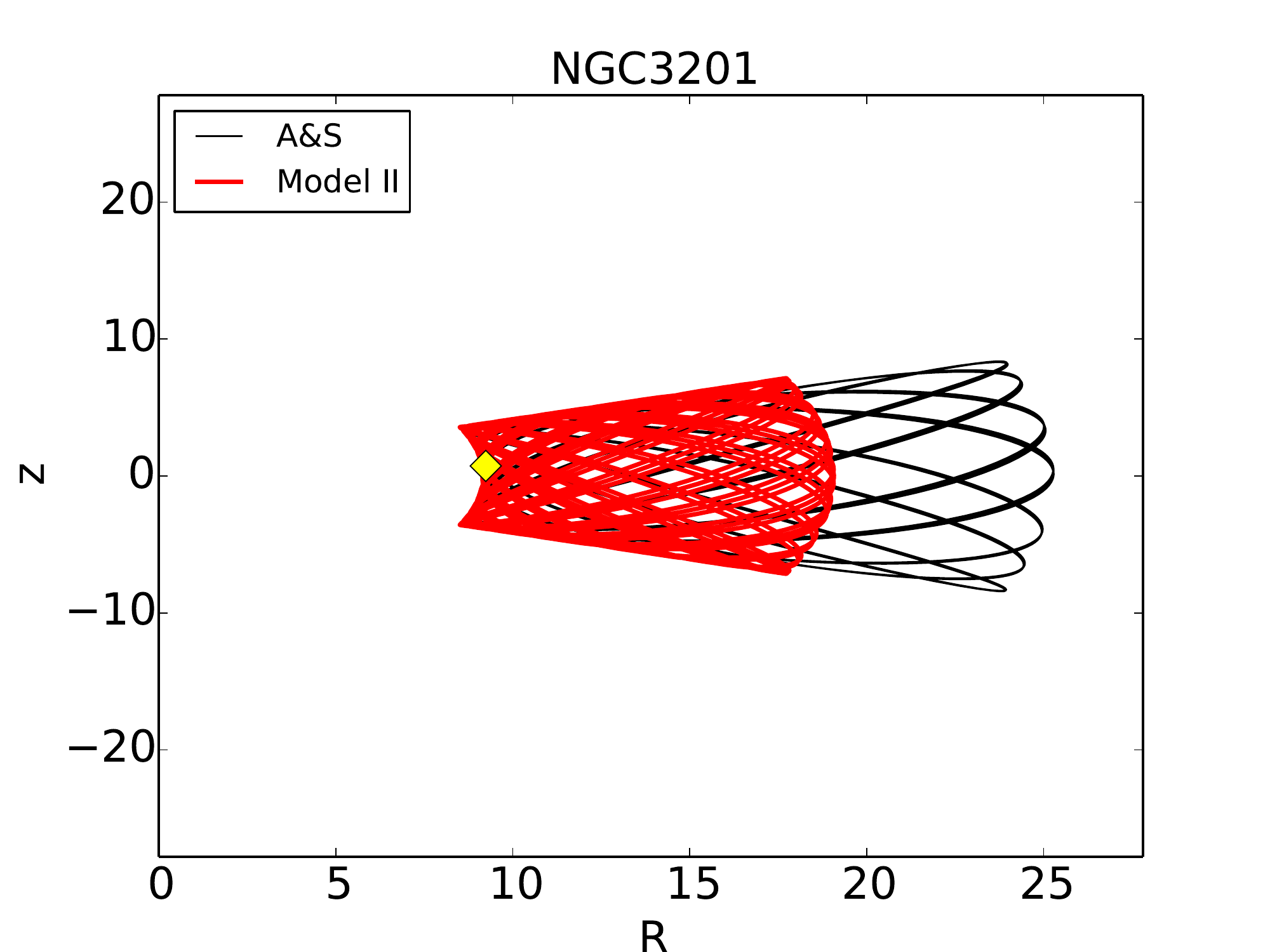}

\includegraphics[trim=0.cm 0cm 2.5cm 0cm, clip=true, width=0.24\textwidth]{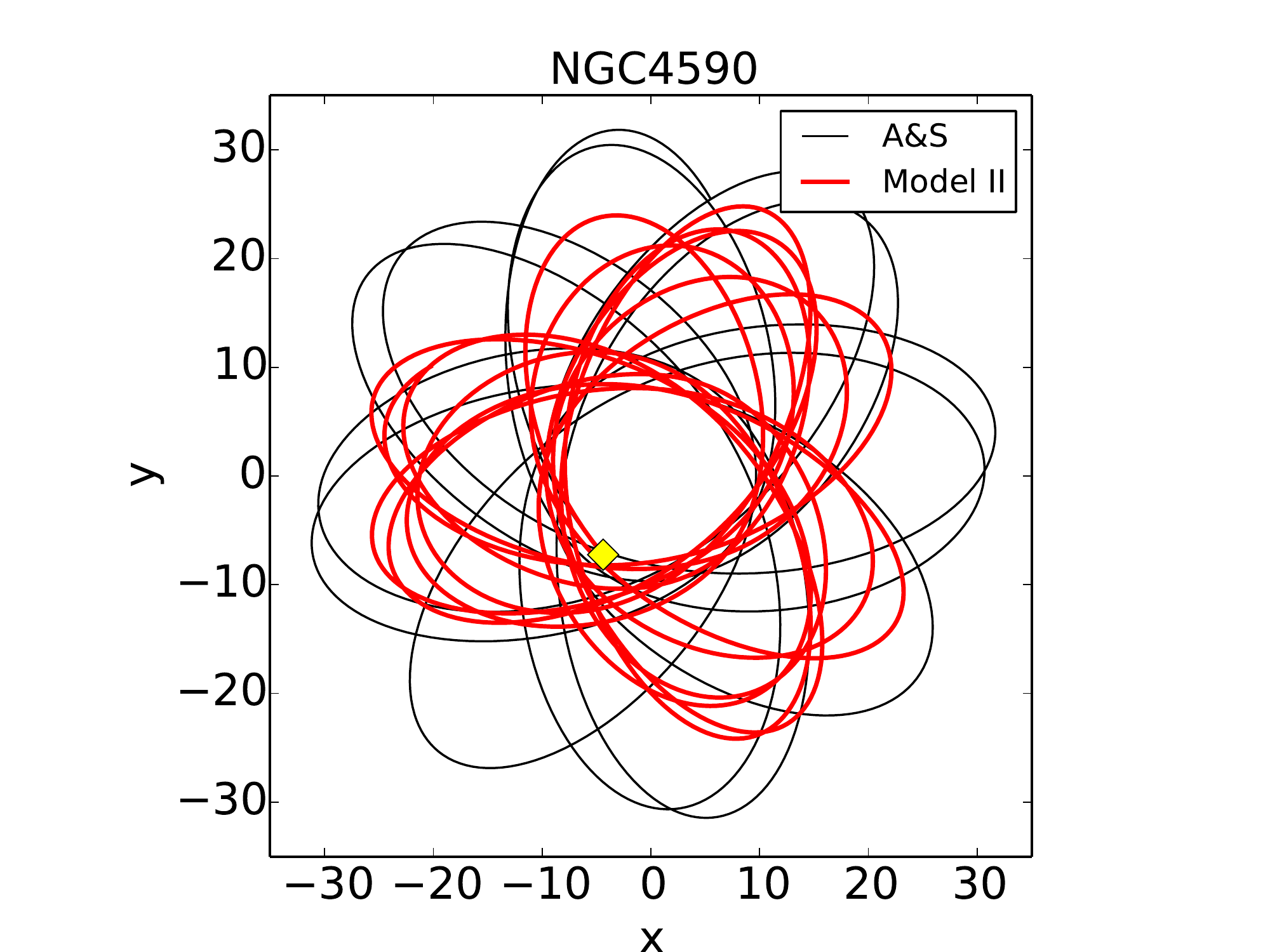}
\includegraphics[trim=0.cm 0cm 2.cm 0cm, clip=true, width=0.24\textwidth]{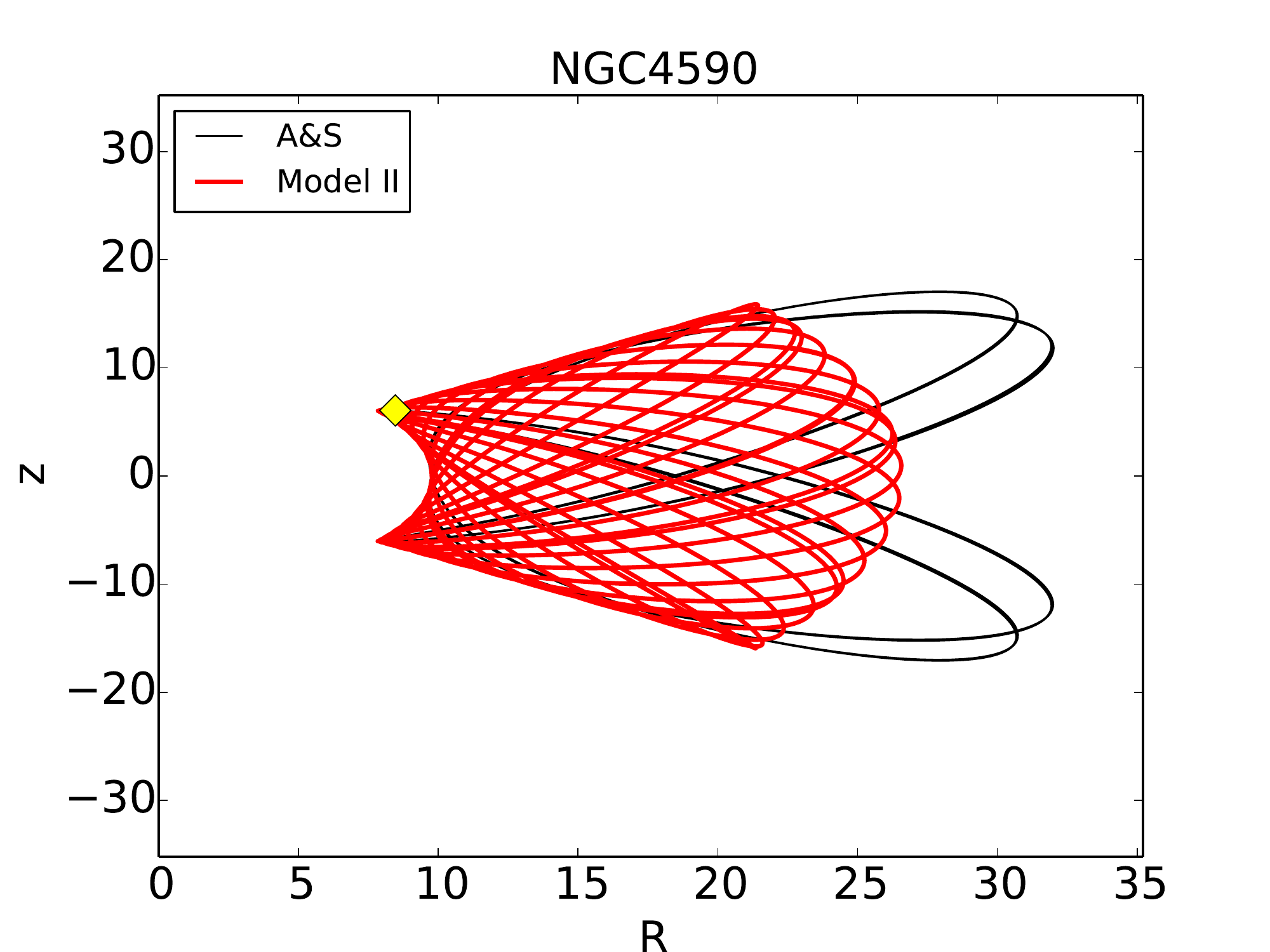}
\includegraphics[trim=0.cm 0cm 2.5cm 0cm, clip=true, width=0.24\textwidth]{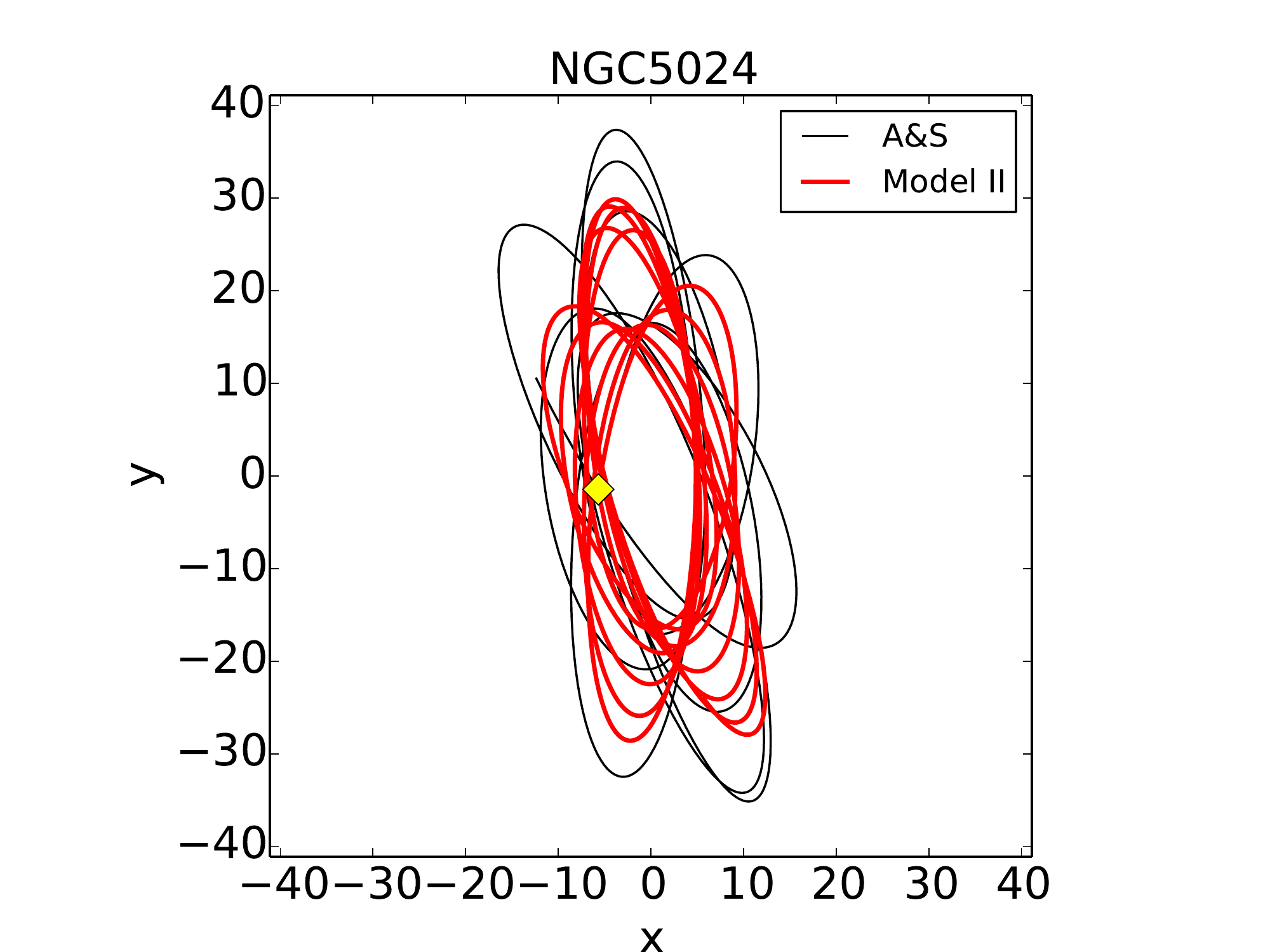}
\includegraphics[trim=0.cm 0cm 2.cm 0cm, clip=true, width=0.24\textwidth]{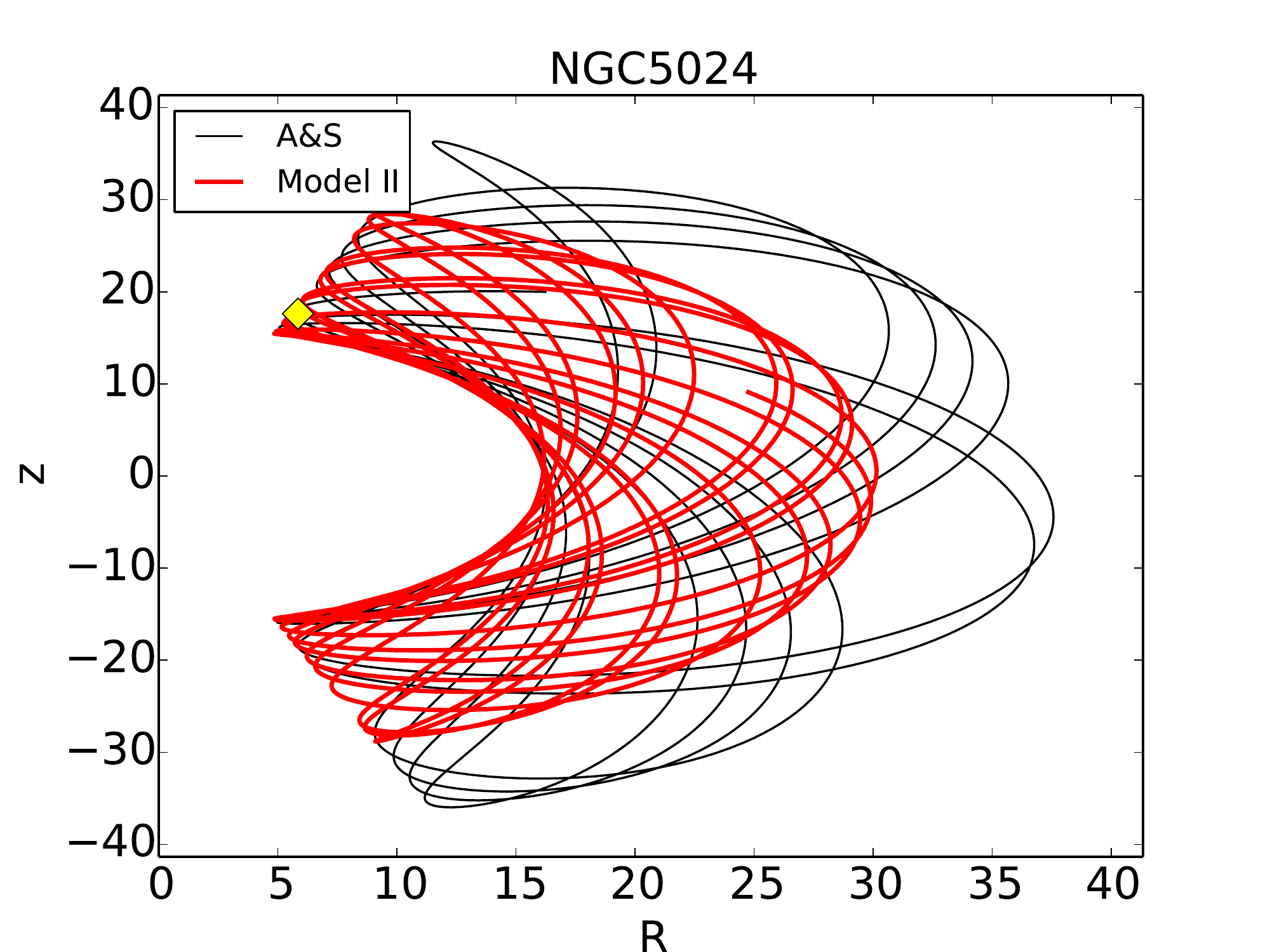}

\includegraphics[trim=0.cm 0cm 2.5cm 0cm, clip=true, width=0.24\textwidth]{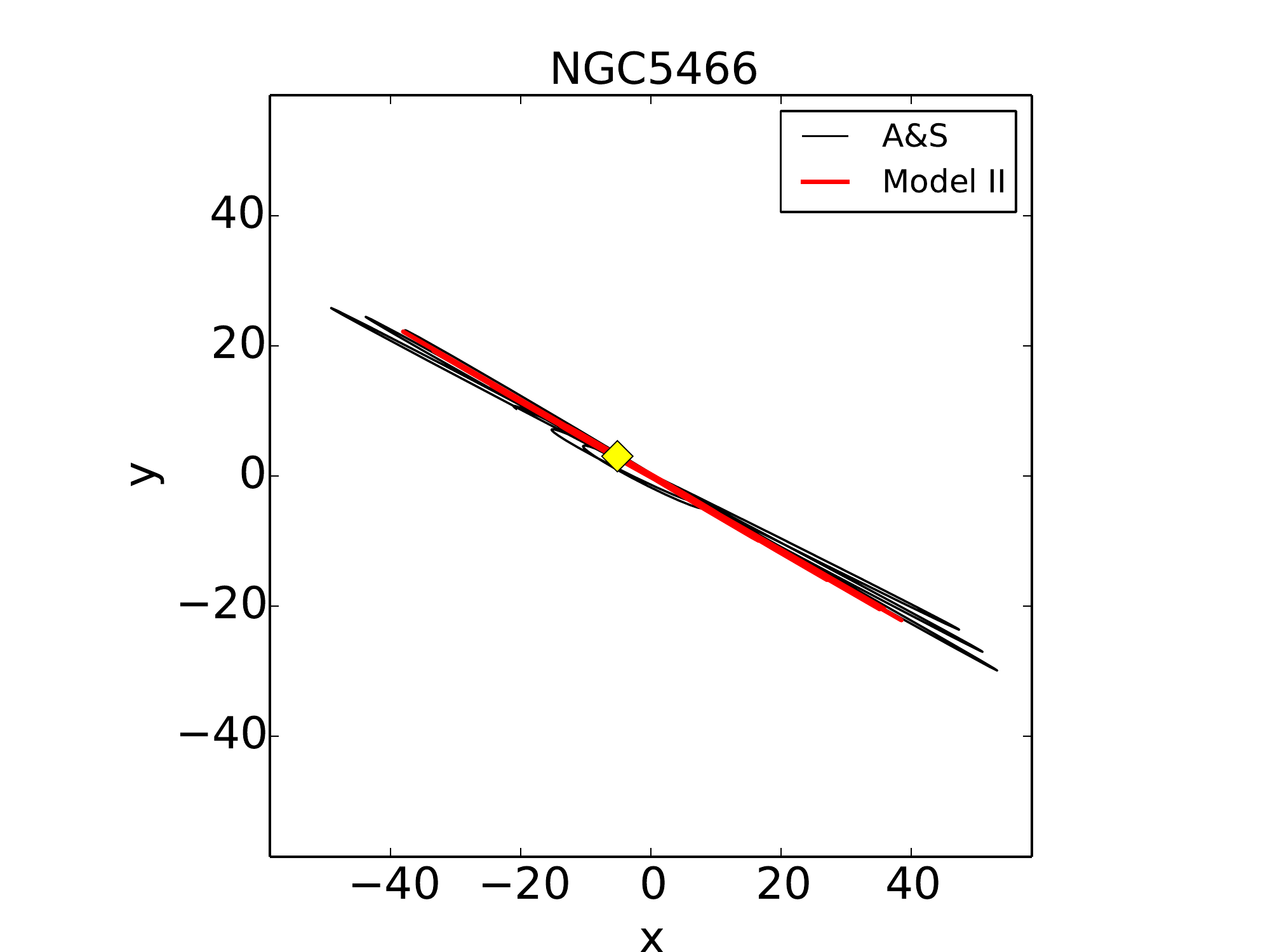}
\includegraphics[trim=0.cm 0cm 2.cm 0cm, clip=true, width=0.24\textwidth]{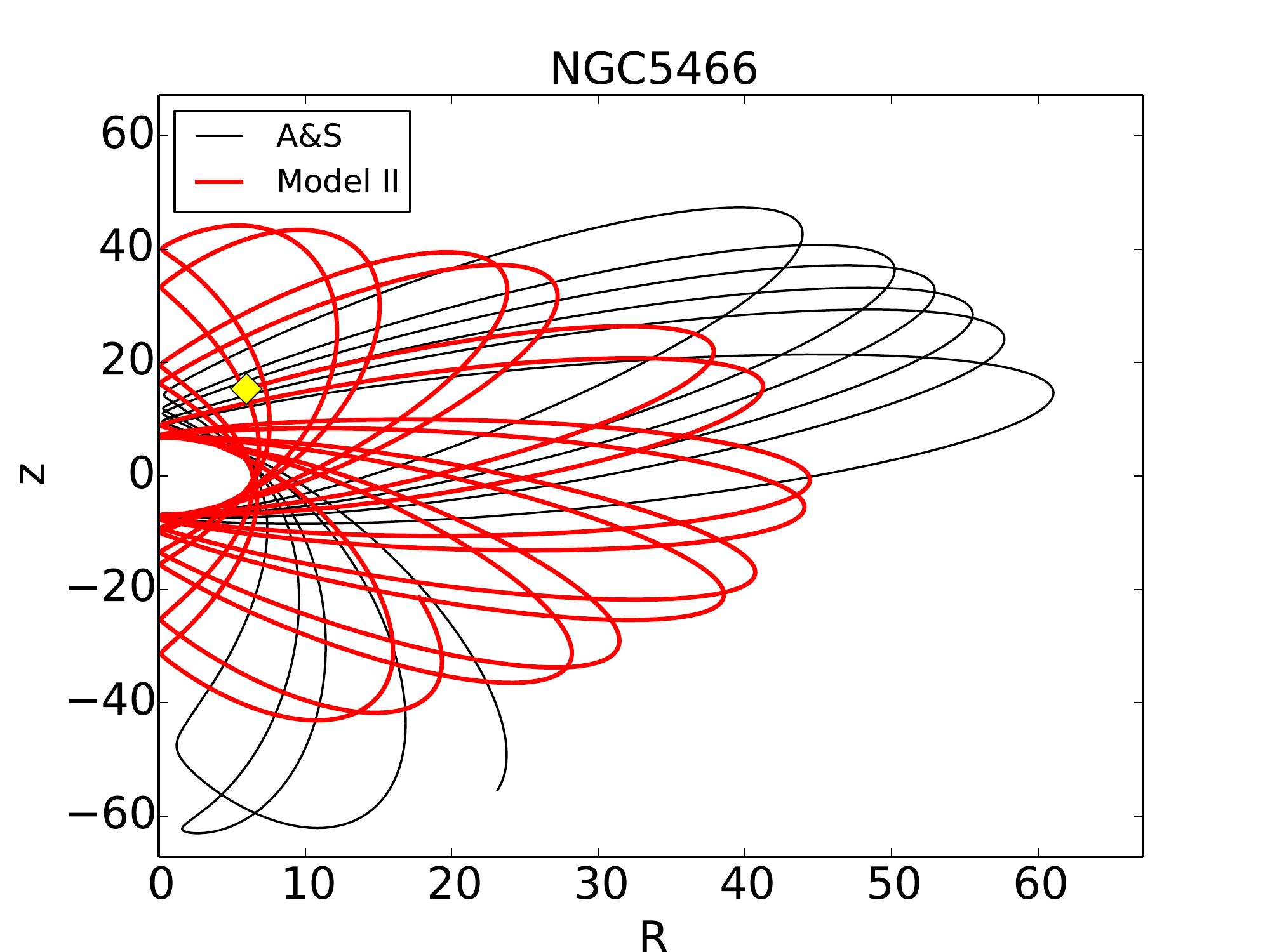}
\includegraphics[trim=0.cm 0cm 2.5cm 0cm, clip=true, width=0.24\textwidth]{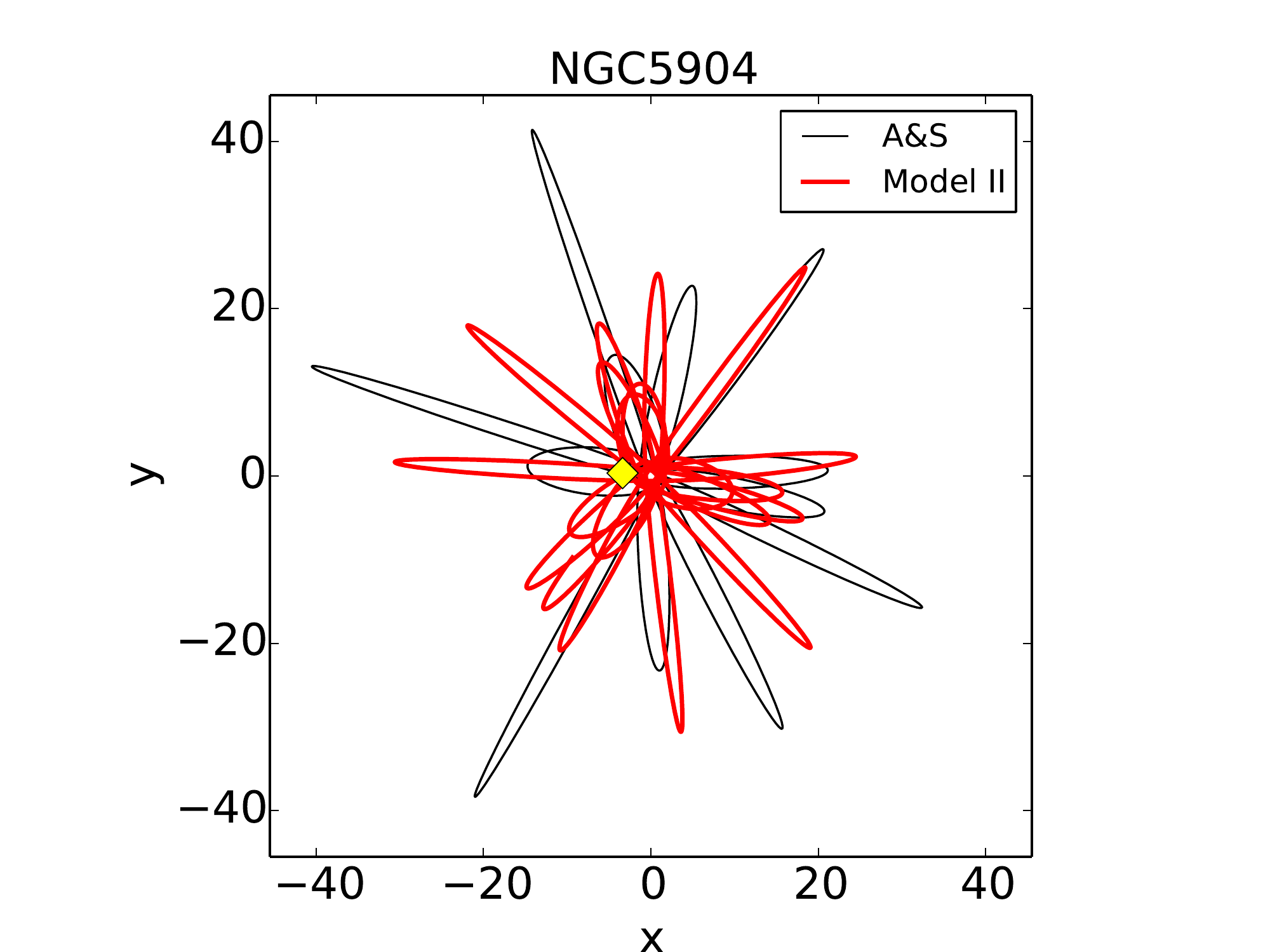}
\includegraphics[trim=0.cm 0cm 2.cm 0cm, clip=true, width=0.24\textwidth]{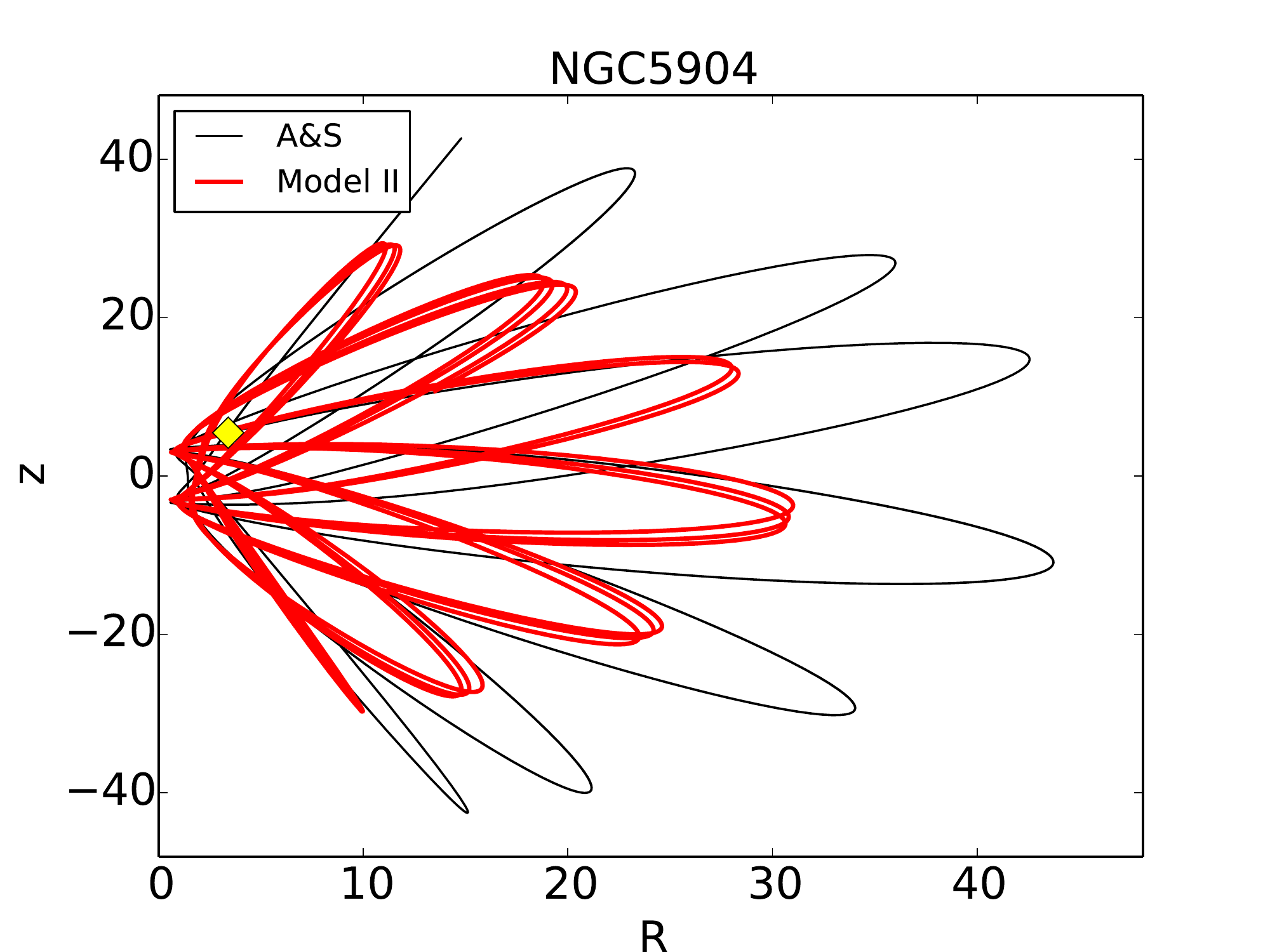}

\includegraphics[trim=0.cm 0cm 2.5cm 0cm, clip=true, width=0.24\textwidth]{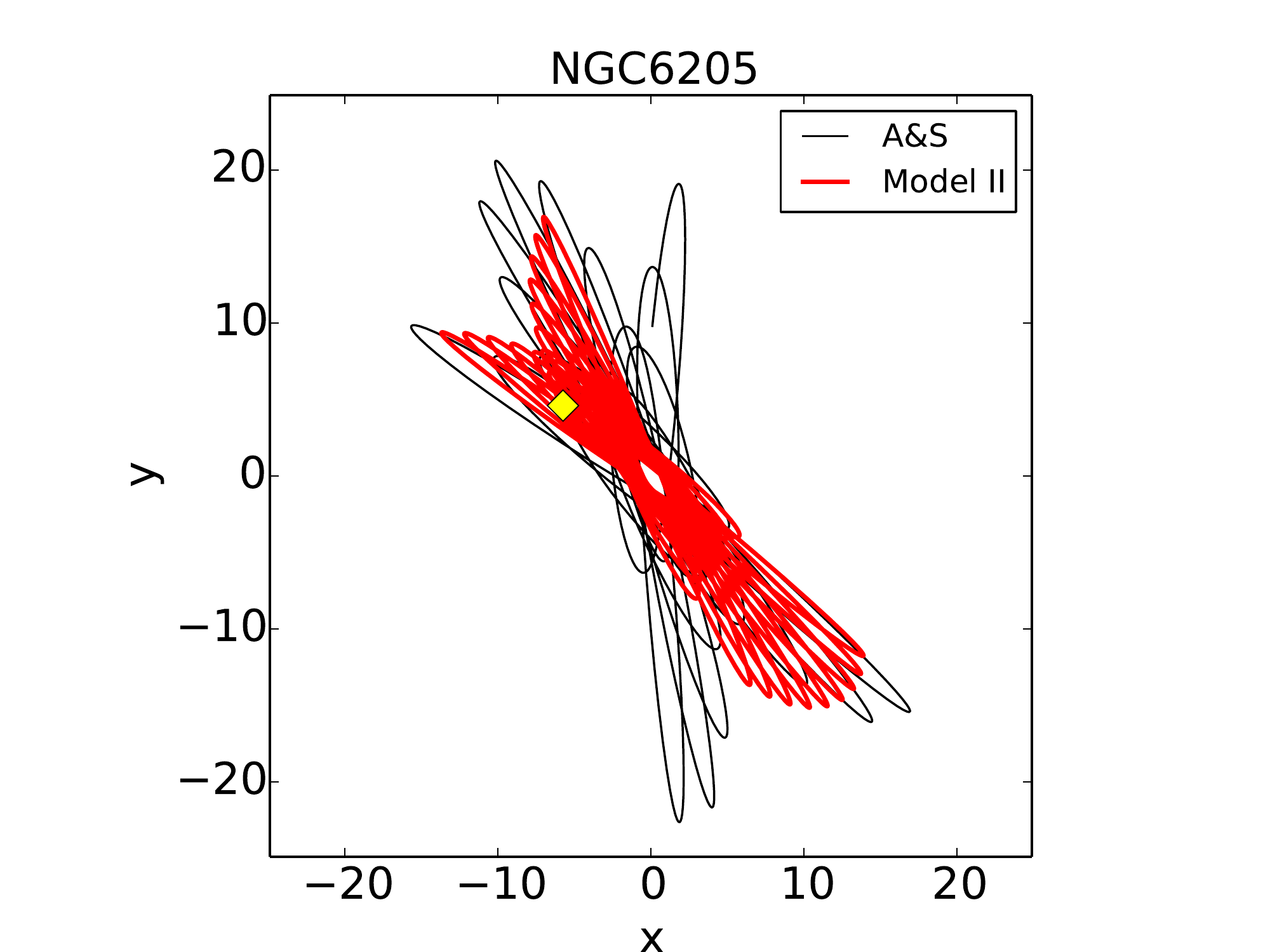}
\includegraphics[trim=0.cm 0cm 2.cm 0cm, clip=true, width=0.24\textwidth]{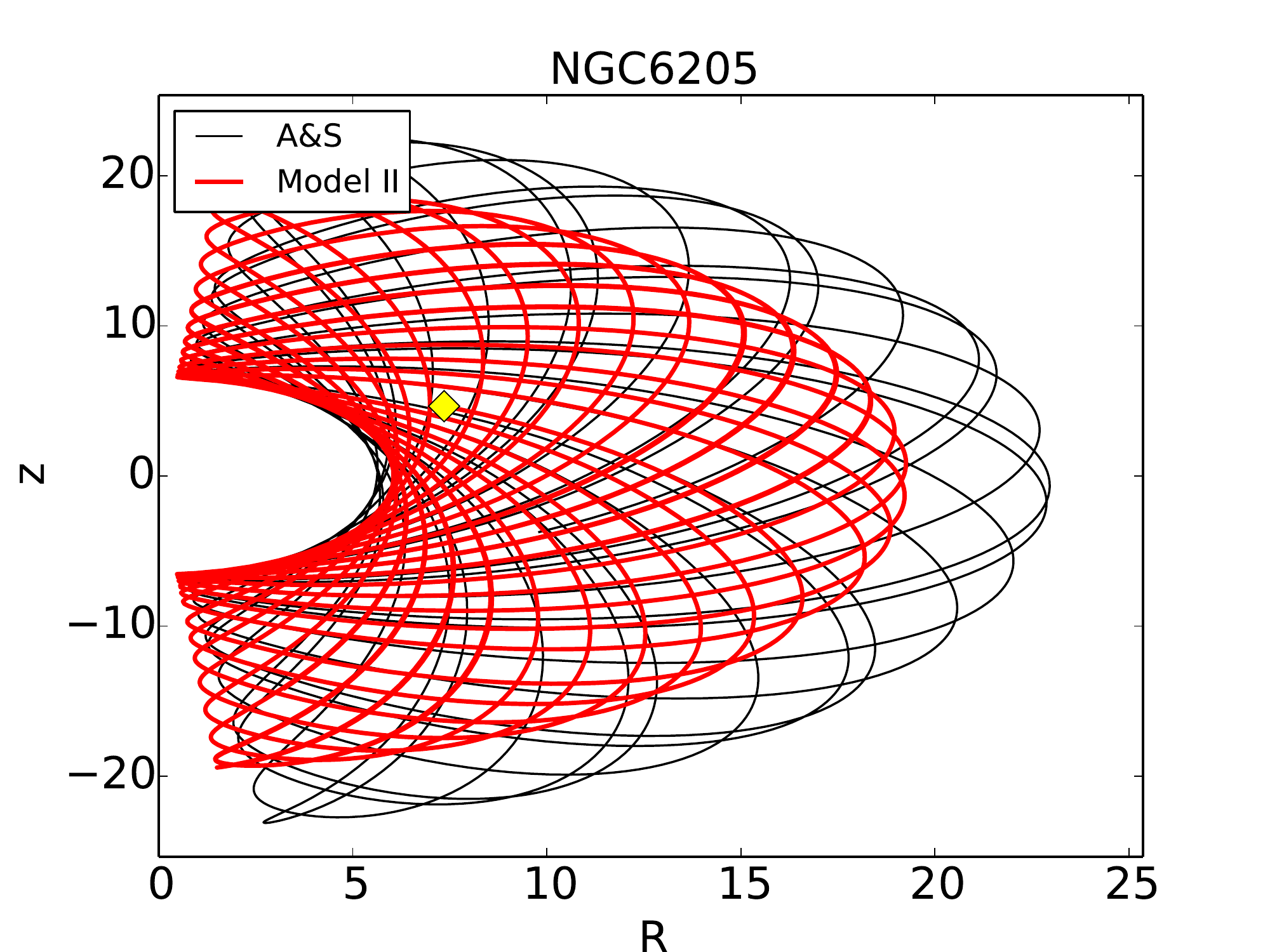}
\includegraphics[trim=0.cm 0cm 2.5cm 0cm, clip=true, width=0.24\textwidth]{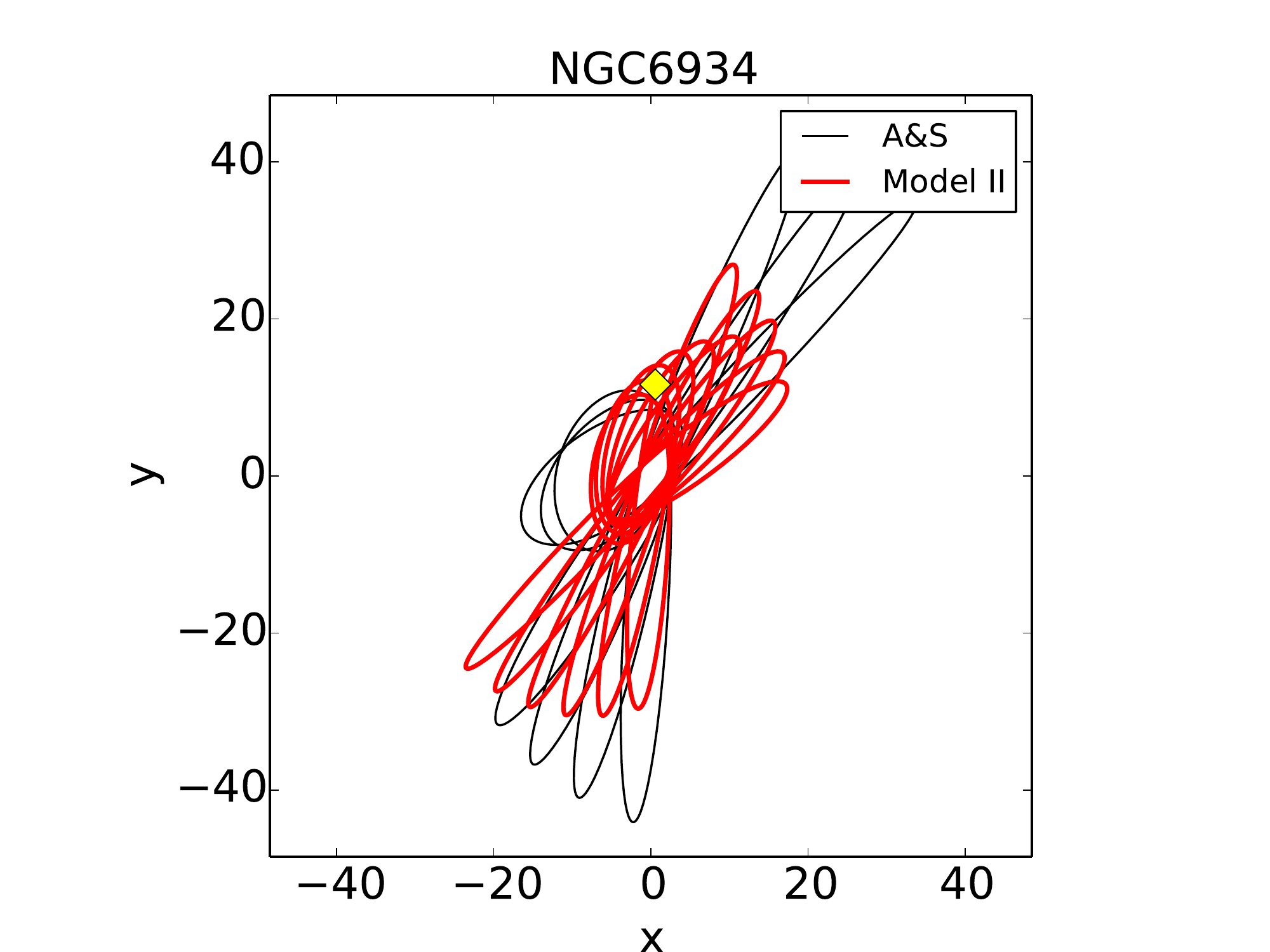}
\includegraphics[trim=0.cm 0cm 2.cm 0cm, clip=true, width=0.24\textwidth]{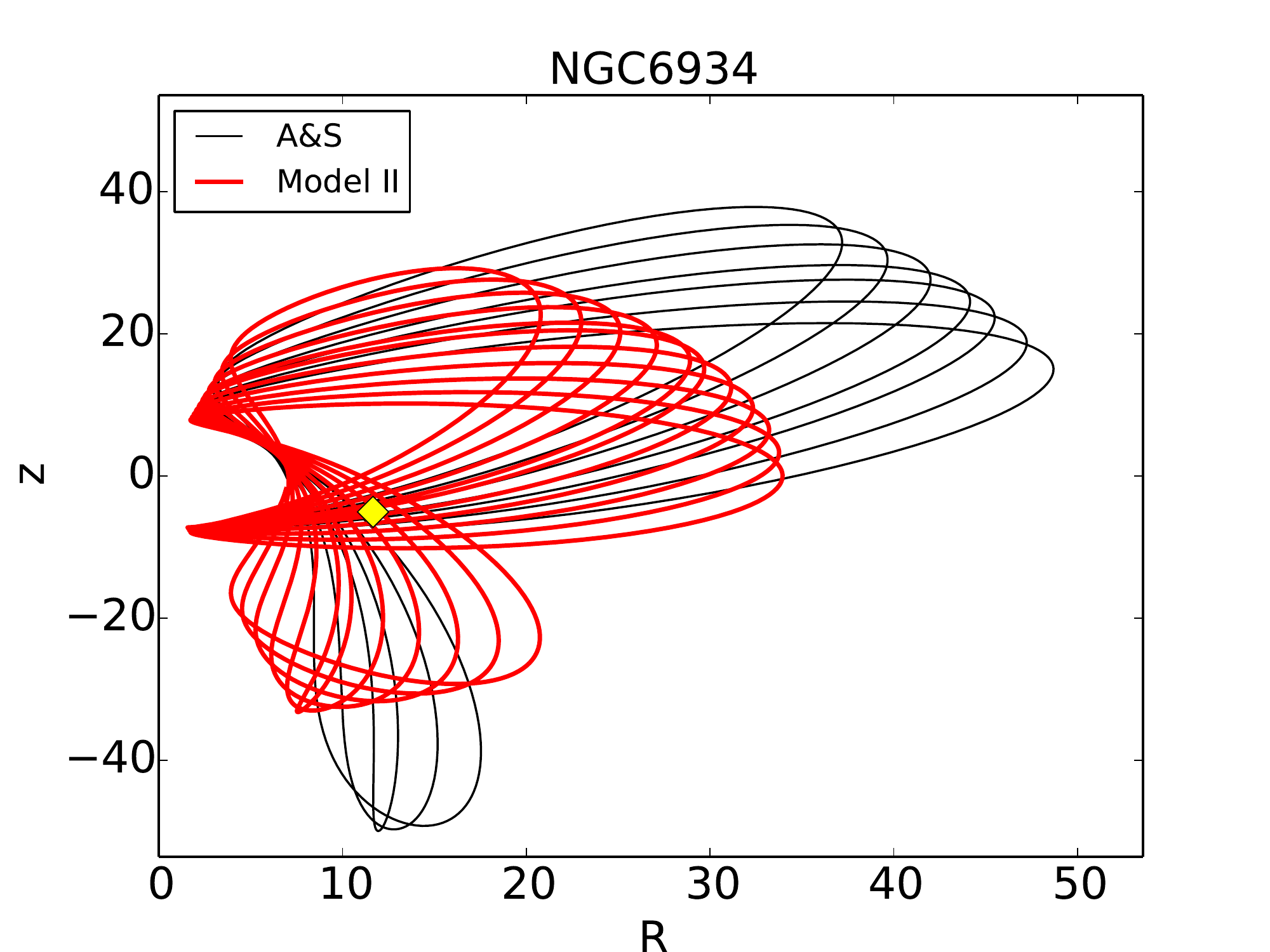}

\includegraphics[trim=0.cm 0cm 2.5cm 0cm, clip=true, width=0.24\textwidth]{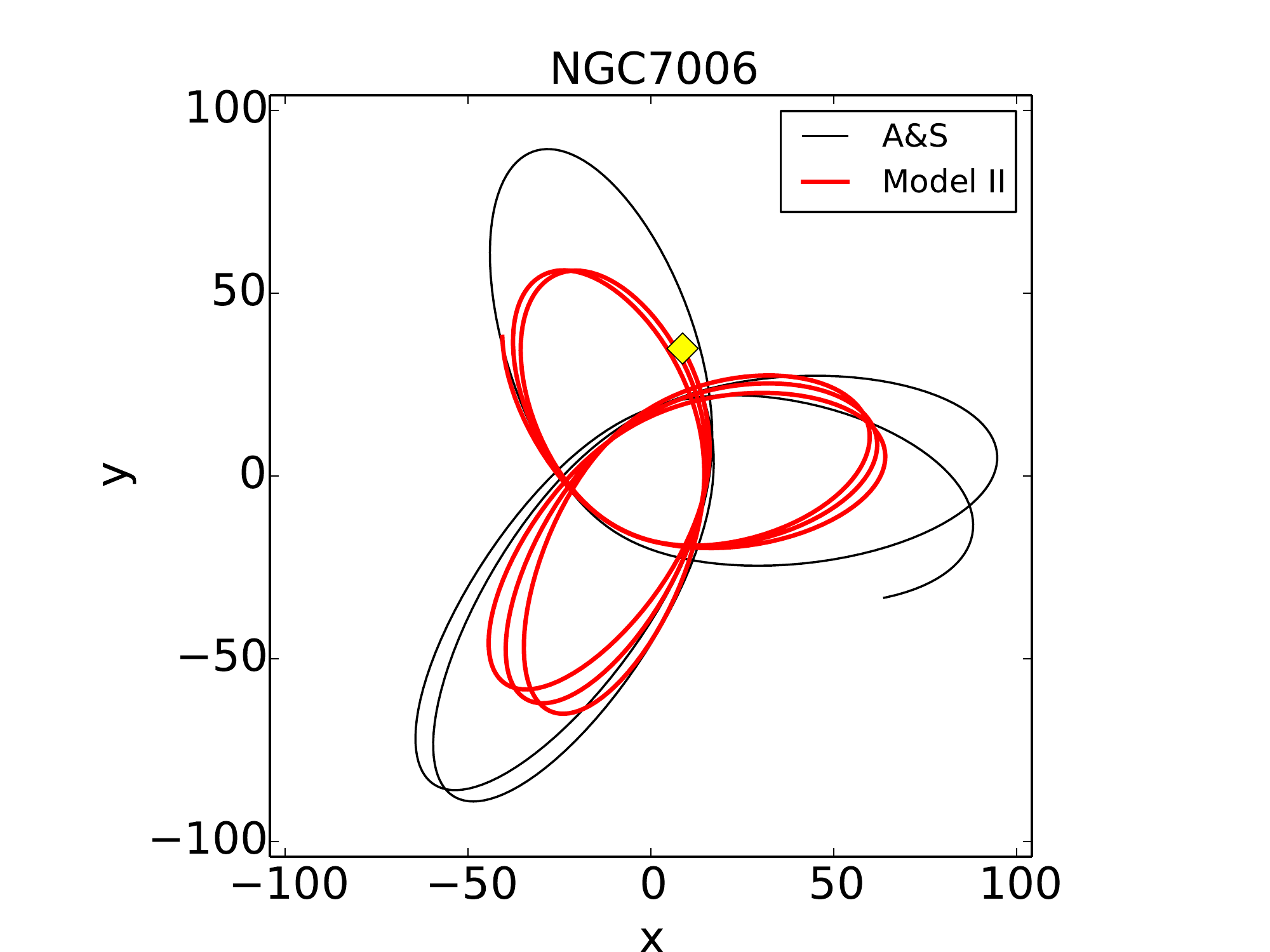}
\includegraphics[trim=0.cm 0cm 2.cm 0cm, clip=true, width=0.24\textwidth]{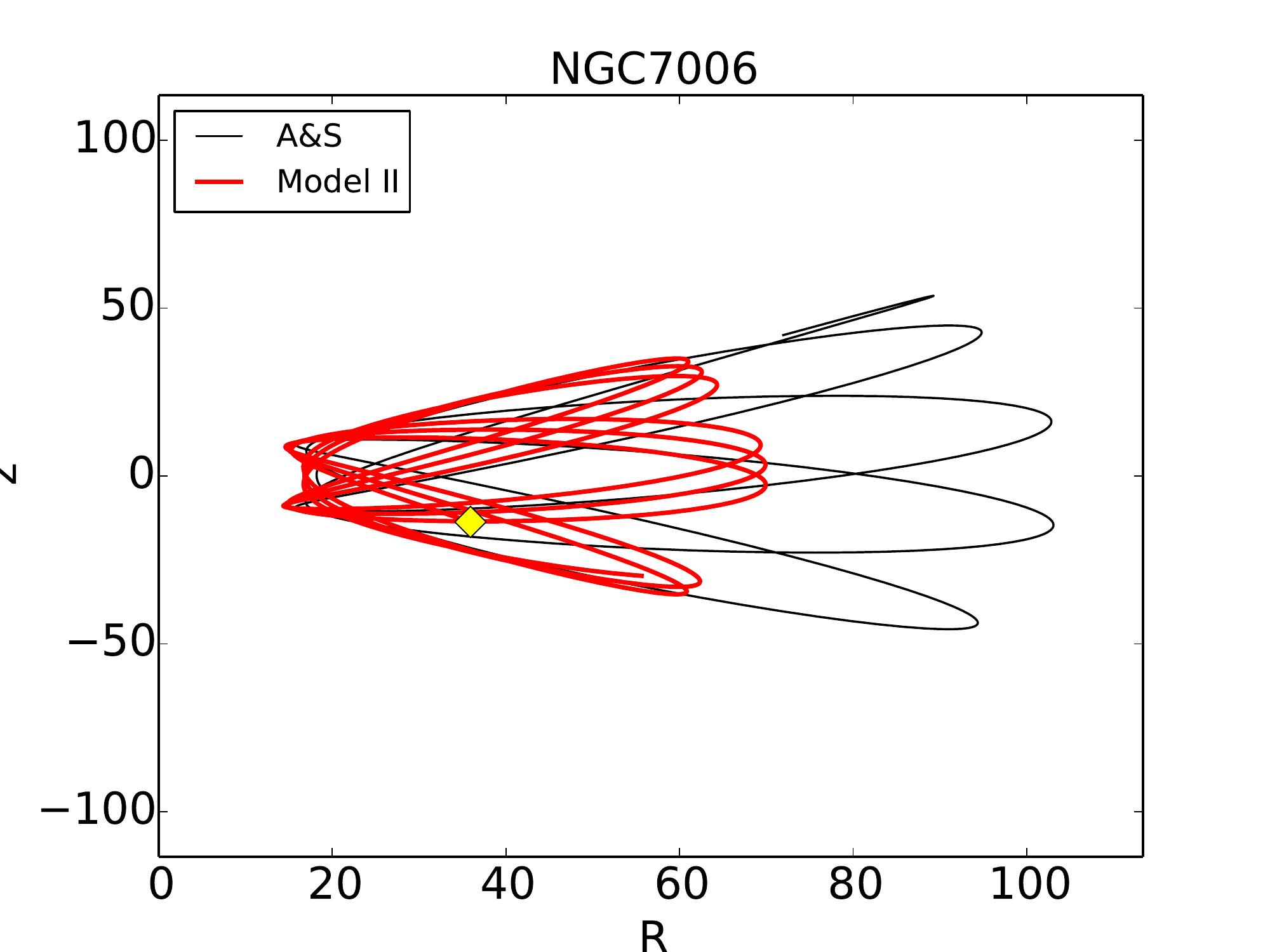}
\includegraphics[trim=0.cm 0cm 2.5cm 0cm, clip=true, width=0.24\textwidth]{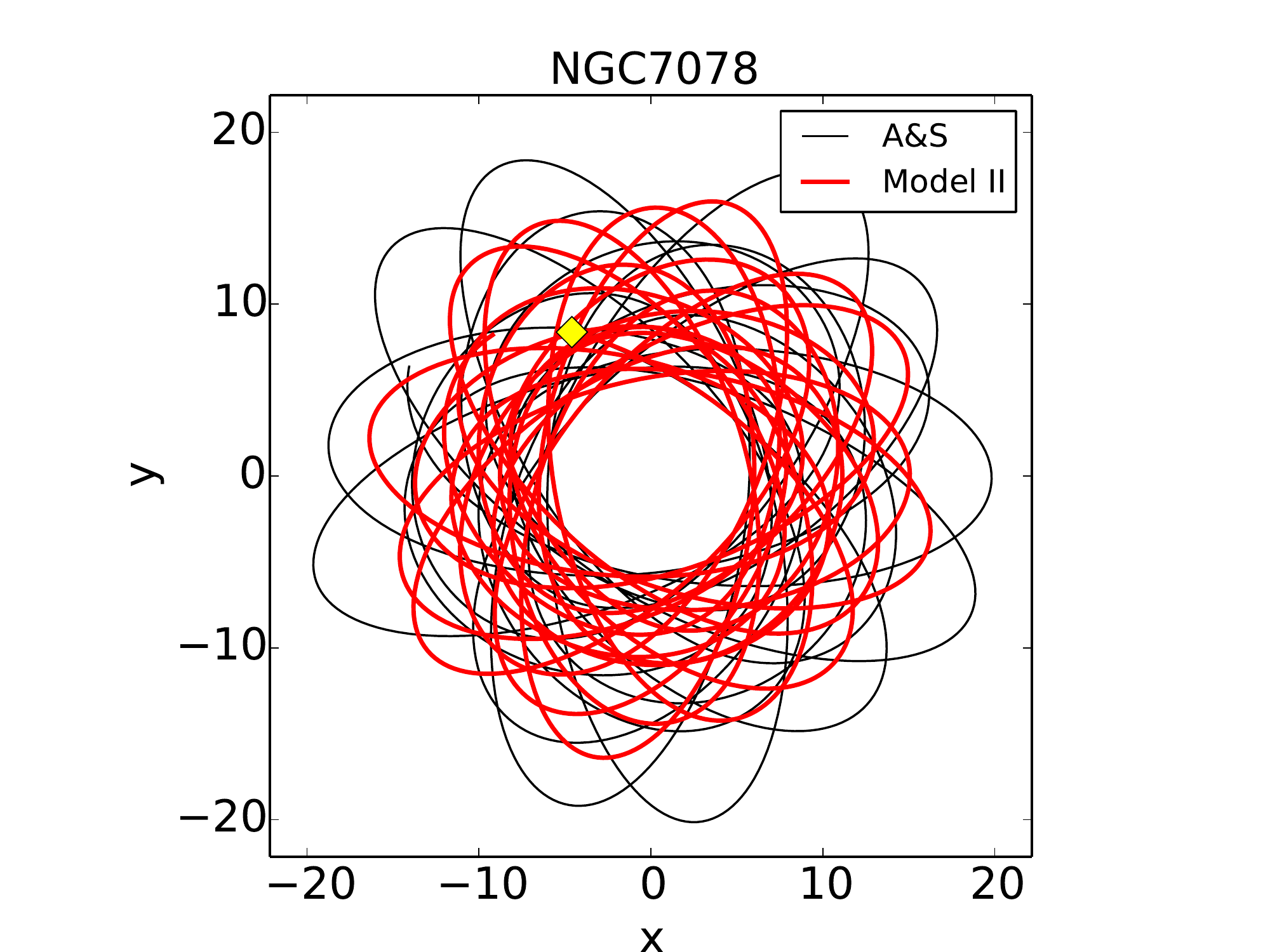}
\includegraphics[trim=0.cm 0cm 2.cm 0cm, clip=true, width=0.24\textwidth]{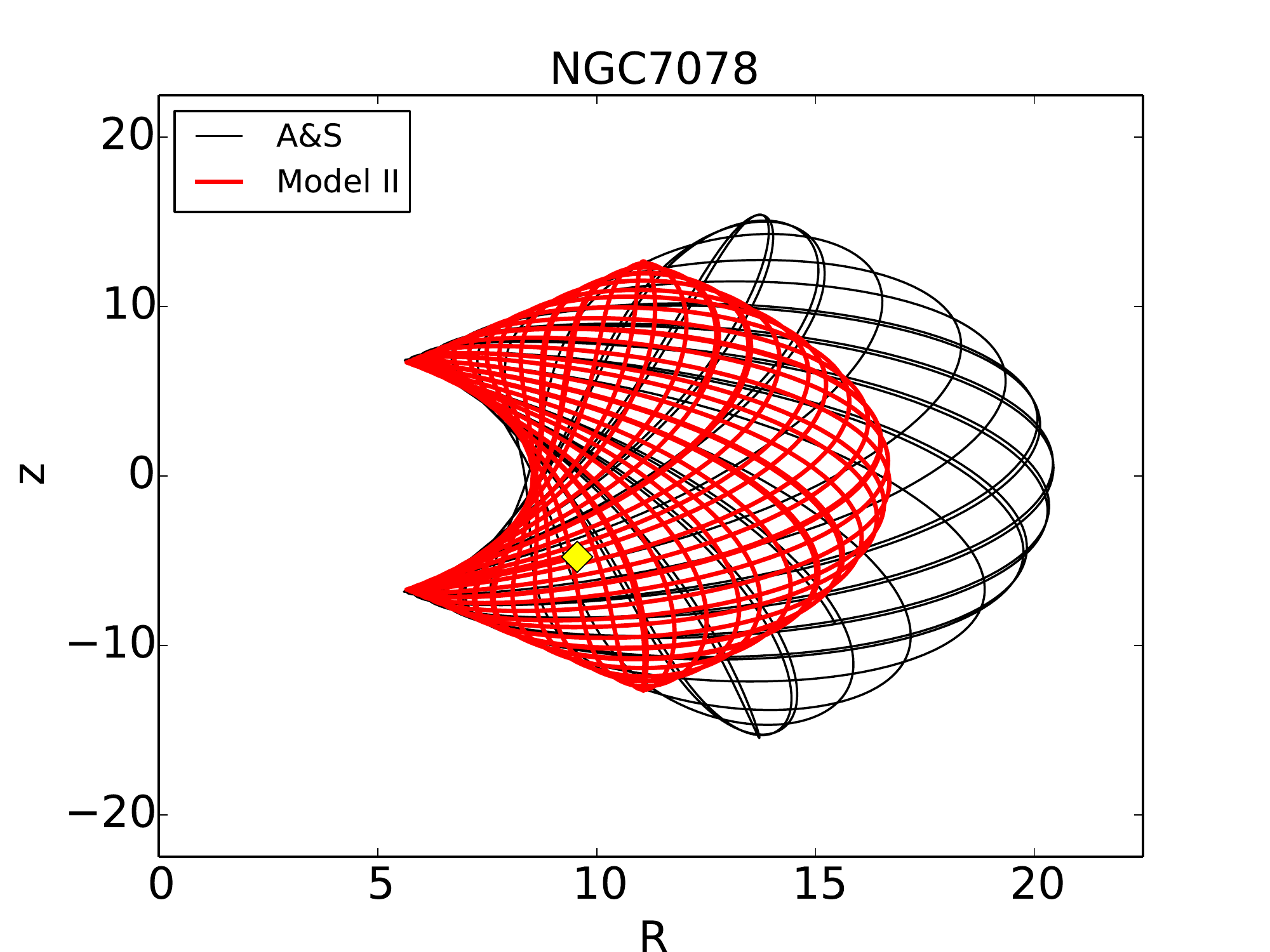}

\includegraphics[trim=0.cm 0cm 2.5cm 0cm, clip=true, width=0.24\textwidth]{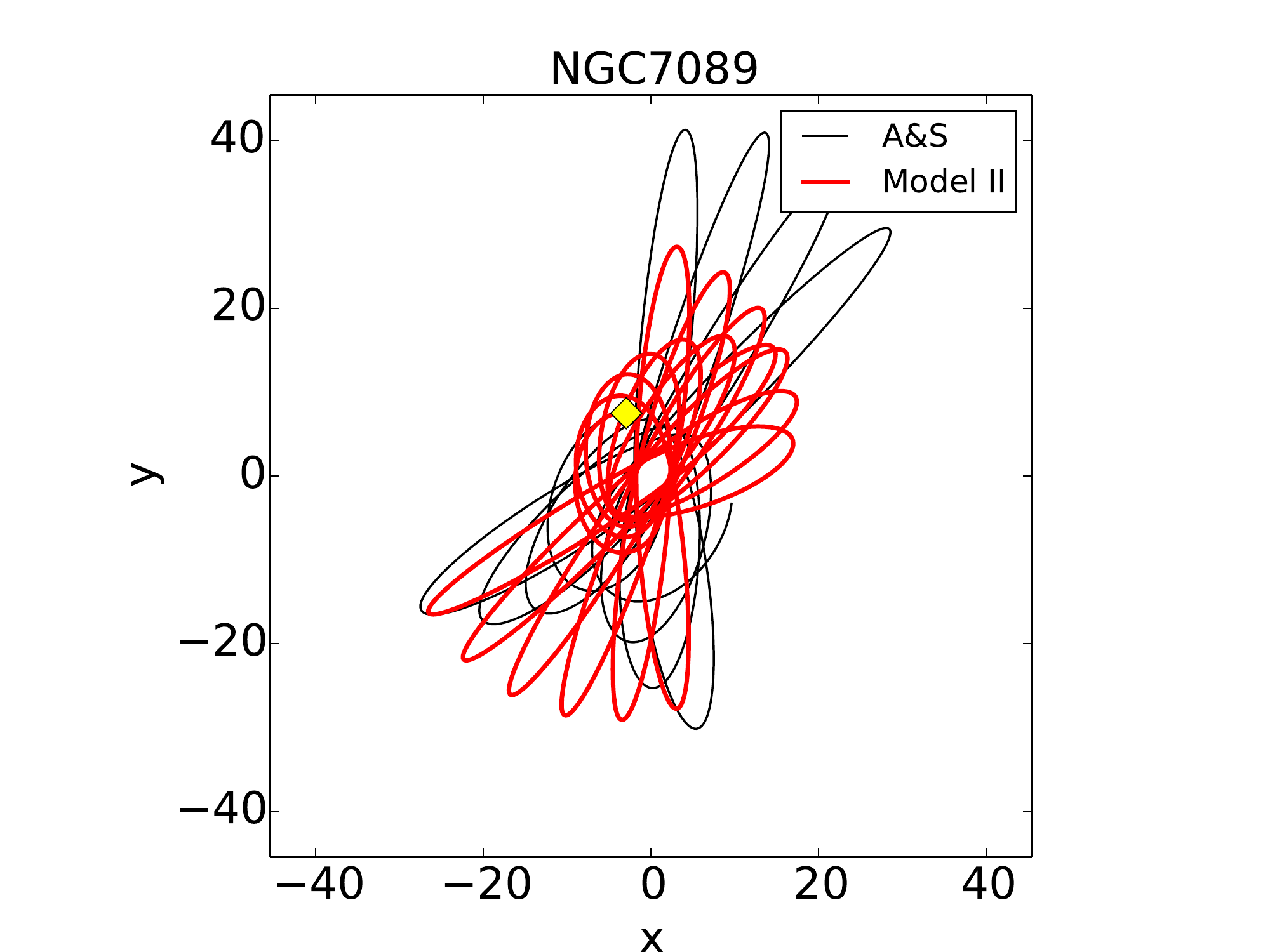}
\includegraphics[trim=0.cm 0cm 2.cm 0cm, clip=true, width=0.24\textwidth]{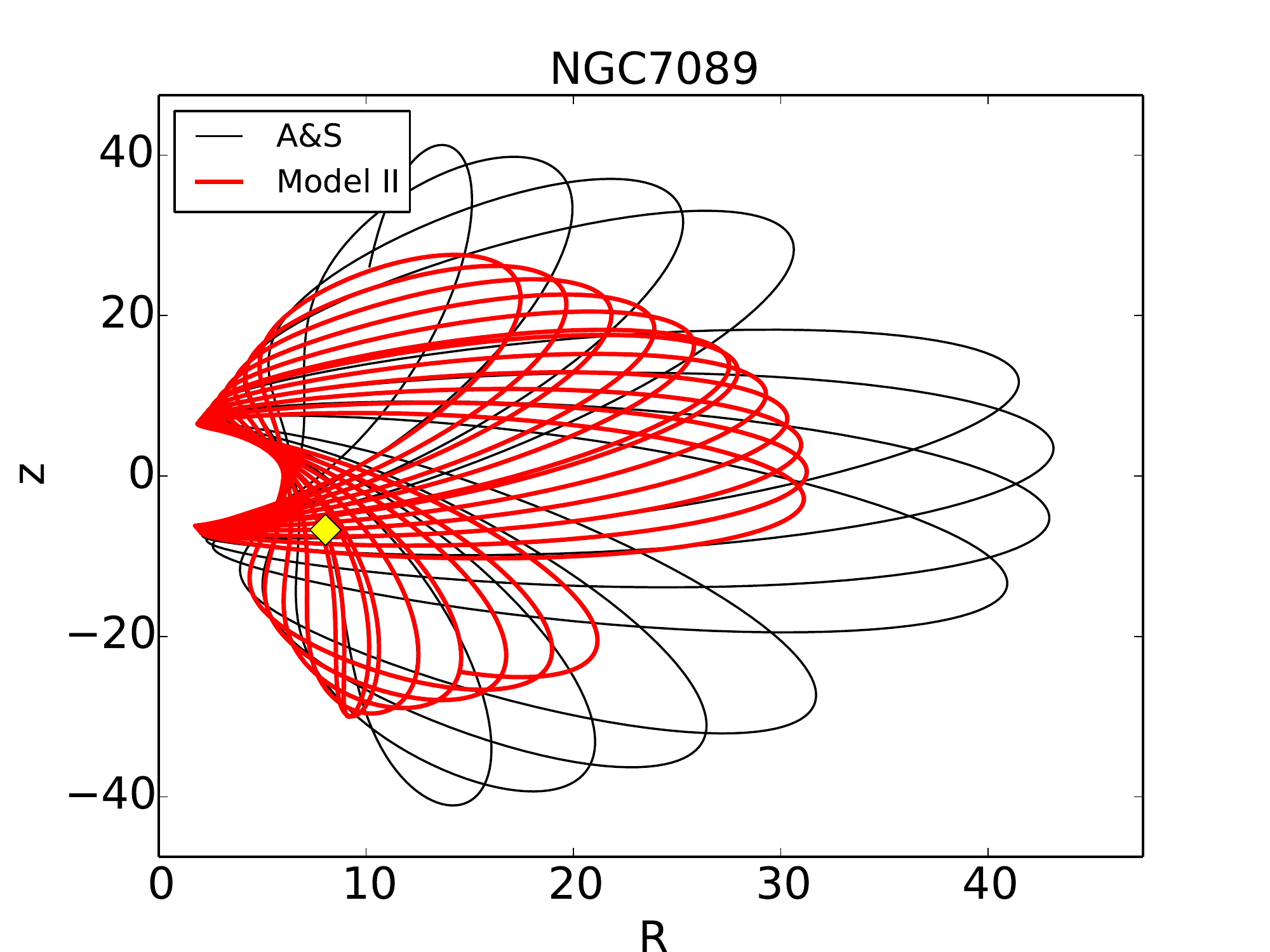}
\includegraphics[trim=0.cm 0cm 2.5cm 0cm, clip=true, width=0.24\textwidth]{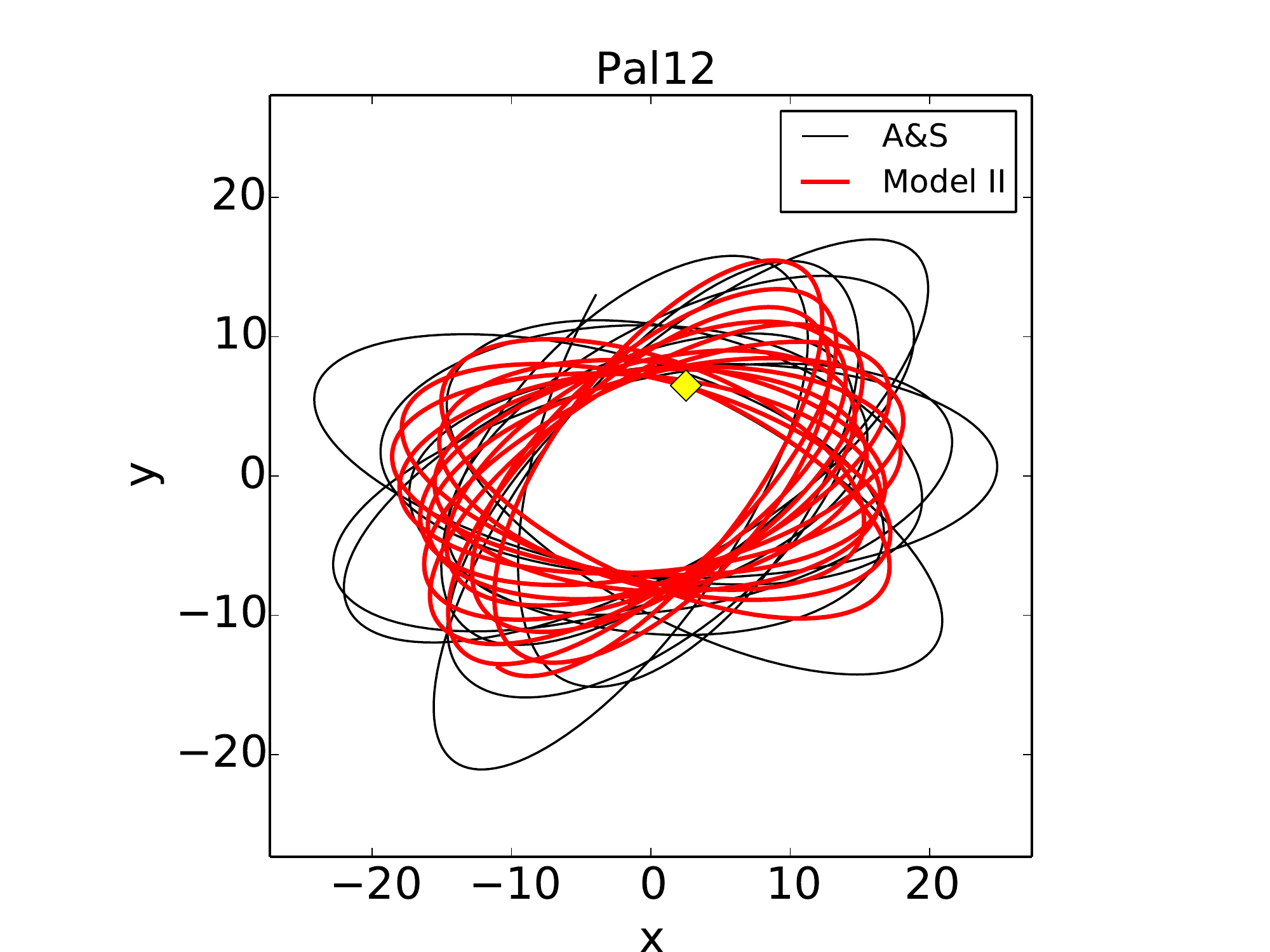}
\includegraphics[trim=0.cm 0cm 2.cm 0cm, clip=true, width=0.24\textwidth]{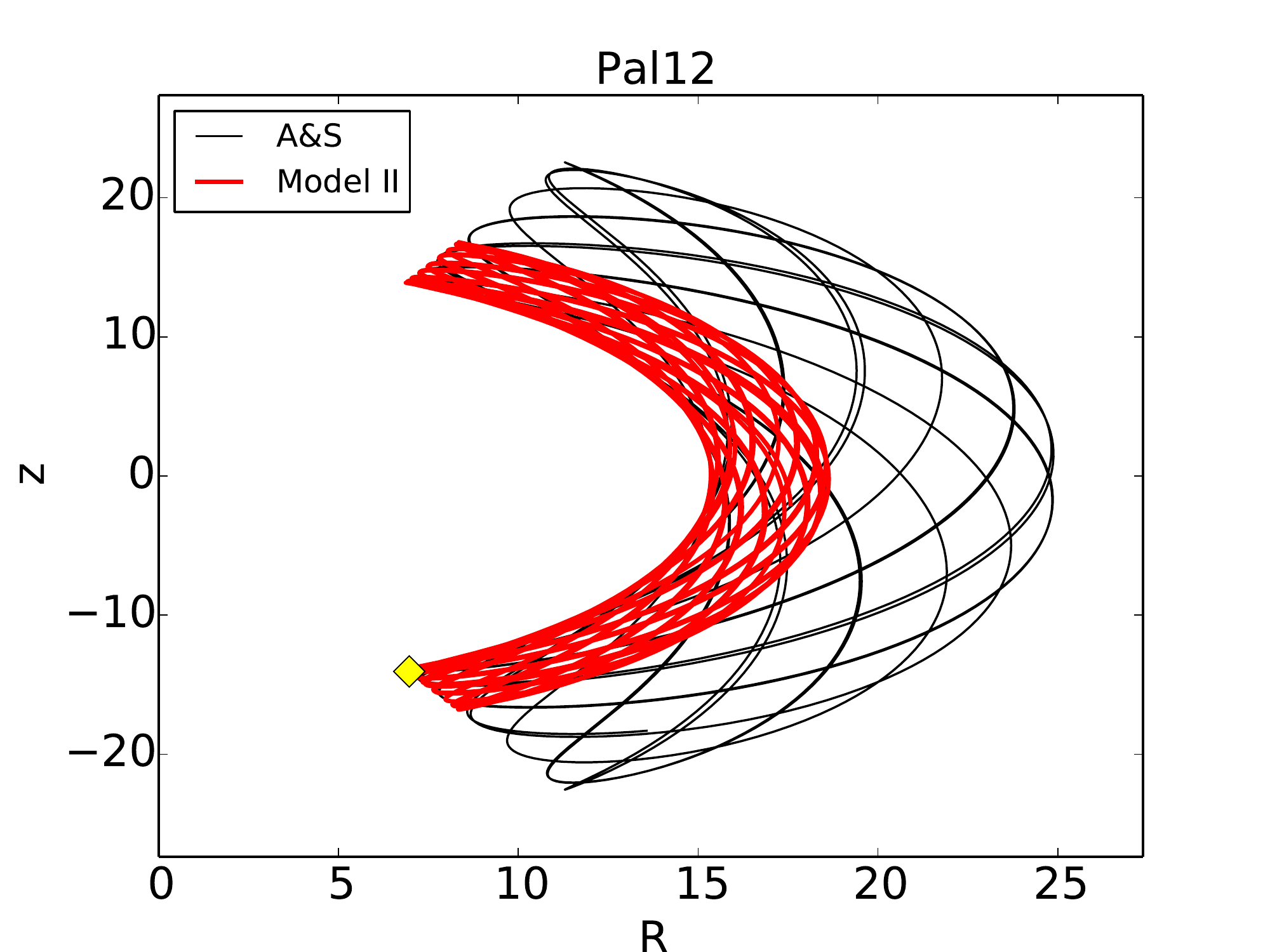}

\caption{Projections in the $x-y$ and in the $R-z$ planes  of the orbits of the Galactic globular clusters NGC~1851, NGC~3201, NGC~4590, NGC~5024, NGC~5466, NGC~5904, NGC~6205, NGC~6934, NGC~7006,  NGC~7078, NGC~7089, and Pal~12. In each plot, the black curve corresponds to the orbit predicted by Model A\&S, the red curve to that predicted by Model II, as indicated in the legend, and the yellow diamond indicates the current position of the cluster.}\label{outerclusters}

\end{figure*}

\begin{figure*}
\includegraphics[trim=0.cm 0cm 2.5cm 0cm, clip=true, width=0.24\textwidth]{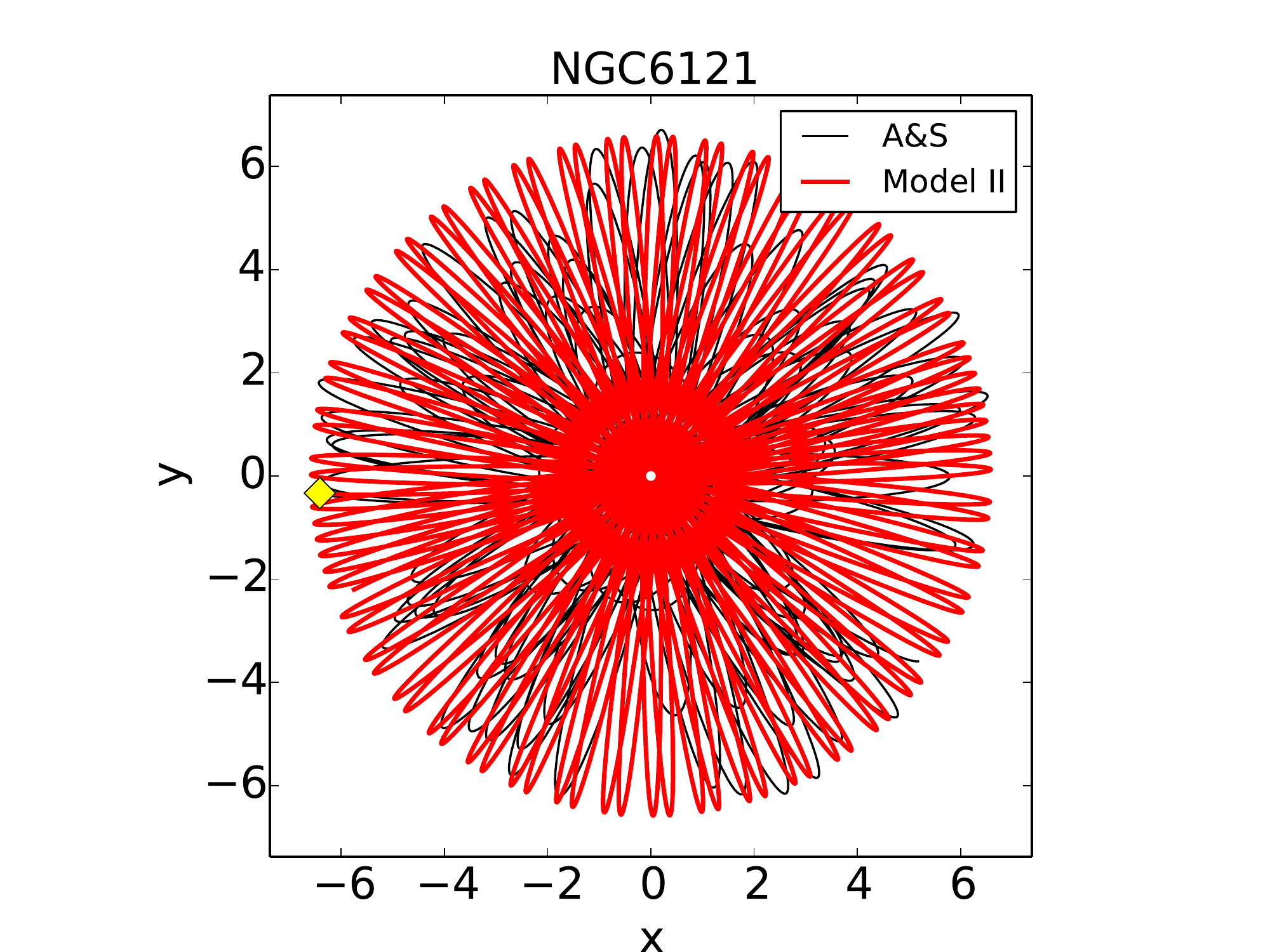}
\includegraphics[trim=0.cm 0cm 2.cm 0cm, clip=true, width=0.24\textwidth]{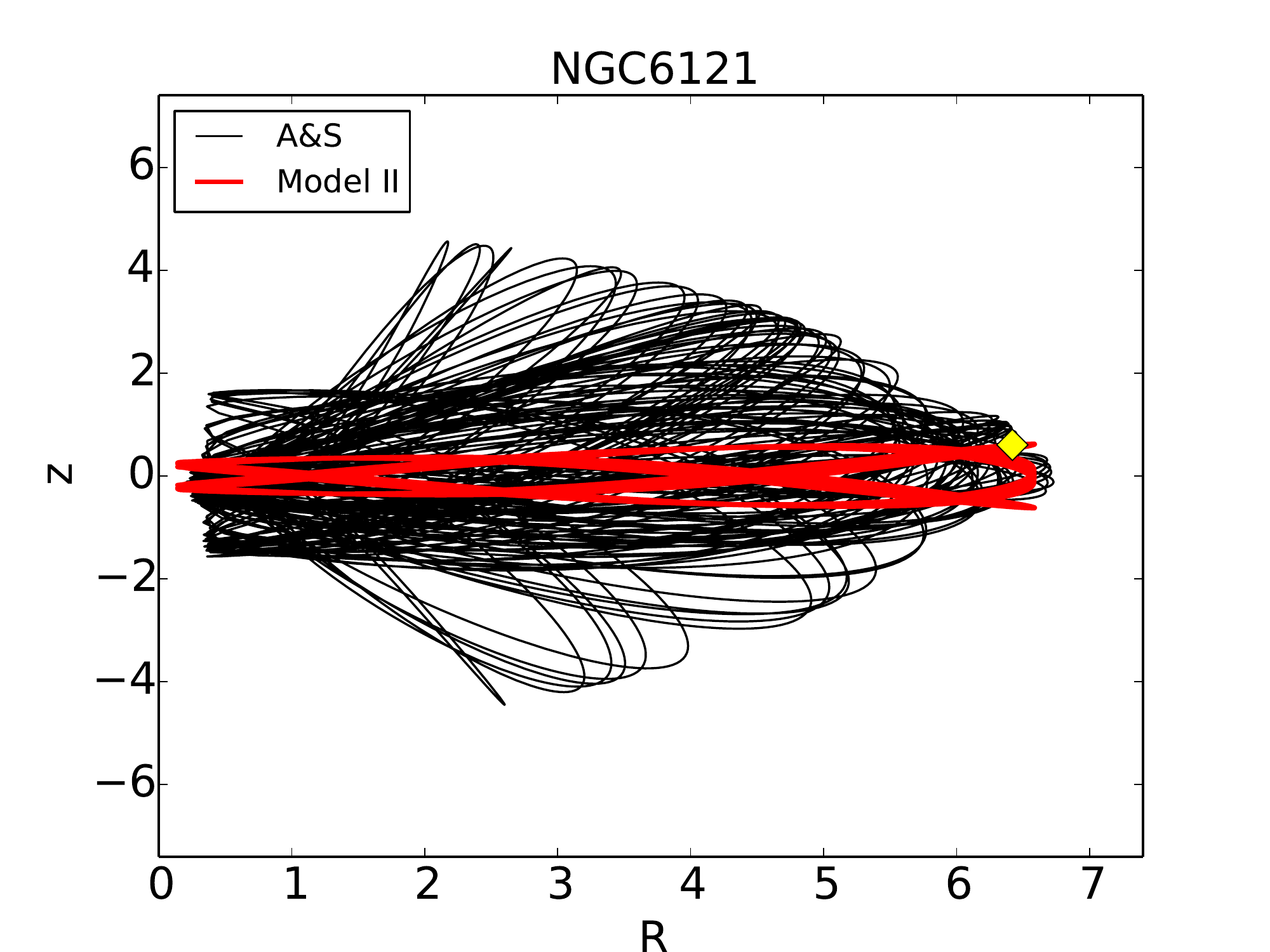}
\includegraphics[trim=0.cm 0cm 2.5cm 0cm, clip=true, width=0.24\textwidth]{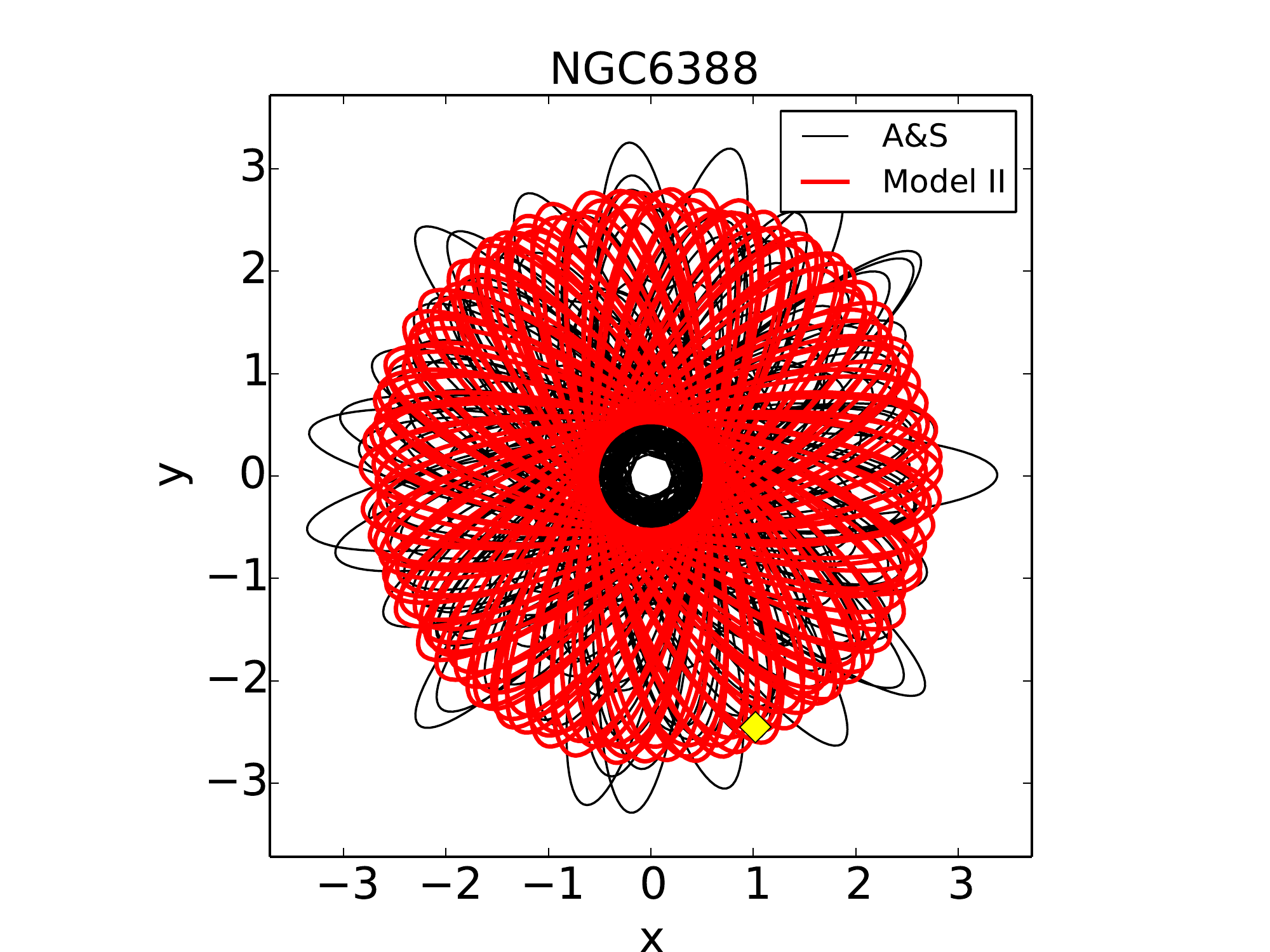}
\includegraphics[trim=0.cm 0cm 2.cm 0cm, clip=true, width=0.24\textwidth]{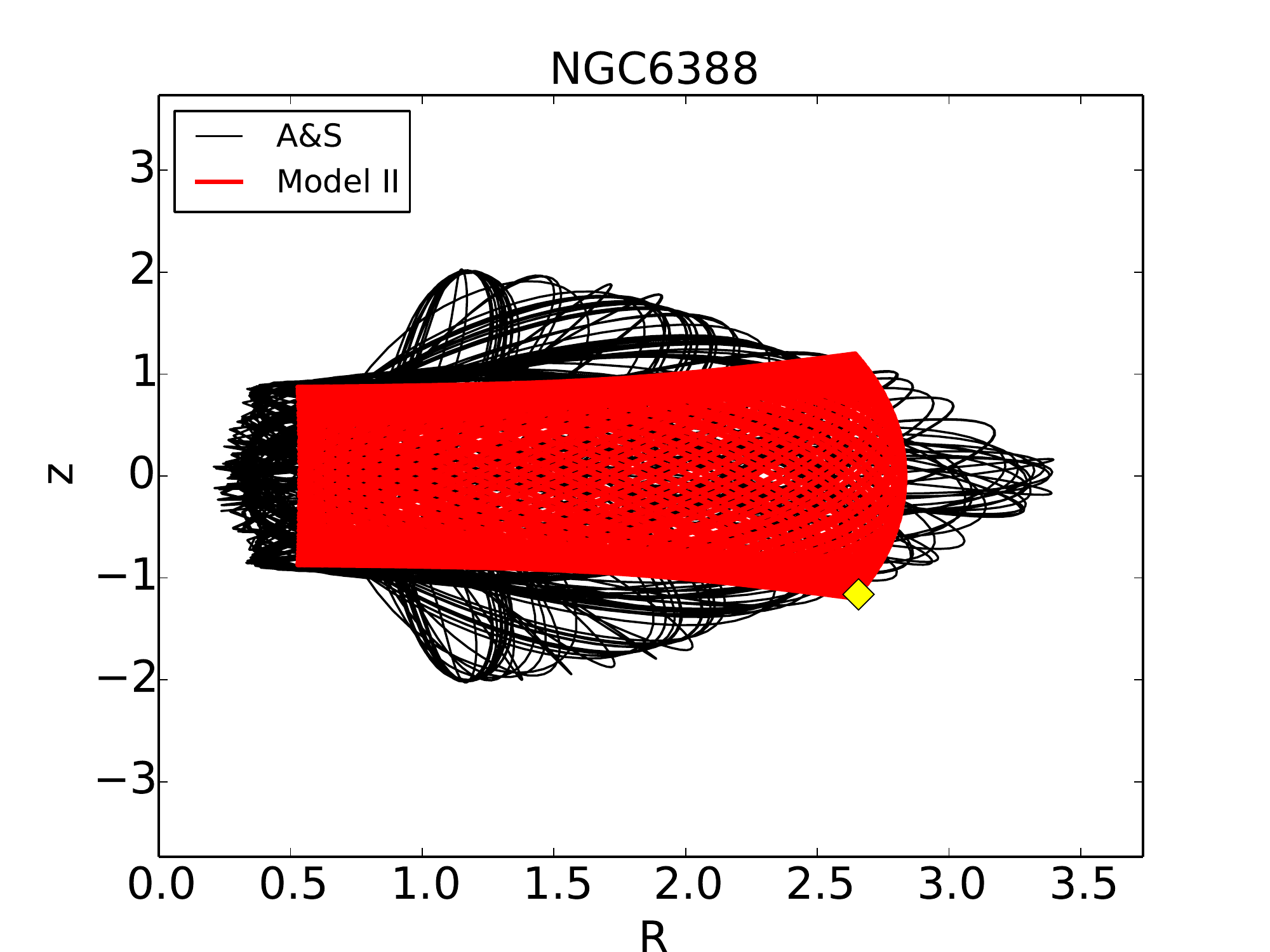}

\includegraphics[trim=0.cm 0cm 2.5cm 0cm, clip=true, width=0.24\textwidth]{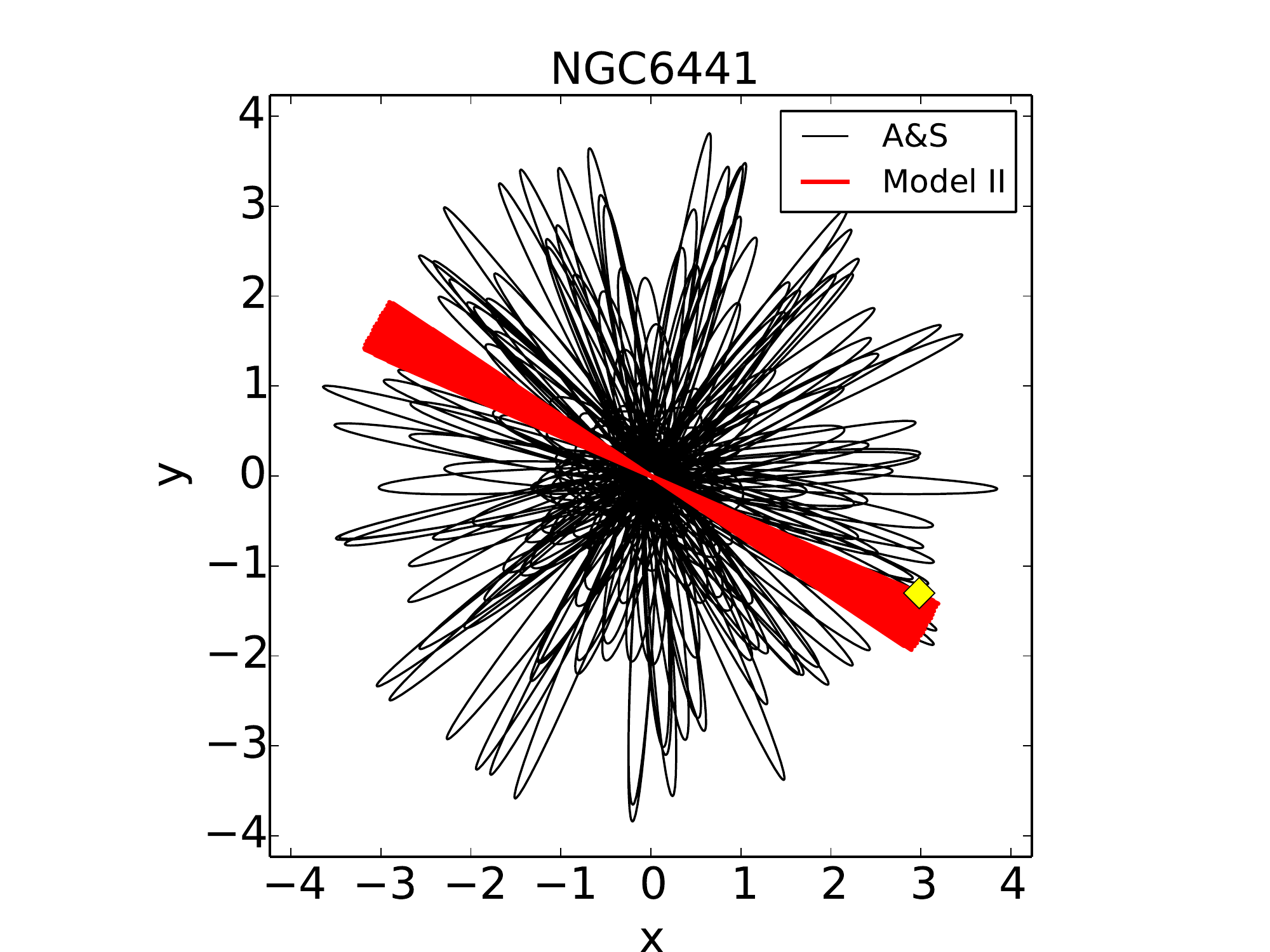}
\includegraphics[trim=0.cm 0cm 2.cm 0cm, clip=true, width=0.24\textwidth]{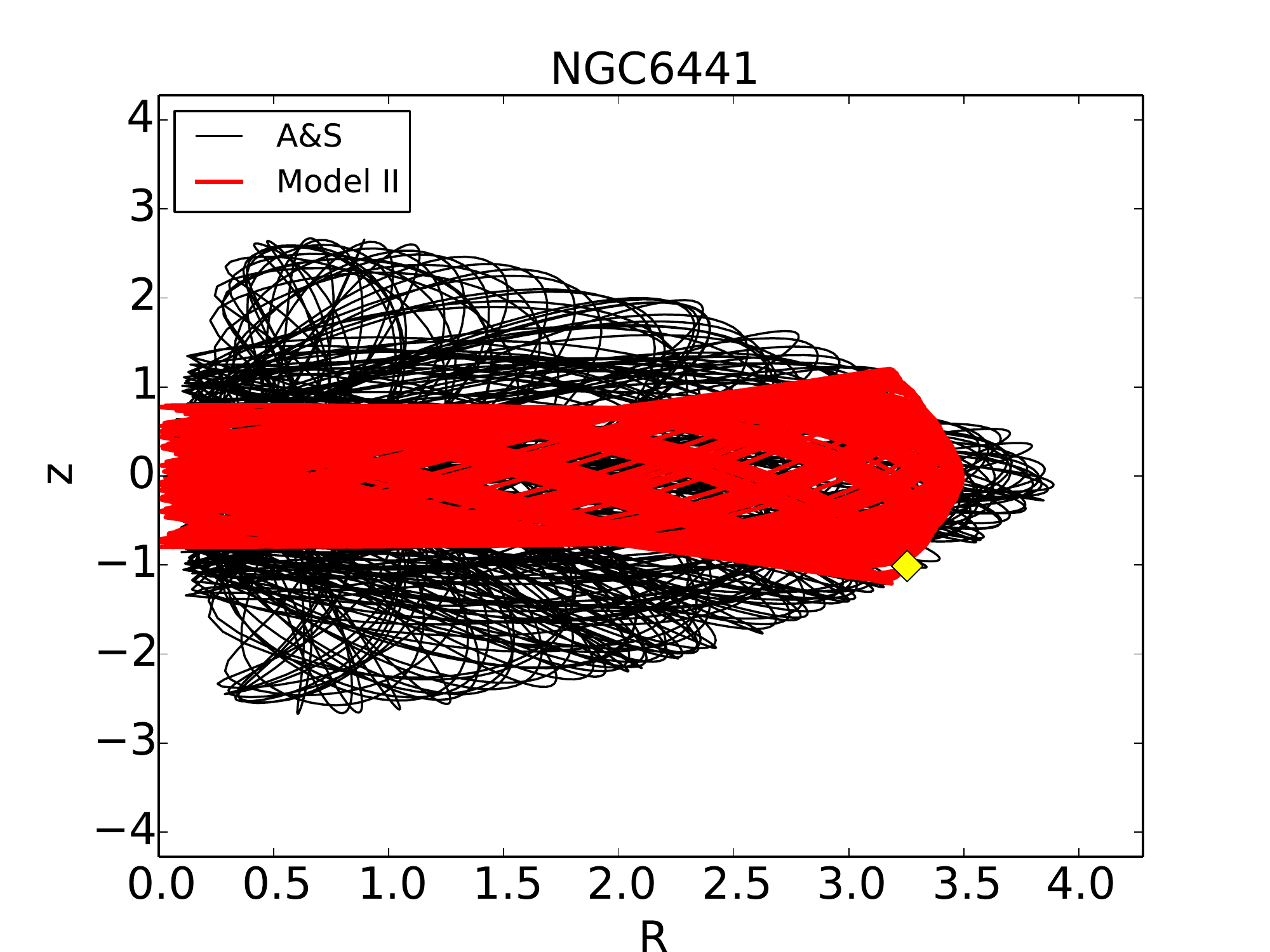}
\includegraphics[trim=0.cm 0cm 2.5cm 0cm, clip=true, width=0.24\textwidth]{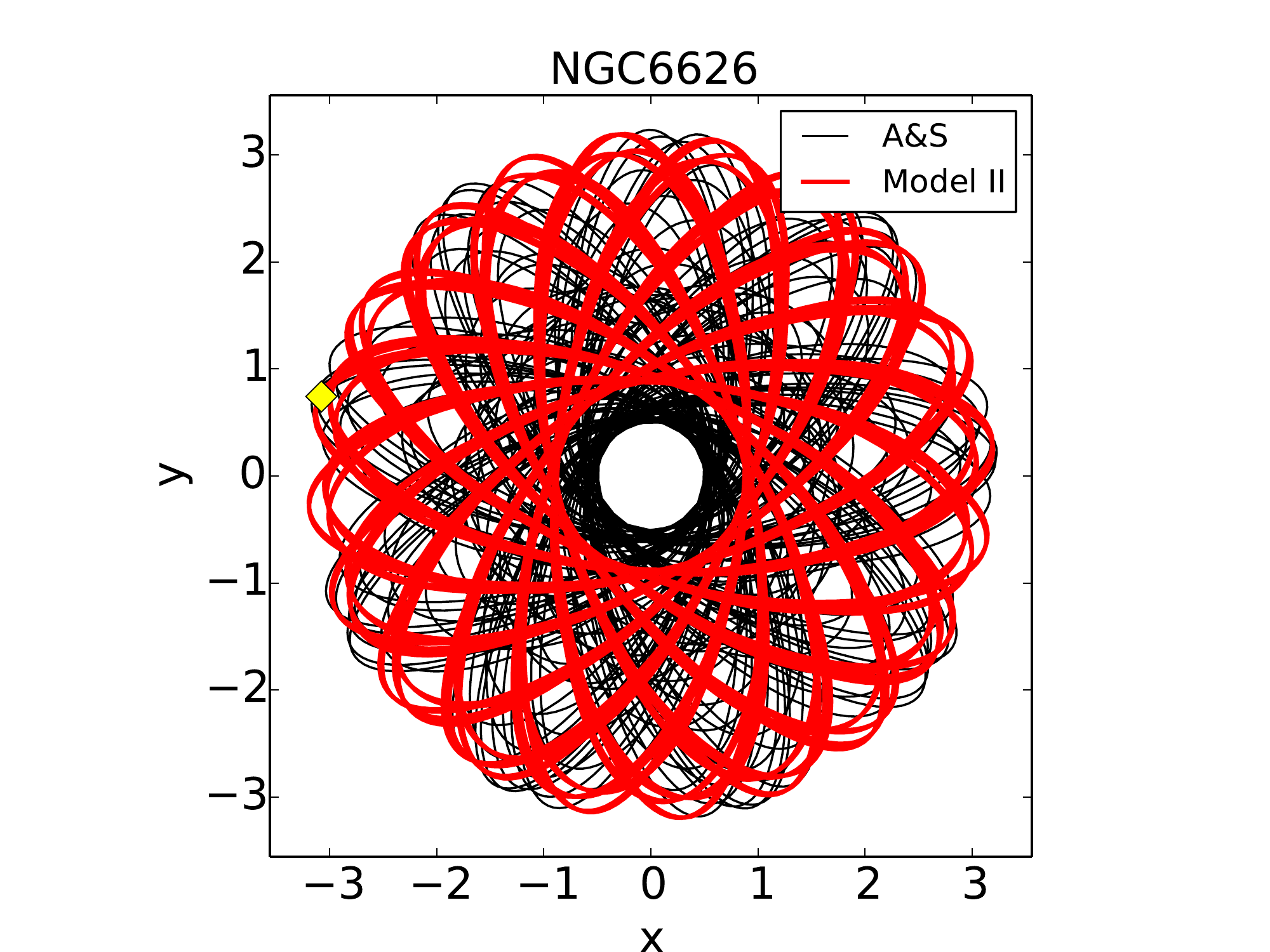}
\includegraphics[trim=0.cm 0cm 2.cm 0cm, clip=true, width=0.24\textwidth]{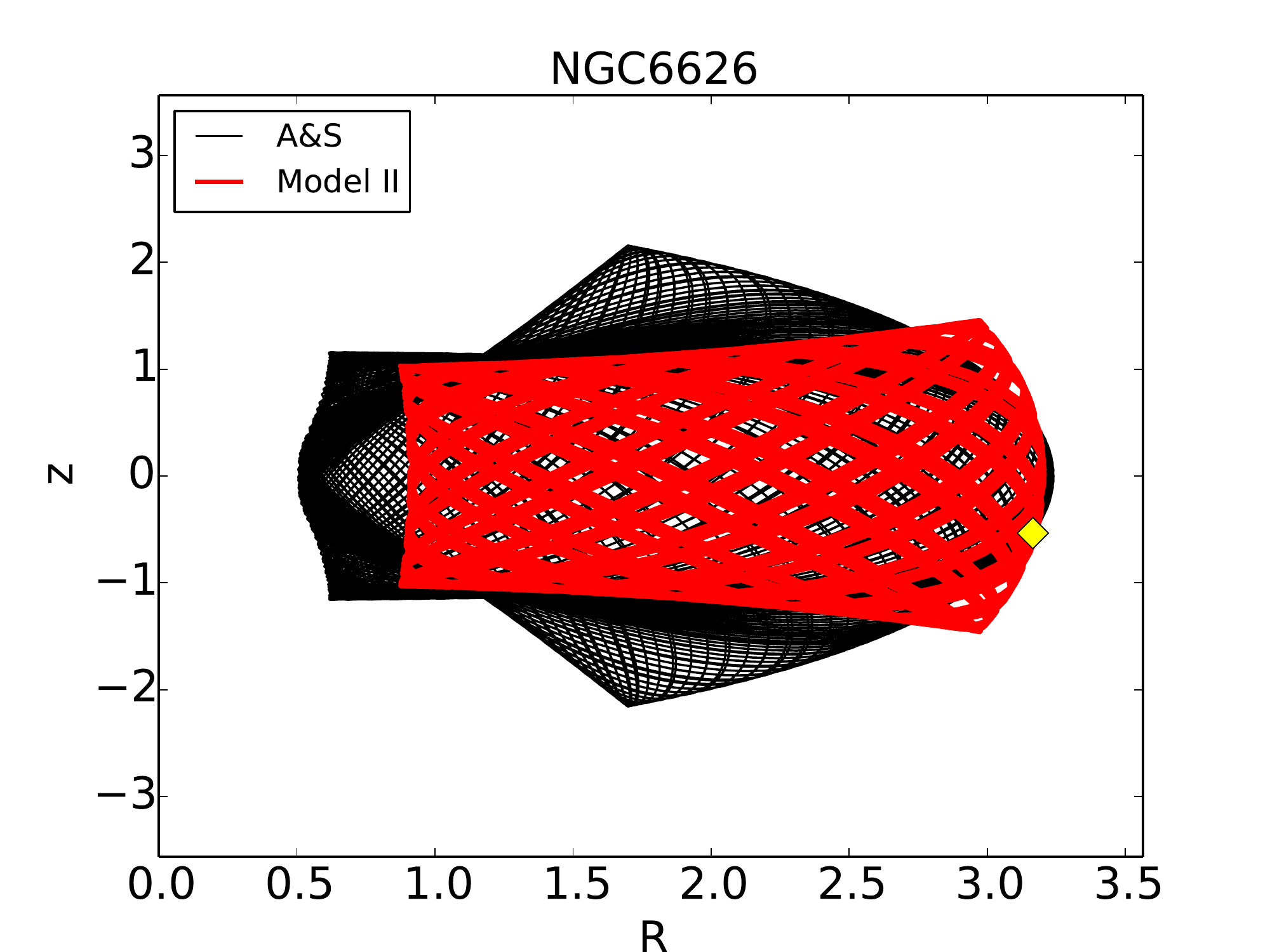}

\includegraphics[trim=0.cm 0cm 2.5cm 0cm, clip=true, width=0.24\textwidth]{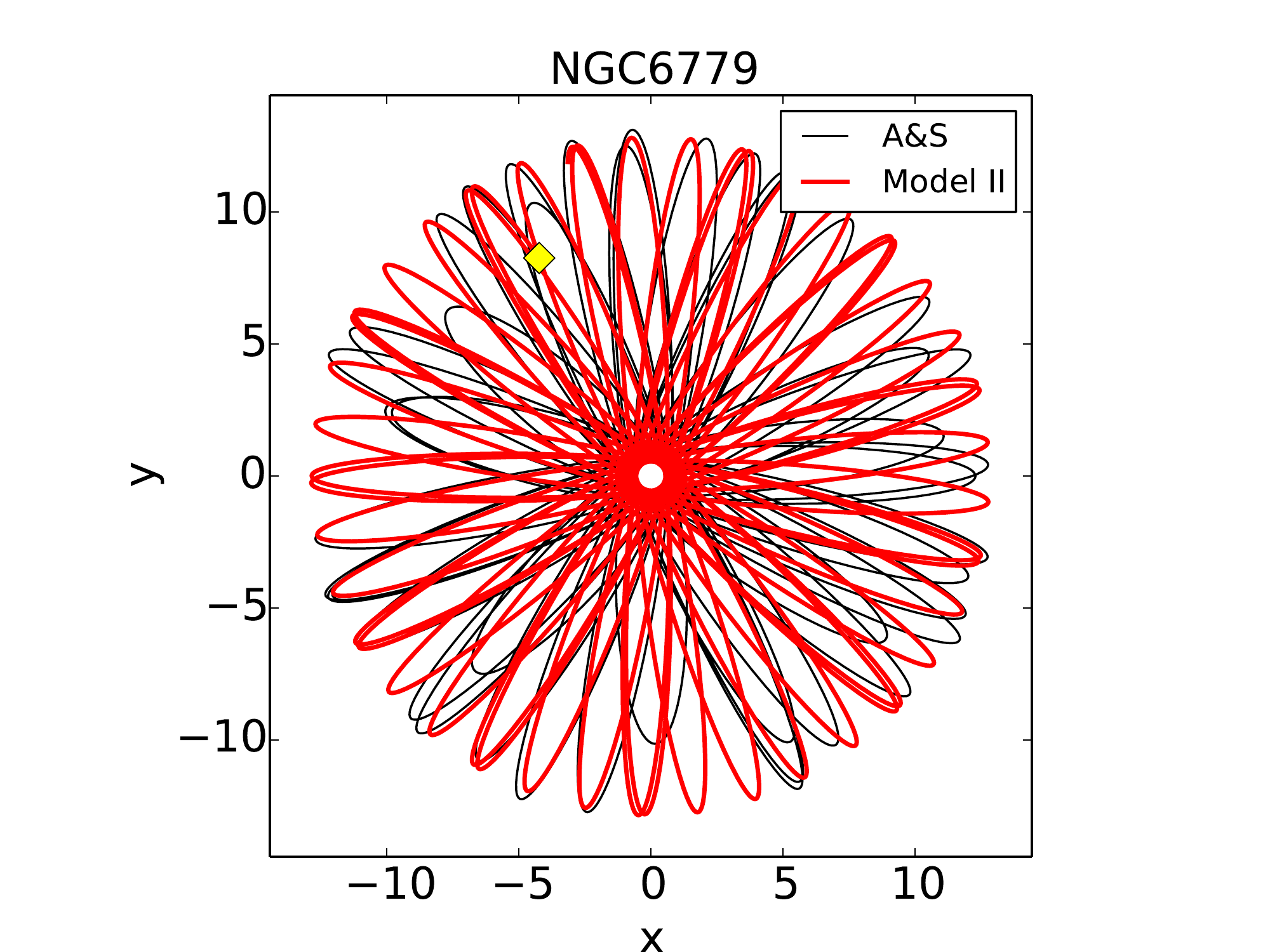}
\includegraphics[trim=0.cm 0cm 1.8cm 0cm, clip=true, width=0.24\textwidth]{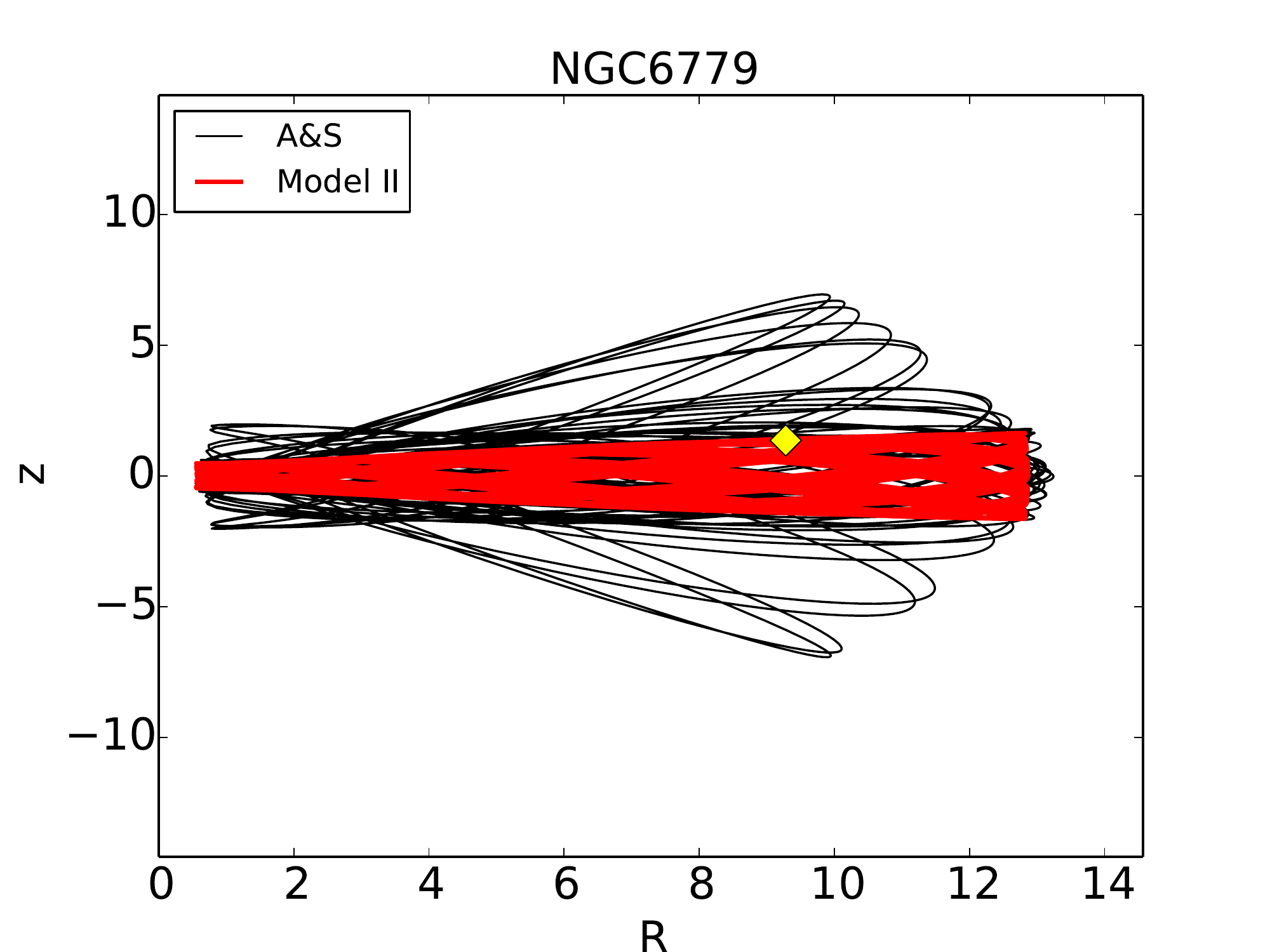}

\caption{Projections in the $x-y$ and in the $R-z$ planes  of the orbits of the Galactic globular clusters NGC~6121, NGC~6388, NGC~6441, NGC~6626, and NGC~6779 . In each plot, the black curve corresponds to the orbit predicted by Model A\&S and the red curve to that predicted by Model II, as indicated in the legend. In each plot, the yellow diamond indicates the current position of the cluster.}\label{bulgeclusters}
\end{figure*}

\begin{figure*}
\centering
\includegraphics[trim=0.5cm 0cm 1.cm 0cm, clip=true, width=0.45\textwidth]{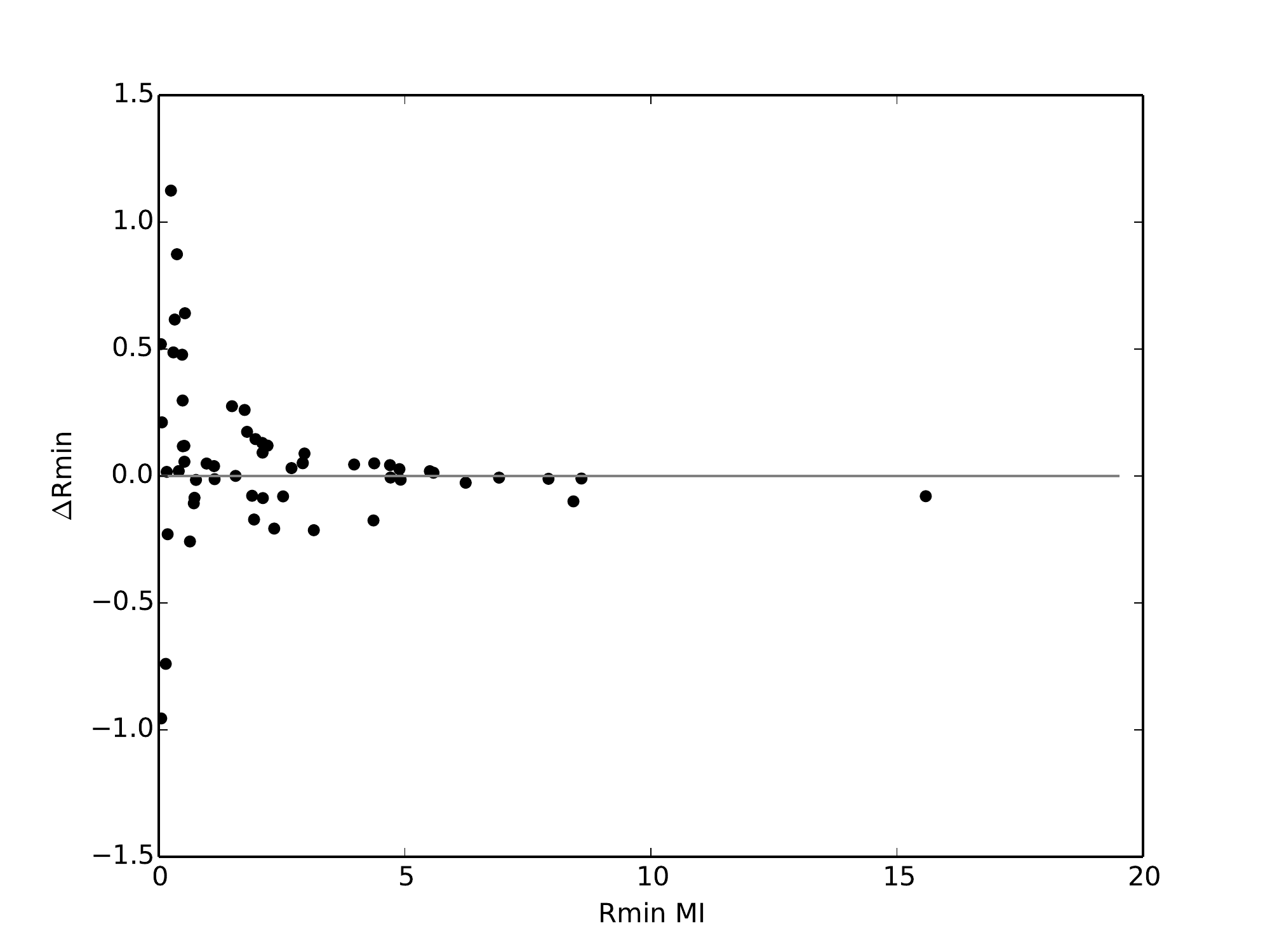}
\includegraphics[trim=0.5cm 0cm 1.cm 0cm, clip=true, width=0.45\textwidth]{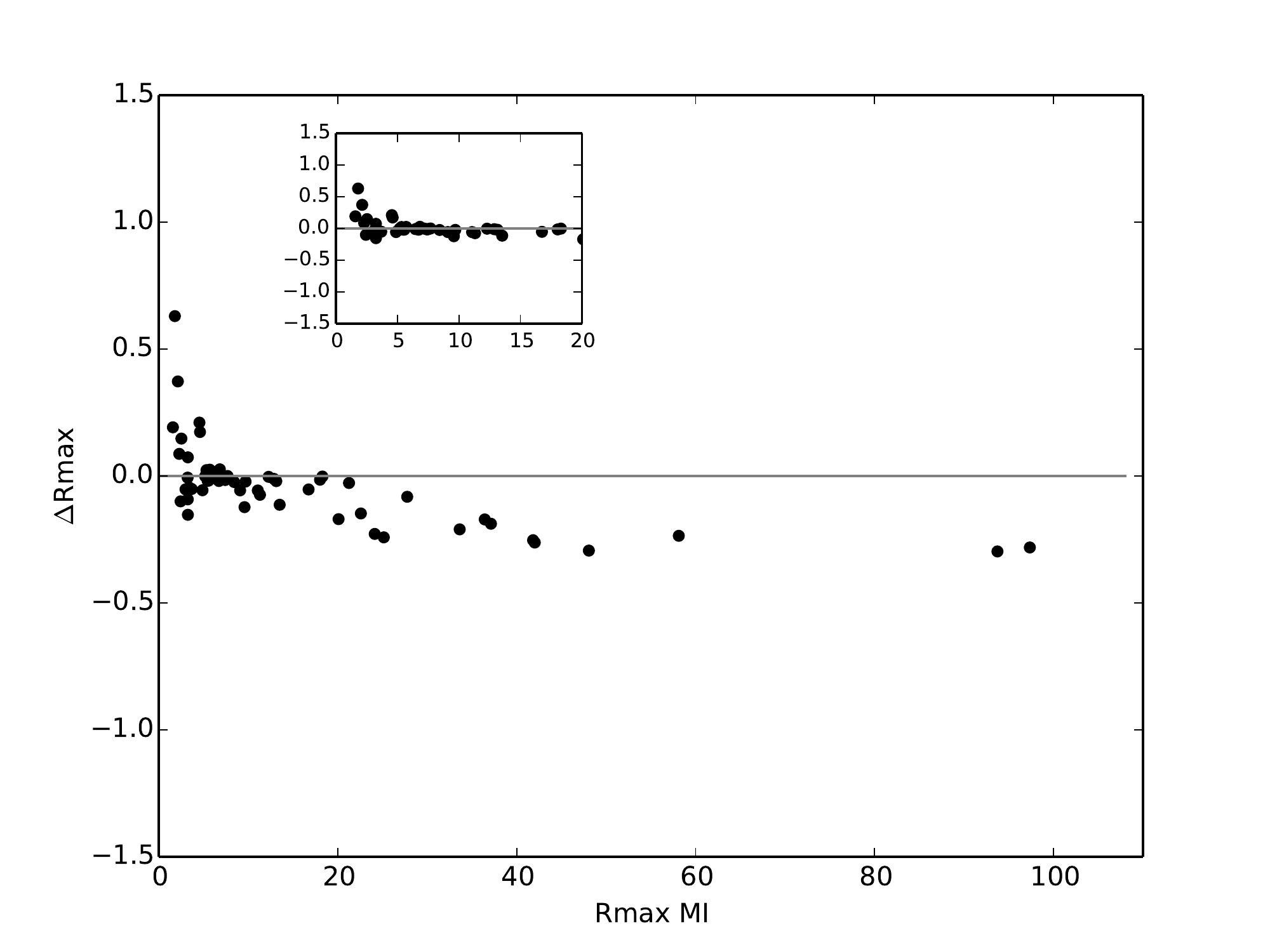}
\includegraphics[trim=0.5cm 0cm 1.cm 0cm, clip=true, width=0.45\textwidth]{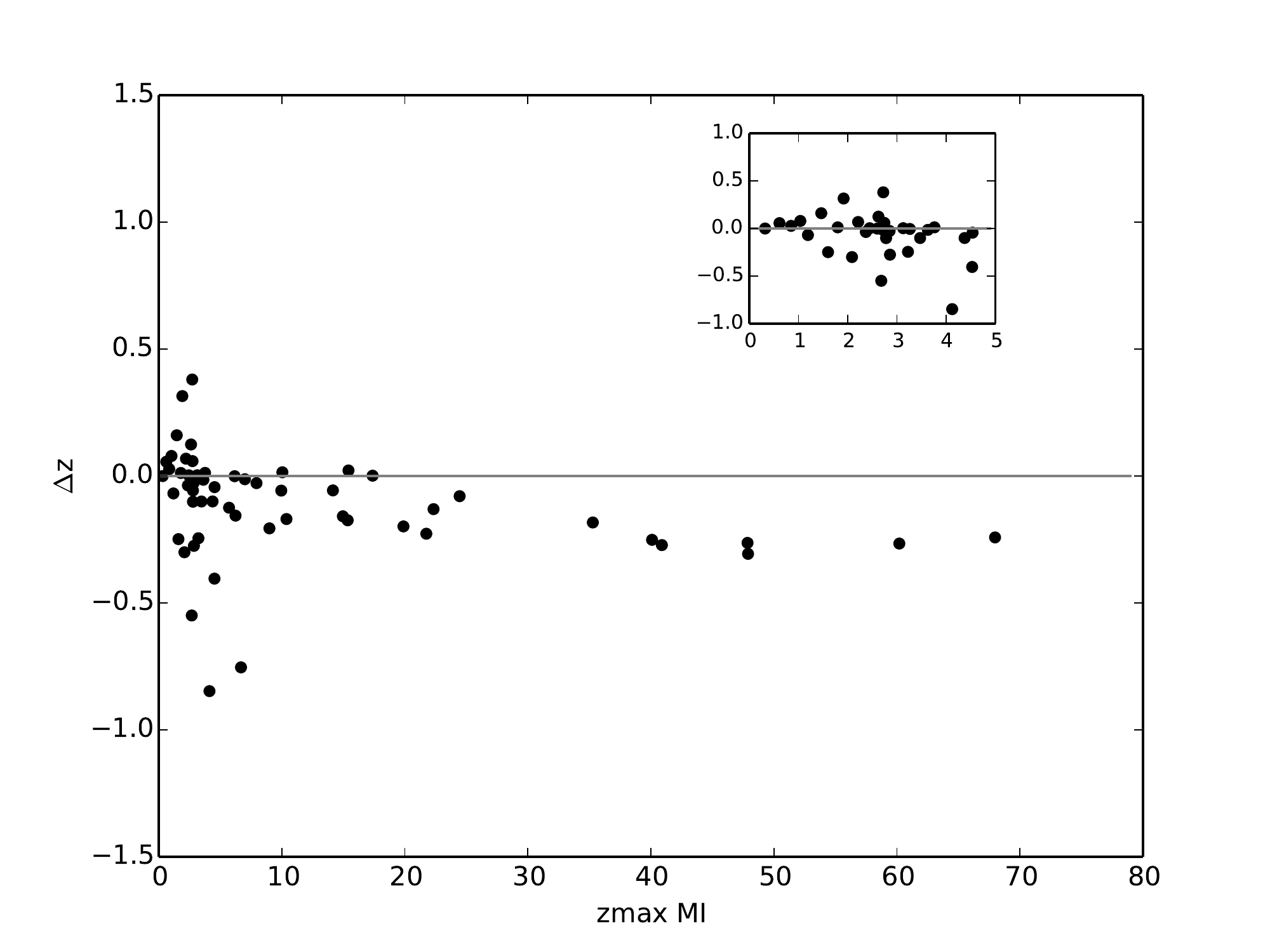}
\includegraphics[trim=0.5cm 0cm 1.cm 0cm, clip=true, width=0.45\textwidth]{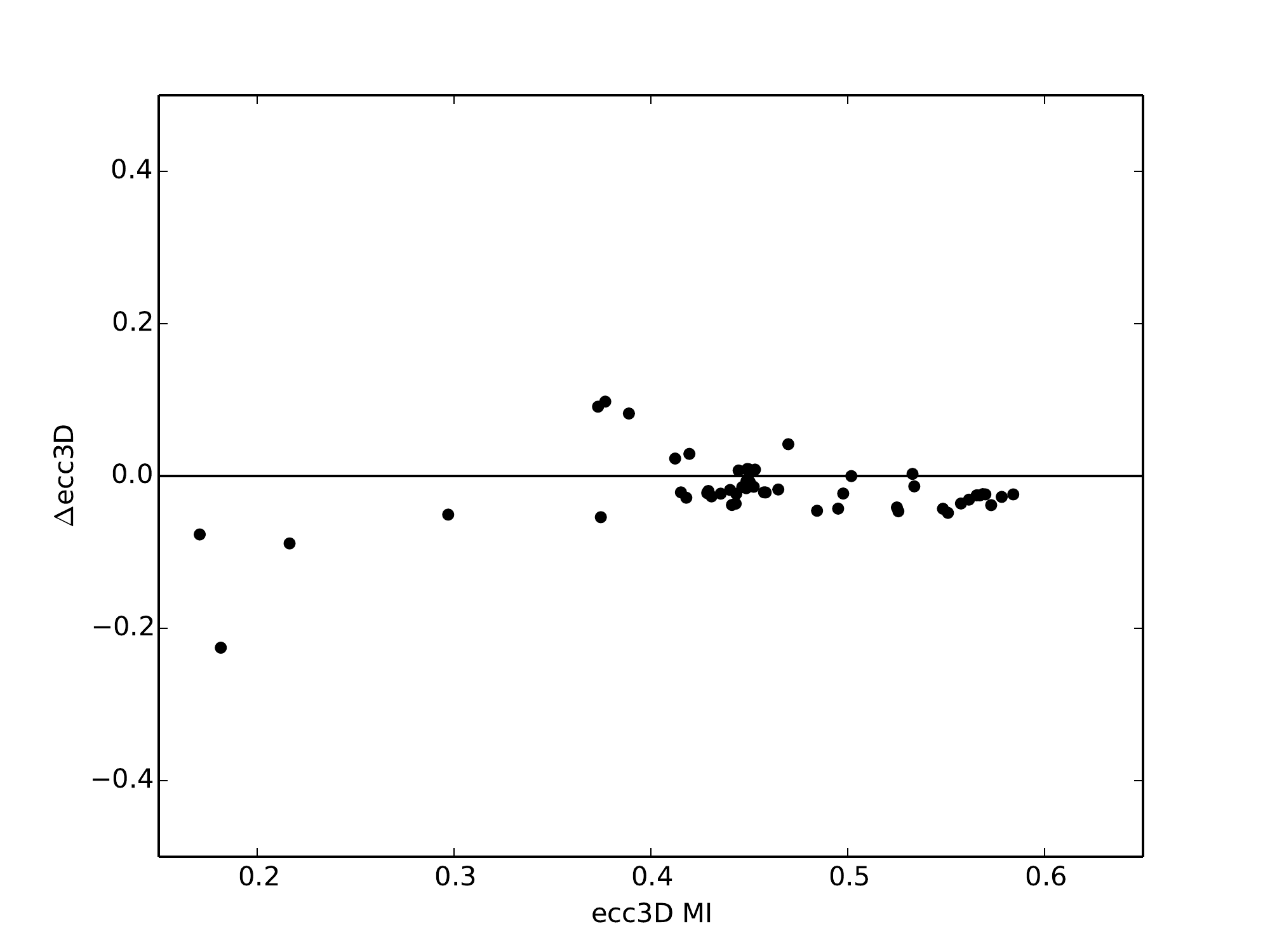}
\includegraphics[trim=0.5cm 0cm 1.cm 0cm, clip=true, width=0.45\textwidth]{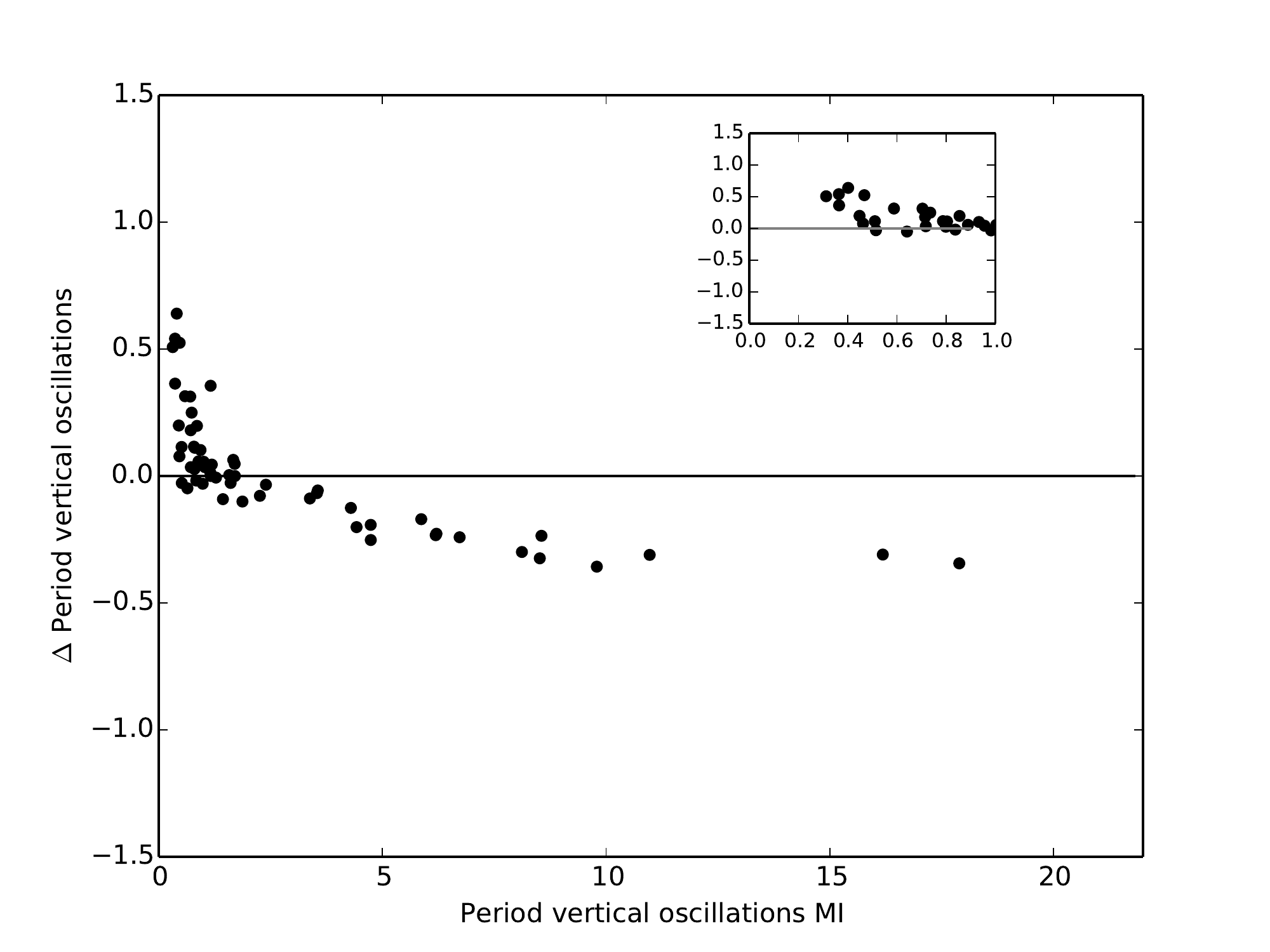}
\includegraphics[trim=0.5cm 0cm 1.cm 0cm, clip=true, width=0.45\textwidth]{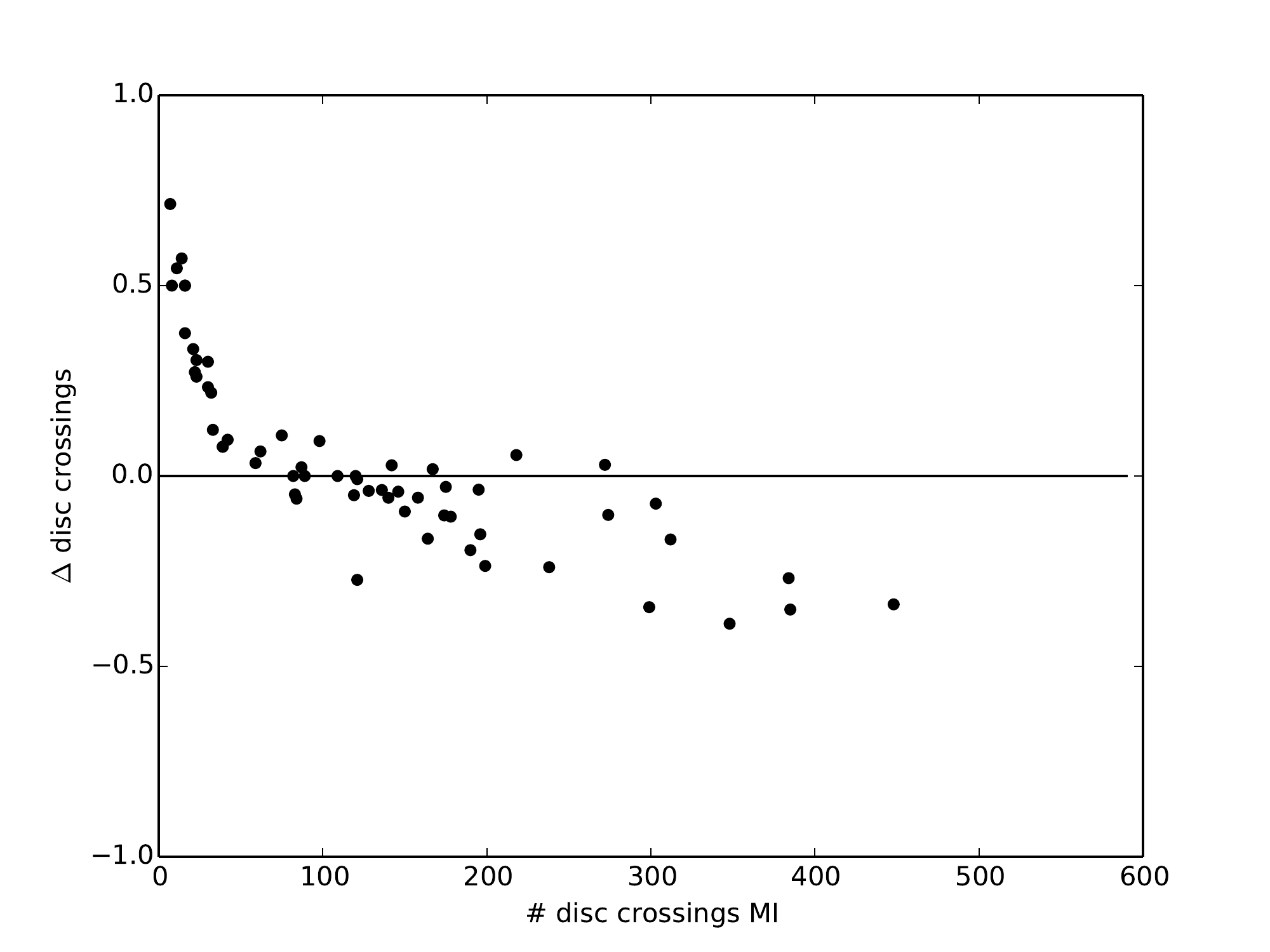}

\caption{Comparison of the orbital characteristics of Galactic globular clusters  integrated in Model I versus Model II. From top-left to bottom-right: Pericentres $R_{min}$, apocentres $R_{max}$, maximum vertical distance from the plane $z_{max}$,  period of vertical oscillations, and 3D eccentricities are given.  The insets in some of the plots show a zoom in the inner regions.  In all the plots, the y-axis shows the difference between Model II and Model I, relative to Model I. Distances are in kpc, time in units of $10^8$ yr. }\label{allGCcomp2}
\end{figure*}

The vast majority of the stars in the \citet{adibek12} sample are thin and thick disc stars currently at the solar vicinity. To sample also the halo population, and the bulge or thick disc, we integrated all the orbits of Galactic globular clusters  in the main catalogue of Casetti-Dinescu, which contains 59  clusters. As for the stars, also the orbits of those clusters were integrated for 7 Gyr in the two thin and thick disc potentials (Model I and Model II) and in the \citet{allen91} potential. Figure~\ref{allGCcomp} shows the comparison in the derived orbital characteristics of the clusters in the three models (see Tables~\ref{table1_orbpar}, \ref{table2_orbpar}, and \ref{table3_orbpar} for the corresponding values). Consistent with the evidence found for stars, the outermost clusters -- those which, in the A\&S model, reach distances beyond 20~kpc from the Galactic centre such as NGC1851, NGC3201, NGC4590,  NGC5024, NGC5466,  NGC 5904, NGC6205, NGC6934, NGC7006, NGC7078, NGC7089, and Pal12 -- have apocentres up to $\sim$30\% smaller in Model II than in A\&S (see Fig.~\ref{outerclusters}), while no significant difference is found for clusters with $R_{max}$ below 20~kpc. Orbits in Model II are less elongated radially, but also vertically, as found when comparing the maximal vertical oscillations above and below the plane: the value of $z_{max}$ can be reduced by as much as 30\% for Model II compared to Model A\&S. Among the clusters which show  $z_{max}$ significantly smaller in Model II than in Model A\&S there are:  NGC5024, NC5466, NGC5904, NGC6934, NGC7006, and NGC7089.  It is also interesting to note the presence of some clusters, like NGC 6121, NGC 6388, and NGC 6441, NGC6626 and NGC6779,  which have a $z_{max}\le 2~$kpc in Model II, but greater than 2~kpc in Model A\&S. Their orbits are shown in Fig.~\ref{bulgeclusters}. Some of them, like NGC6388 and NGC6441, are bulge clusters confined in the innermost few kpcs of the Galaxy, where they are perturbed and scattered by the central spheroid  in Model A\&S. When the central spheroid is not present, as in the case of Model II, their vertical excursions are reduced and their orbits become more disc-like. 
Finally, differences in the orbital properties ($R_{max}, z_{max}, ..$) are reflected also in the number of disc crossings and in the related period of vertical oscillations that clusters experience over time.
As shown in the bottom panels of Fig.~\ref{allGCcomp}, while Model I predicts values very similar to those of the A\&S model, Model II can depart significantly from the A\&S estimates: there are clusters like NGC6266, NGC6273, NGC6293, NGC6304, NGC6316, NGC6342, NGC6388, and NGC6441 for which Model II predicts a number of disc crossings up to 30\% less than those found in the A\&S model.  These clusters are thus characterized by longer periods of vertical oscillations in Model II than in Model A\&S. There are also clusters, like Pal13, NGC5466, NGC6934, and NGC7006, where the situation is the opposite, and for which Model II predicts periods of disc crossings significantly shorter than those found in Model A\&S. 
Finally, in Fig.~\ref{allGCcomp2} we show the comparison between the orbital parameters of Galactic globular clusters in Models I and II. As in Fig.~\ref{allstarscomp2}, for each quantity (pericentre, apocentre, maximum height above the plane), we have quantified the differences found in the two models with the quantity $\Delta$, defined as the difference between the value attained by this quantity in Model II and the corresponding value for Model I, normalized to Model I.  The largest differences between the two models  are found at large radii and large heights above the plane, where Model II predicts smaller values of $R_{max}$ and smaller $z_{max}$, as expected from its deeper gravitational potential. Lower values of $z_{max}$ are also reflected in shorter periods of vertical oscillations for these clusters, and consequently a larger number  of disc crossings. Halo cluster orbits thus tend to be less extended both radially and vertically in Model II than in Model I, and consequently suffer -- over a fixed time interval - more frequent disc crossings.

\begin{figure}
\includegraphics[trim=0.cm 0cm 0.5cm 0cm, clip=true, width=0.5\textwidth]{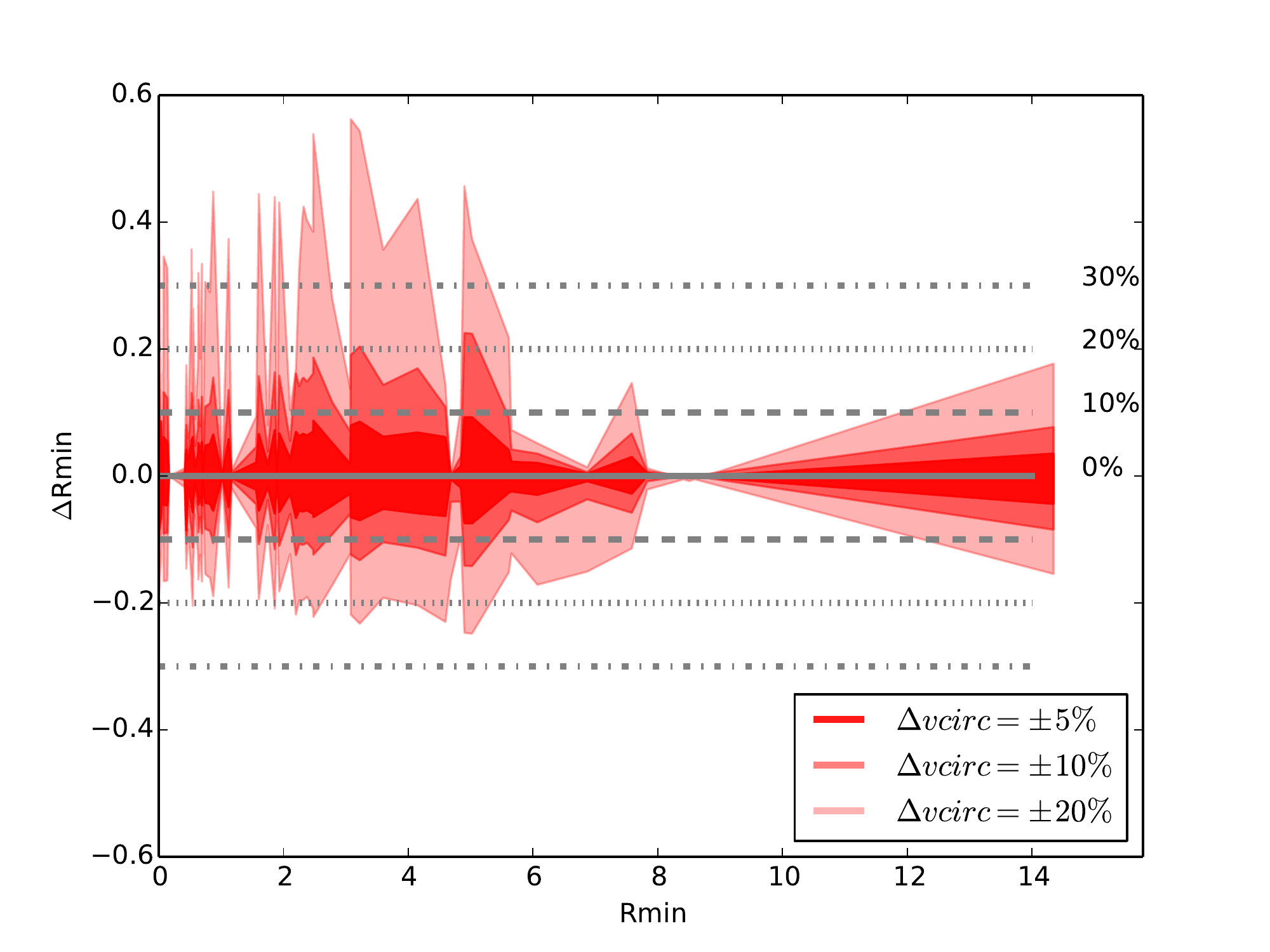}
\includegraphics[trim=0.cm 0cm 0.5cm 0cm, clip=true, width=0.5\textwidth]{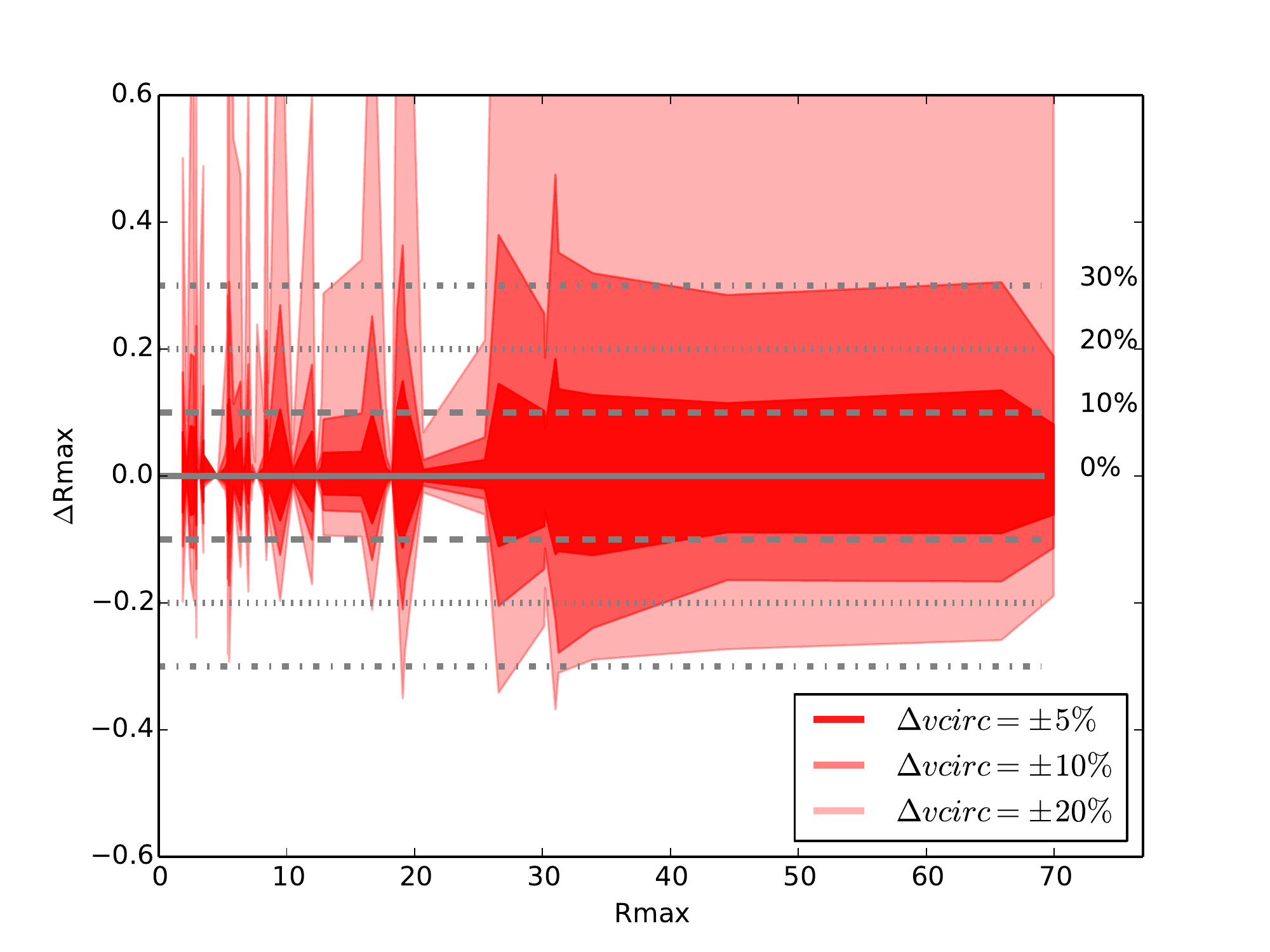}
\includegraphics[trim=0.cm 0cm 0.5cm 0cm, clip=true, width=0.5\textwidth]{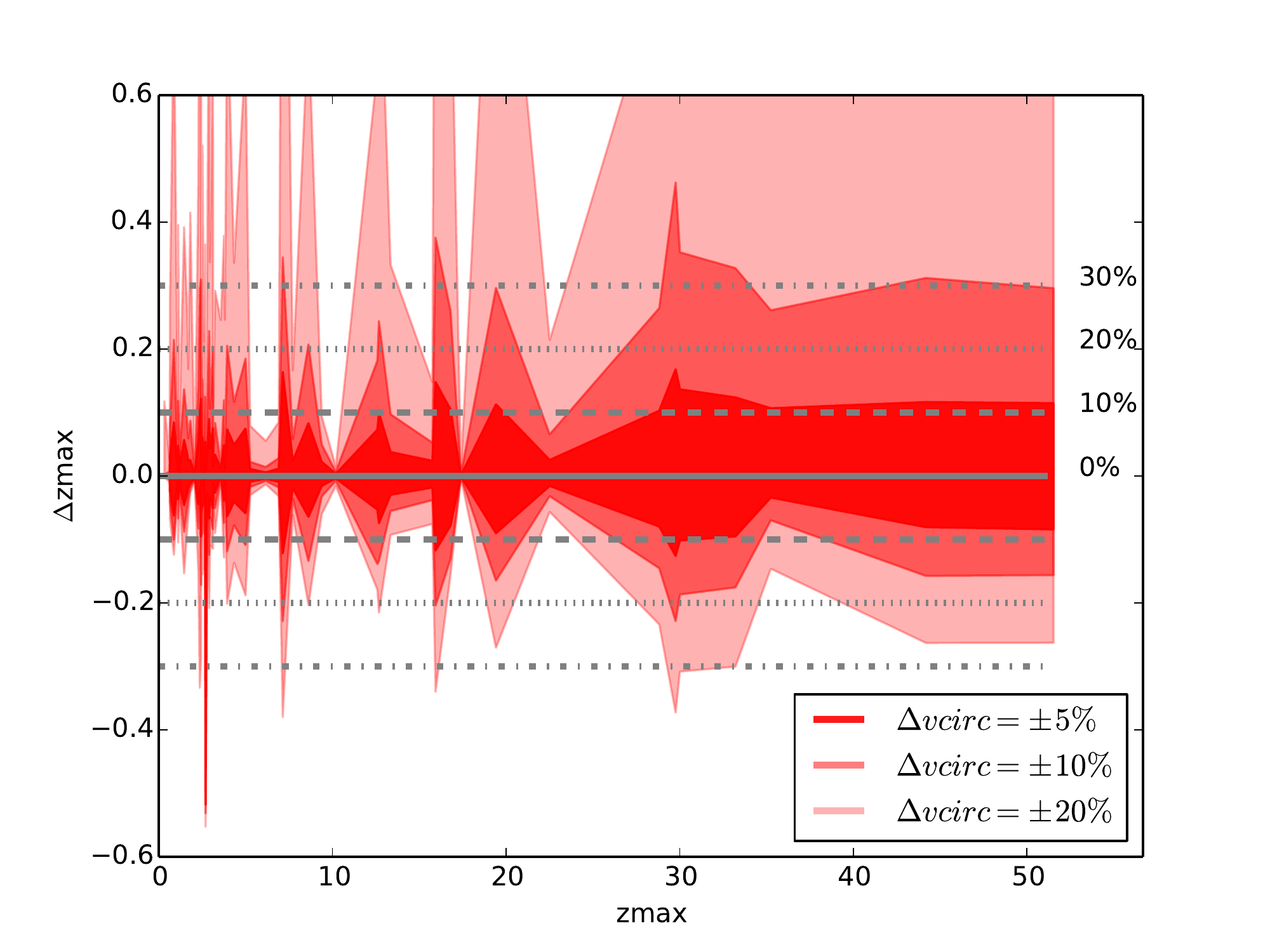}
\caption{From top to bottom: Expected relative uncertainties in the pericentres $R_{min}$, apocentres $R_{max}$ , and maximal
 heights from the plane $z_{max}$ for all Galactic globular clusters in the main catalogue of Casetti-Dinescu, assuming $\pm$ 5\%, $\pm$10\%, and $\pm$ 20\% offsets in the circular velocity of Model II. The uncertainties are defined relatively to the parameters obtained assuming the rotation curve of Model II.}\label{knowrotcurve}
\end{figure}

Overall we have seen that the uncertainties that affect our knowledge of the Galactic rotation curve are still considerable, both in the inner few kpc of the Galaxy, and at the solar radius and beyond (cf. Fig.~\ref{rotcurveMI-II}). To quantify how a given uncertainty in the rotation curve is reflected in the corresponding uncertainties in the orbital parameters of globular clusters, for each cluster in the main catalogue of Casetti-Dinescu, we have integrated its orbit for 7~Gyr in a potential similar to that of Model II, but with the dark matter halo mass changed so as to generate a rotation curve that differs from that of Model II by $\pm$5\%, $\pm$10\%, and $\pm$20\%.  In Fig.~\ref{knowrotcurve}, we report the value of the expected uncertainties in the pericentres $R_{min}$, apocentres $R_{max}$ , and maximal heights from the plane $z_{max}$ for all these clusters, assuming $\pm$5\%, $\pm$10\%, and $\pm$20\% offsets in the circular velocity of Model II. The uncertainties are defined relatively to the parameters obtained assuming the rotation curve of Model II. As can be seen, for a given offset in the velocity curve, the largest uncertainties are found in the values of $R_{max}$  and $z_{max}$: uncertainties as high as 30\% in these quantities can be reached with typical offsets in the rotation curve of $\pm$10\%. It is thus clear that these effects need to be taken into account for all integrations of stars and stellar systems in the Galaxy, because they are not negligible.

\section{Conclusions}\label{conclusions}

Recent observational results suggest the need to reconsider the mass budget of the stellar populations in the Milky Way and their relative weights. In particular there is growing evidence of a massive and centrally concentrated $\alpha-$enhanced thick disc in the Galaxy, and a limited -- or non existent -- classical bulge. \\
Motivated by these findings, we have built two new mass models for the Galaxy which include a massive and centrally concentrated thick disc, characteristics disregarded in all mass models proposed so far for orbits' computation.\\
We have shown that :
\begin{itemize}
\item these mass models can satisfy a number of observational constraints : Galactic rotation curve(s), disc(s) scale lengths and heights, baryon density at the solar vicinity,  perpendicular force $Kz$ at 1.1~kpc from the Galactic plane; 
\item the inclusion of a massive thick disc, the absence of a classical spheroid, and the presence of a more massive dark matter halo in one of the two models, all these changes have an impact on the reconstructed orbits of stars and globular clusters in the Milky Way. In particular, some Galactic globular clusters show less extended orbits in Model II than what is predicted by the \citet{allen91} model. 
\item when a classical bulge is not present,  because of the absence of its scattering effect, some inner clusters tend to have more flattened, disc-like orbits. 
\item these changes have implications for the period of vertical oscillations of clusters in the Galaxy, and their frequency of disc crossings, with possible consequences for their tidal disruption times. 
\end{itemize}

Overall, we find that the large uncertainties still affecting our current knowledge of the Galactic rotation curve have a major impact on the Galactic mass models (i.e. the presence or not of a central spheroid, dark matter content), and on the derived orbital properties of stars and star clusters in the Galaxy. We have seen that, depending on the mass model adopted, for some stars and globular clusters the differences in orbital parameters can be as high as 30\% with respect to the A\&S model and can be reached with typical offsets in the rotation curve of $\pm$10\%. These uncertainties are comparable to those found when taking into account only the current errors in the estimate of distances, proper motions, and radial velocities \citep[see for example][]{dinescu99}. Before new and more refined mass models for the Galaxy will be developed and constrained by the new data from the Gaia mission, this suggests that those uncertainties need to be considered for reconstructing the orbits of stars and stellar systems in the Galaxy. 

\section{Acknowledgments}
E.~Pouliasis warmly thanks the GEPI laboratory for  providing financial support to his Master internship. 
The authors thank M. Reid for sending a tabular form of the data reported in Fig.~\ref{vrot} and also the referee, A. Quillen, for valuable comments and suggestions on the manuscript.  The ANR (Agence Nationale de la Recherche) is acknowledged for its financial support through the MOD4Gaia project (ANR-15-CE31-0007, P. I.: P. Di Matteo).

\begin{table*}
\begin{threeparttable}
\caption{Orbital parameters for the 59 globular clusters considered in this study for the case of the Allen \& Santillan model.}
\label{table1_orbpar}  
\small
\begin{tabular}{lrrrrrrrrr}
\hline
       ID &   R$_{min}$ &   R$_{max}$ &   z$_{max}$ &   ecc3D &         E &   L$_{min}$ &   L$_{max}$ &      L$_z$ &    P$_z$ \\
       (1)&      (2)&         (3)&          (4)&         (5)&           (6)&        (7)&       (8)&            (9)&       (10) \\
\hline
NGC~104 &       5.187 &       7.648 &       3.609 &   0.162 & -1272.930 &     132.576 &     157.815 & -117.741 &  1.603 \\
NGC~288 &       2.124 &      12.158 &       9.991 &   0.582 & -1162.000 &     106.014 &     172.323 &   63.315 &  6.331 \\
NGC~362 &       0.681 &      11.330 &       7.476 &   0.884 & -1256.750 &      34.565 &     113.901 &   28.708 & 17.672 \\
NGC~1851 &       5.540 &      36.756 &      15.131 &   0.718 &  -753.223 &     235.684 &     266.943 & -215.174 &  6.107 \\
NGC~1904 &       4.043 &      21.070 &       9.565 &   0.656 &  -964.715 &     160.831 &     201.057 & -143.888 &  3.494 \\
NGC~2298 &       2.453 &      17.939 &      14.713 &   0.639 & -1019.120 &     136.613 &     205.487 &   80.379 &  3.546 \\
NGC~2808 &       2.256 &      12.425 &       4.454 &   0.668 & -1195.120 &      87.232 &     122.203 &  -84.371 &  1.672 \\
NGC~3201 &       8.672 &      25.278 &       8.412 &   0.468 &  -862.928 &     287.383 &     307.072 &  270.785 &  2.452 \\
NGC~4147 &       4.498 &      27.951 &      24.809 &   0.503 &  -822.096 &     286.731 &     334.392 &  136.844 &  4.494 \\
NGC~4372 &       2.907 &       7.579 &       2.078 &   0.435 & -1374.420 &      89.639 &     115.379 &  -86.411 &  1.747 \\
NGC~4590 &       7.940 &      32.010 &      17.076 &   0.559 &  -757.453 &     321.932 &     350.639 & -261.381 &  5.976 \\
NGC~4833 &       0.271 &       8.882 &       5.612 &   0.941 & -1411.470 &      15.802 &      85.886 &  -15.795 &  1.085 \\
NGC~5024 &       4.925 &      37.577 &      36.320 &   0.404 &  -684.709 &     457.585 &     492.269 & -138.614 &  6.769 \\
NGC~5139 &       1.224 &       6.639 &       3.175 &   0.675 & -1491.870 &      45.758 &      86.927 &   44.834 &  0.899 \\
NGC~5272 &       2.794 &      16.792 &      14.228 &   0.556 & -1030.610 &     154.190 &     217.468 &  -85.374 &  4.408 \\
NGC~5466 &       0.163 &      61.053 &      63.002 &   0.815 &  -554.531 &     262.227 &     331.610 &    6.696 &  8.837 \\
Pal~ 5 &       3.325 &      18.173 &      17.417 &   0.322 &  -946.962 &     247.600 &     295.466 &  -84.942 &  0.872 \\
NGC~5897 &       0.802 &       9.249 &       8.053 &   0.797 & -1298.390 &      46.079 &     127.600 &  -25.333 &  3.452 \\
NGC~5904 &       0.530 &      43.739 &      42.673 &   0.939 &  -690.671 &      71.967 &     193.125 &  -23.566 & 11.769 \\
NGC~5927 &       4.705 &       5.848 &       0.863 &   0.108 & -1414.670 &     109.827 &     116.720 & -108.637 &  1.745 \\
NGC~5986 &       0.054 &       5.555 &       4.024 &   0.975 & -1627.280 &       3.046 &      71.446 &   -3.032 &  8.985 \\
NGC~6093 &       0.533 &       3.210 &       3.779 &   0.238 & -1557.770 &      54.909 &      80.937 &   -9.619 &  0.763 \\
NGC~6121 &       0.203 &       6.748 &       4.330 &   0.940 & -1542.600 &      11.664 &      73.181 &   11.652 &  0.729 \\
NGC~6144 &       1.121 &       2.597 &       2.644 &   0.214 & -1661.100 &      45.480 &      63.599 &   20.961 &  0.987 \\
NGC~6171 &       1.850 &       3.422 &       2.604 &   0.248 & -1602.920 &      53.575 &      75.987 &  -37.698 &  0.739 \\
NGC~6205 &       0.672 &      22.960 &      23.113 &   0.611 &  -915.090 &     185.159 &     256.360 &   21.505 &  0.741 \\
NGC~6218 &       2.105 &       5.585 &       3.266 &   0.431 & -1477.710 &      65.028 &      99.148 &  -52.841 &  0.829 \\
NGC~6254 &       2.390 &       5.243 &       2.854 &   0.357 & -1492.170 &      68.345 &      97.288 &  -57.577 &  4.936 \\
NGC~6266 &       1.238 &       2.476 &       1.047 &   0.303 & -1815.910 &      35.297 &      44.192 &  -33.207 &  1.053 \\
NGC~6273 &       0.390 &       1.743 &       1.800 &   0.261 & -1847.930 &      30.539 &      40.956 &    8.320 &  1.003 \\
NGC~6284 &       6.291 &      10.158 &       3.569 &   0.215 & -1176.730 &     167.234 &     188.475 & -156.794 &  0.443 \\
NGC~6287 &       0.020 &       6.189 &       4.622 &   0.992 & -1580.190 &       1.155 &      75.138 &    1.150 &  0.444 \\
NGC~6293 &       0.007 &       3.340 &       2.396 &   0.988 & -1860.450 &       0.353 &      40.667 &    0.352 &  1.884 \\
NGC~6304 &       1.827 &       3.395 &       0.616 &   0.304 & -1721.480 &      48.265 &      55.085 &  -47.495 &  0.793 \\
NGC~6316 &       0.304 &       2.471 &       1.259 &   0.720 & -1922.460 &      13.866 &      24.104 &   13.462 &  0.404 \\
NGC~6333 &       0.751 &       5.352 &       2.831 &   0.715 & -1594.650 &      30.000 &      66.106 &  -29.304 &  0.453 \\
NGC~6341 &       0.440 &      10.965 &       5.409 &   0.877 & -1284.140 &      20.922 &      87.658 &  -20.530 &  0.311 \\
NGC~6342 &       0.406 &       1.669 &       1.457 &   0.433 & -1944.470 &      23.426 &      33.235 &  -10.723 &  0.693 \\
NGC~6356 &       2.757 &       7.586 &       3.114 &   0.467 & -1377.560 &      84.976 &     114.294 &  -77.650 &  1.333 \\
NGC~6362 &       1.963 &       5.892 &       2.451 &   0.470 & -1497.590 &      62.007 &      90.208 &  -59.612 &  0.342 \\
NGC~6388 &       0.207 &       3.419 &       2.029 &   0.886 & -1844.980 &      10.858 &      39.583 &   10.853 &  1.186 \\
NGC~6397 &       2.895 &       6.674 &       2.553 &   0.377 & -1413.520 &      85.406 &     111.714 &  -81.744 &  0.879 \\
NGC~6441 &       0.056 &       3.853 &       2.671 &   0.964 & -1794.370 &       3.061 &      47.847 &   -3.052 &  0.381 \\
NGC~6584 &       0.712 &      14.117 &       9.371 &   0.897 & -1162.130 &      37.564 &     121.276 &  -28.918 &  1.053 \\
NGC~6626 &       0.505 &       3.240 &       2.159 &   0.730 & -1772.420 &      22.370 &      46.540 &  -17.415 &  0.477 \\
NGC~6656 &       3.020 &      10.023 &       1.984 &   0.532 & -1280.480 &     101.766 &     122.468 &  -99.741 &  2.197 \\
NGC~6712 &       0.313 &       6.732 &       3.818 &   0.773 & -1478.880 &      13.999 &      57.884 &  -13.340 &  0.519 \\
NGC~6723 &       0.161 &       2.115 &       2.656 &   0.144 & -1681.040 &      43.599 &      56.949 &   -2.749 &  1.194 \\
NGC~6752 &       4.705 &       5.813 &       1.811 &   0.100 & -1390.930 &     108.623 &     123.510 & -103.457 &  0.860 \\
NGC~6779 &       0.532 &      13.245 &       6.948 &   0.921 & -1217.940 &      29.062 &     104.946 &   29.017 &  0.711 \\
NGC~6809 &       0.498 &       7.035 &       5.954 &   0.801 & -1435.310 &      31.803 &     101.539 &  -15.078 &  1.051 \\
NGC~6838 &       4.937 &       7.148 &       0.320 &   0.183 & -1357.780 &     126.527 &     128.180 & -126.399 &  1.313 \\
NGC~6934 &       1.567 &      48.690 &      49.933 &   0.766 &  -629.106 &     263.661 &     327.372 &   61.029 &  1.249 \\
NGC~7006 &      15.748 &     103.098 &      53.675 &   0.703 &  -349.532 &     682.987 &     698.208 & -585.158 &  0.601 \\
NGC~7078 &       5.595 &      20.430 &      15.469 &   0.425 &  -930.612 &     241.086 &     284.847 & -159.915 & 10.387 \\
NGC~7089 &       1.959 &      43.153 &      41.266 &   0.743 &  -690.596 &     243.589 &     309.707 &   72.553 & 19.008 \\
NGC~7099 &       2.319 &       7.282 &       6.226 &   0.372 & -1317.520 &      93.732 &     142.743 &   55.235 &  4.460 \\
Pal~12 &       6.946 &      24.882 &      22.545 &   0.244 &  -805.656 &     377.298 &     411.862 & -168.957 &  8.434 \\
Pal~13 &       8.564 &     100.660 &      69.227 &   0.780 &  -370.875 &     506.007 &     535.811 &  339.931 &  1.691 \\
\hline
\end{tabular}
\begin{tablenotes}
\item \textbf{Note.} -- Column 1: Identifier. Column 2: pericentre (kpc). Column 3: apocentre (kpc). Column 4: maximum vertical distance from the plane (kpc). Column 5: 3D eccentricity. Column 6: energy ($\rm{100 km^2/s^2}$). Column 7: minimum value of the total angular momentum (10 $\rm{km s^{-1} kpc^{-1}}$). Column 8: maximum value of the total angular momentum (10 $\rm{km s^{-1} kpc^{-1}}$). Column 9: z-component of the angular momentum (10 $\rm{km s^{-1} kpc^{-1}}$). Column 10: period of disc plane crossing ($\rm{10^{8}}$yr).
\end{tablenotes}
\end{threeparttable}
\end{table*}

\begin{table*}
\begin{threeparttable}
\caption{Orbital parameters for the 59 globular clusters considered in this study for the case of Model I.}
\label{table2_orbpar} 
\small 
\begin{tabular}{lrrrrrrrrr}
\hline
       ID &   R$_{min}$ &   R$_{max}$ &   z$_{max}$ &   ecc3D &         E &   L$_{min}$ &   L$_{max}$ &      L$_z$ &    P$_z$ \\
       (1)&      (2)&         (3)&          (4)&         (5)&           (6)&        (7)&       (8)&            (9)&       (10) \\
\hline
NGC~104 &       4.891 &       7.692 &       3.628 &   0.185 & -1332.350 &     133.933 &     153.830 & -118.690 &  1.577 \\
NGC~288 &       1.937 &      12.282 &      10.047 &   0.599 & -1223.470 &     104.881 &     158.051 &   62.088 &  6.190 \\
NGC~362 &       0.712 &      11.069 &       6.999 &   0.877 & -1319.620 &      35.621 &      96.039 &   27.937 & 16.184 \\
NGC~1851 &       5.509 &      36.435 &      14.964 &   0.716 &  -808.579 &     238.069 &     260.169 & -217.000 &  6.207 \\
NGC~1904 &       3.970 &      21.264 &       9.958 &   0.654 & -1018.560 &     165.331 &     195.369 & -146.195 &  3.555 \\
NGC~2298 &       2.346 &      18.025 &      15.421 &   0.617 & -1071.750 &     147.867 &     199.326 &   78.550 &  3.536 \\
NGC~2808 &       2.103 &      12.871 &       4.537 &   0.710 & -1245.760 &      86.948 &     116.684 &  -85.303 &  1.665 \\
NGC~3201 &       8.589 &      25.153 &       8.996 &   0.466 &  -916.618 &     288.707 &     303.257 &  269.657 &  2.396 \\
NGC~4147 &       4.363 &      27.768 &      24.457 &   0.510 &  -881.584 &     284.597 &     319.218 &  135.448 &  4.739 \\
NGC~4372 &       2.927 &       7.569 &       2.781 &   0.437 & -1428.840 &      93.659 &     113.223 &  -87.205 &  1.697 \\
NGC~4590 &       7.920 &      33.633 &      19.889 &   0.549 &  -814.227 &     324.902 &     347.311 & -262.008 &  5.868 \\
NGC~4833 &       0.325 &       8.430 &       4.528 &   0.896 & -1466.750 &      16.697 &      71.785 &  -16.493 &  1.155 \\
NGC~5024 &       4.915 &      37.127 &      35.282 &   0.392 &  -740.378 &     462.513 &     486.733 & -139.424 &  6.725 \\
NGC~5139 &       1.135 &       6.658 &       2.859 &   0.685 & -1552.470 &      44.936 &      75.861 &   44.072 &  0.983 \\
NGC~5272 &       2.698 &      16.745 &      14.156 &   0.560 & -1090.430 &     156.231 &     203.581 &  -86.378 &  4.295 \\
NGC~5466 &       0.143 &      58.119 &      60.195 &   0.805 &  -612.915 &     269.379 &     317.664 &    5.959 &  8.554 \\
Pal~5 &       3.151 &      18.289 &      17.382 &   0.340 & -1007.750 &     245.946 &     281.831 &  -83.848 &  0.837 \\
NGC~5897 &       0.756 &       9.096 &       7.945 &   0.759 & -1367.600 &      50.311 &     116.581 &  -25.073 &  3.379 \\
NGC~5904 &       0.520 &      42.025 &      40.894 &   0.924 &  -760.278 &      82.894 &     175.756 &  -24.049 & 10.973 \\
NGC~5927 &       4.710 &       5.336 &       0.849 &   0.061 & -1483.770 &     110.327 &     114.877 & -108.937 &  1.703 \\
NGC~5986 &       0.064 &       4.888 &       2.779 &   0.796 & -1701.880 &       3.038 &      38.618 &   -2.915 &  8.517 \\
NGC~6093 &       0.489 &       3.243 &       3.762 &   0.261 & -1642.910 &      54.218 &      81.106 &   -9.497 &  0.799 \\
NGC~6121 &       0.181 &       6.721 &       4.123 &   0.948 & -1601.050 &      10.738 &      64.650 &   10.735 &  0.588 \\
NGC~6144 &       1.124 &       2.530 &       2.624 &   0.199 & -1751.850 &      45.222 &      63.392 &   20.959 &  0.933 \\
NGC~6171 &       1.745 &       3.396 &       2.599 &   0.257 & -1686.220 &      54.039 &      75.619 &  -38.072 &  0.641 \\
NGC~6205 &       0.634 &      22.589 &      22.334 &   0.582 &  -974.760 &     198.255 &     248.186 &   20.684 &  0.714 \\
NGC~6218 &       1.965 &       5.461 &       3.262 &   0.430 & -1550.570 &      66.056 &      95.643 &  -53.464 &  0.787 \\
NGC~6254 &       2.211 &       5.188 &       2.851 &   0.371 & -1563.950 &      69.171 &      94.414 &  -58.234 &  4.738 \\
NGC~6266 &       1.488 &       2.293 &       1.037 &   0.213 & -1917.150 &      37.588 &      44.840 &  -33.464 &  1.003 \\
NGC~6273 &       0.406 &       1.802 &       1.917 &   0.312 & -1935.870 &      29.114 &      42.152 &    8.345 &  0.956 \\
NGC~6284 &       6.237 &       9.589 &       3.471 &   0.188 & -1243.010 &     166.995 &     182.423 & -155.855 &  0.448 \\
NGC~6287 &       0.043 &       4.611 &       4.375 &   0.849 & -1670.110 &      19.041 &      71.063 &    1.253 &  0.468 \\
NGC~6293 &       0.009 &       3.255 &       2.369 &   0.989 & -1955.270 &       0.483 &      41.266 &    0.482 &  1.871 \\
NGC~6304 &       1.794 &       3.252 &       0.613 &   0.289 & -1816.940 &      48.715 &      53.942 &  -47.873 &  0.854 \\
NGC~6316 &       0.322 &       2.441 &       1.191 &   0.710 & -2023.730 &      14.187 &      24.473 &   13.725 &  0.402 \\
NGC~6333 &       0.973 &       4.548 &       2.744 &   0.648 & -1688.320 &      36.686 &      66.471 &  -29.415 &  0.462 \\
NGC~6341 &       0.370 &      11.312 &       6.240 &   0.936 & -1343.570 &      21.392 &      83.631 &  -21.386 &  0.312 \\
NGC~6342 &       0.475 &       1.582 &       1.462 &   0.348 & -2036.820 &      24.152 &      33.714 &  -10.744 &  0.735 \\
NGC~6356 &       2.526 &       7.425 &       3.128 &   0.478 & -1442.250 &      84.610 &     108.904 &  -76.755 &  1.160 \\
NGC~6362 &       2.110 &       5.564 &       2.445 &   0.444 & -1568.310 &      66.727 &      88.425 &  -59.973 &  0.364 \\
NGC~6388 &       0.248 &       2.991 &       1.600 &   0.757 & -1939.700 &      11.365 &      27.125 &   10.999 &  1.164 \\
NGC~6397 &       2.962 &       6.605 &       2.696 &   0.364 & -1467.650 &      90.502 &     110.854 &  -82.661 &  0.889 \\
NGC~6441 &       0.049 &       3.684 &       2.681 &   0.968 & -1886.320 &       2.648 &      48.811 &   -2.625 &  0.365 \\
NGC~6584 &       0.727 &      13.514 &      10.381 &   0.877 & -1231.180 &      39.076 &     119.952 &  -28.371 &  1.094 \\
NGC~6626 &       0.532 &       3.228 &       2.087 &   0.717 & -1862.230 &      22.655 &      45.640 &  -17.855 &  0.515 \\
NGC~6656 &       2.926 &       9.704 &       2.211 &   0.526 & -1339.690 &     103.146 &     121.602 & -100.510 &  2.261 \\
NGC~6712 &       0.297 &       6.824 &       3.224 &   0.841 & -1558.450 &      13.962 &      56.700 &  -13.666 &  0.510 \\
NGC~6723 &       0.162 &       2.135 &       2.722 &   0.175 & -1767.850 &      43.078 &      58.134 &   -2.777 &  1.279 \\
NGC~6752 &       4.379 &       5.708 &       1.796 &   0.121 & -1456.340 &     109.446 &     121.201 & -104.199 &  0.804 \\
NGC~6779 &       0.522 &      13.136 &       6.686 &   0.923 & -1268.600 &      28.499 &      88.693 &   28.413 &  0.704 \\
NGC~6809 &       0.485 &       6.460 &       5.713 &   0.821 & -1516.190 &      30.015 &      93.036 &  -15.593 &  1.034 \\
NGC~6838 &       4.699 &       7.146 &       0.321 &   0.206 & -1407.540 &     127.431 &     128.430 & -127.302 &  1.435 \\
NGC~6934 &       1.561 &      48.070 &      47.898 &   0.749 &  -689.816 &     276.812 &     321.123 &   61.110 &  1.186 \\
NGC~7006 &      15.586 &      97.353 &      47.856 &   0.686 &  -401.967 &     683.780 &     694.245 & -583.918 &  0.717 \\
NGC~7078 &       5.583 &      20.096 &      15.358 &   0.408 &  -989.720 &     246.692 &     278.062 & -160.572 &  9.792 \\
NGC~7089 &       1.897 &      41.836 &      40.095 &   0.739 &  -751.476 &     244.909 &     292.406 &   72.134 & 17.893 \\
NGC~7099 &       2.116 &       7.330 &       6.168 &   0.387 & -1386.490 &      93.111 &     134.445 &   54.724 &  4.421 \\
Pal~12 &       6.915 &      24.138 &      21.748 &   0.223 &  -867.352 &     378.288 &     402.841 & -168.597 &  8.117 \\
Pal~13 &       8.426 &      93.733 &      67.975 &   0.768 &  -422.770 &     508.325 &     528.904 &  338.854 &  1.607 \\
\hline
\end{tabular}
\begin{tablenotes}
\item \textbf{Note.} -- Column 1: Identifier. Column 2: pericentre (kpc). Column 3: apocentre (kpc). Column 4: maximum vertical distance from the plane (kpc). Column 5: 3D eccentricity. Column 6: energy ($\rm{100 km^2/s^2}$). Column 7: minimum value of the total angular momentum (10 $\rm{km s^{-1} kpc^{-1}}$). Column 8: maximum value of the total angular momentum (10 $\rm{km s^{-1} kpc^{-1}}$). Column 9: z-component of the angular momentum (10 $\rm{km s^{-1} kpc^{-1}}$). Column 10: period of disc plane crossing ($\rm{10^{8}}$yr).
\end{tablenotes}
\end{threeparttable}
\end{table*}

\begin{table*}
\begin{threeparttable}
\caption{Orbital parameters for the 59 globular clusters considered in this study for the case of Model II.}
\label{table3_orbpar}  
\small
\begin{tabular}{lrrrrrrrrr}
\hline
       ID &   R$_{min}$ &   R$_{max}$ &   z$_{max}$ &   ecc3D &         E &   L$_{min}$ &   L$_{max}$ &      L$_z$ &    P$_z$ \\
       (1)&      (2)&         (3)&          (4)&         (5)&           (6)&        (7)&       (8)&            (9)&       (10) \\
\hline
NGC~104 &       5.023 &       7.692 &       3.575 &   0.168 & -1808.770 &     140.127 &     157.485 & -124.392 &  1.582 \\
NGC~288 &       1.605 &      12.242 &      10.195 &   0.642 & -1696.660 &      95.130 &     149.237 &   54.705 &  4.747 \\
NGC~362 &       0.636 &      10.439 &       6.909 &   0.857 & -1807.370 &      30.795 &     102.843 &   23.302 & 11.176 \\
NGC~1851 &       5.613 &      30.202 &      12.590 &   0.662 & -1247.590 &     250.749 &     268.598 & -227.976 &  4.795 \\
NGC~1904 &       4.150 &      20.677 &       9.385 &   0.634 & -1443.670 &     179.606 &     203.653 & -160.071 &  3.352 \\
NGC~2298 &       1.860 &      17.760 &      15.758 &   0.645 & -1514.920 &     142.338 &     192.026 &   67.556 &  3.298 \\
NGC~2808 &       2.377 &      12.726 &       4.338 &   0.673 & -1704.850 &      96.665 &     121.597 &  -90.903 &  1.771 \\
NGC~3201 &       8.504 &      19.075 &       7.141 &   0.353 & -1415.160 &     283.466 &     295.671 &  262.871 &  2.314 \\
NGC~4147 &       3.599 &      25.495 &      22.513 &   0.541 & -1309.490 &     269.995 &     301.467 &  127.053 &  3.544 \\
NGC~4372 &       3.079 &       7.527 &       2.623 &   0.404 & -1911.060 &      98.051 &     116.033 &  -91.985 &  1.778 \\
NGC~4590 &       7.832 &      26.565 &      15.940 &   0.464 & -1270.200 &     331.678 &     349.705 & -265.778 &  4.870 \\
NGC~4833 &       0.525 &       8.230 &       2.698 &   0.873 & -1960.220 &      20.964 &      60.670 &  -20.693 &  1.169 \\
NGC~5024 &       4.844 &      30.144 &      28.821 &   0.306 & -1151.340 &     479.688 &     498.593 & -144.298 &  5.102 \\
NGC~5139 &       1.121 &       6.666 &       2.072 &   0.690 & -2047.190 &      40.284 &      64.485 &   39.487 &  0.953 \\
NGC~5272 &       2.781 &      15.858 &      13.353 &   0.528 & -1548.740 &     166.812 &     206.635 &  -92.416 &  3.755 \\
NGC~5466 &       0.037 &      44.428 &      44.175 &   0.749 & -1046.830 &     278.261 &     318.088 &    1.531 &  6.539 \\
 Pal~5 &       2.478 &      18.249 &      17.403 &   0.427 & -1451.550 &     233.004 &     267.018 &  -77.264 &  0.822 \\
 NGC~5897 &       0.744 &       8.581 &       7.721 &   0.727 & -1868.680 &      48.027 &     116.102 &  -23.509 &  3.080 \\
 NGC~5904 &       0.582 &      31.012 &      29.767 &   0.890 & -1253.800 &      85.214 &     170.032 &  -26.953 &  7.564 \\
 NGC~5927 &       4.682 &       5.460 &       0.873 &   0.075 & -1969.940 &     112.165 &     116.624 & -110.738 &  1.702 \\
 NGC~5986 &       0.078 &       4.613 &       2.498 &   0.956 & -2200.360 &       2.575 &      59.374 &   -2.210 &  5.755 \\
 NGC~6093 &       0.546 &       2.945 &       3.808 &   0.232 & -2144.660 &      51.037 &      76.851 &   -8.764 &  0.821 \\
 NGC~6121 &       0.139 &       6.589 &       0.629 &   0.935 & -2090.420 &       5.245 &      29.999 &    5.222 &  0.773 \\
 NGC~6144 &       1.168 &       2.903 &       2.950 &   0.332 & -2226.820 &      38.681 &      66.707 &   20.944 &  1.029 \\
 NGC~6171 &       2.199 &       3.242 &       2.596 &   0.142 & -2165.130 &      57.180 &      76.762 &  -40.325 &  0.610 \\
 NGC~6205 &       0.471 &      19.256 &      19.419 &   0.526 & -1448.100 &     204.645 &     246.046 &   15.750 &  0.843 \\
 NGC~6218 &       2.251 &       5.358 &       3.247 &   0.359 & -2038.320 &      70.408 &      98.626 &  -57.212 &  0.878 \\
 NGC~6254 &       2.475 &       5.183 &       2.776 &   0.319 & -2051.410 &      72.717 &      96.480 &  -62.183 &  3.826 \\
 NGC~6266 &       1.897 &       2.492 &       1.119 &   0.144 & -2358.080 &      38.686 &      48.538 &  -35.011 &  1.060 \\
 NGC~6273 &       0.414 &       2.937 &       2.520 &   0.728 & -2310.850 &      12.909 &      55.015 &    8.490 &  0.998 \\
 NGC~6284 &       6.071 &       8.412 &       3.123 &   0.136 & -1755.230 &     161.512 &     174.654 & -150.210 &  0.537 \\
 NGC~6287 &       0.065 &       5.410 &       3.937 &   0.971 & -2100.750 &       2.623 &      77.310 &    1.878 &  0.713 \\
 NGC~6293 &       0.062 &       2.758 &       2.281 &   0.954 & -2349.350 &       1.845 &      49.557 &    1.258 &  1.682 \\
 NGC~6304 &       2.107 &       3.492 &       0.648 &   0.248 & -2282.690 &      51.010 &      56.515 &  -50.145 &  1.023 \\
 NGC~6316 &       0.821 &       2.197 &       1.109 &   0.468 & -2473.620 &      17.384 &      33.453 &   15.303 &  0.659 \\
 NGC~6333 &       1.020 &       5.505 &       2.905 &   0.668 & -2101.290 &      35.122 &      74.916 &  -30.080 &  0.498 \\
 NGC~6341 &       0.693 &      10.475 &       5.266 &   0.865 & -1816.880 &      30.635 &      90.249 &  -26.534 &  0.471 \\
NGC~ 6342 &       0.702 &       1.885 &       1.697 &   0.478 & -2438.130 &      16.007 &      37.305 &  -10.876 &  0.919 \\
 NGC~6356 &       2.323 &       7.308 &       3.137 &   0.495 & -1943.110 &      78.837 &     104.605 &  -71.373 &  1.572 \\
 NGC~6362 &       2.304 &       5.462 &       2.452 &   0.385 & -2057.200 &      69.209 &      91.612 &  -62.145 &  0.561 \\
 NGC~6388 &       0.527 &       2.834 &       1.202 &   0.690 & -2419.160 &      13.047 &      36.096 &   11.877 &  1.163 \\
 NGC~6397 &       3.224 &       6.582 &       2.673 &   0.320 & -1947.910 &      96.232 &     115.087 &  -88.175 &  0.940 \\
 NGC~6441 &       0.002 &       3.495 &       1.208 &   0.998 & -2373.870 &       0.065 &      33.419 &   -0.061 &  0.498 \\
 NGC~6584 &       0.664 &      11.983 &       8.622 &   0.860 & -1738.450 &      35.666 &     112.482 &  -25.081 &  1.133 \\
 NGC~6626 &       0.874 &       3.207 &       1.460 &   0.571 & -2346.390 &      22.841 &      46.235 &  -20.501 &  0.500 \\
 NGC~6656 &       3.073 &       9.491 &       2.362 &   0.501 & -1816.360 &     107.897 &     119.216 & -105.132 &  2.084 \\
 NGC~6712 &       0.442 &       7.004 &       2.434 &   0.881 & -2035.480 &      16.616 &      57.461 &  -15.629 &  0.569 \\
 NGC~6723 &       0.164 &       2.931 &       3.756 &   0.616 & -2222.450 &      21.842 &      67.432 &   -2.944 &  1.271 \\
 NGC~6752 &       4.597 &       5.851 &       1.818 &   0.109 & -1935.250 &     114.194 &     124.696 & -108.665 &  0.893 \\
 NGC~6779 &       0.551 &      12.870 &       1.646 &   0.917 & -1734.980 &      24.980 &      56.285 &   24.776 &  0.924 \\
 NGC~6809 &       0.630 &       6.396 &       5.000 &   0.765 & -2011.820 &      28.264 &      92.275 &  -18.686 &  1.070 \\
 NGC~6838 &       4.900 &       7.145 &       0.321 &   0.186 & -1882.290 &     132.866 &     133.747 & -132.732 &  1.304 \\
 NGC~6934 &       1.562 &      33.946 &      33.209 &   0.657 & -1179.890 &     278.197 &     314.778 &   61.597 &  1.239 \\
 NGC~7006 &      14.346 &      69.946 &      35.247 &   0.613 &  -762.341 &     681.637 &     690.413 & -576.457 &  0.742 \\
 NGC~7078 &       5.655 &      16.681 &      12.678 &   0.327 & -1470.520 &     250.219 &     275.201 & -164.525 &  6.295 \\
 NGC~7089 &       1.749 &      31.256 &      30.015 &   0.678 & -1228.060 &     239.586 &     280.374 &   69.611 & 11.736 \\
 NGC~7099 &       1.932 &       7.264 &       6.163 &   0.416 & -1889.240 &      88.410 &     129.727 &   51.653 &  3.531 \\
Pal~12 &       6.870 &      18.631 &      16.803 &   0.100 & -1324.790 &     374.303 &     393.486 & -166.429 &  5.687 \\
Pal~13 &       7.584 &      65.879 &      51.527 &   0.709 &  -796.896 &     508.846 &     525.906 &  332.374 &  1.563 \\
\hline
\end{tabular}
\begin{tablenotes}
\item \textbf{Note.} -- Column 1: Identifier. Column 2: pericentre (kpc). Column 3: apocentre (kpc). Column 4: maximum vertical distance from the plane (kpc). Column 5: 3D eccentricity. Column 6: energy ($\rm{100 km^2/s^2}$). Column 7: minimum value of the total angular momentum (10 $\rm{km s^{-1} kpc^{-1}}$). Column 8: maximum value of the total angular momentum (10 $\rm{km s^{-1} kpc^{-1}}$). Column 9: z-component of the angular momentum (10 $\rm{km s^{-1} kpc^{-1}}$). Column 10: period of disc plane crossing ($\rm{10^{8}}$yr).
\end{tablenotes}
\end{threeparttable}
\end{table*}

\end{document}